\documentclass[12pt]{iopart}
\usepackage{iopams}
\expandafter\let\csname equation*\endcsname\relax
\expandafter\let\csname endequation*\endcsname\relax
\usepackage{amsmath}
\usepackage{graphicx}
\usepackage{bm}
\allowdisplaybreaks[1]
\usepackage{verbatim}
\usepackage{xcolor}
\usepackage{units}
\usepackage{amsbsy}
\usepackage{esint}
\usepackage{perpage}
\MakePerPage{footnote}
\usepackage{natbib}

\begin{document}

\title[Amplitude limits and nonlinear damping of shear-Alfv\'en waves]{Amplitude limits and nonlinear damping of shear-Alfv\'en waves in high-beta low-collisionality plasmas}

\author{J.~Squire}
\address{Theoretical Astrophysics, 350-17, California Institute of Technology, Pasadena, CA 91125, USA}
\address{Walter Burke Institute for Theoretical Physics, Pasadena, CA 91125, USA}
\author{A.~A.~Schekochihin}
\address{The Rudolf Peierls Centre for Theoretical Physics, University of Oxford, 1 Keble Road, Oxford, OX1 3NP, UK}
\address{Merton College, Oxford OX1 4JD, UK}
\author{E.~Quataert}
\address{Astronomy Department and Theoretical Astrophysics Center, University of California, Berkeley, CA 94720, USA}

\begin{abstract}
This work, which extends Squire et al.~[ApJL, 830 L25 (2016)], explores the effect
of self-generated pressure anisotropy on linearly polarized shear-Alfv\'en fluctuations in low-collisionality plasmas.
Such anisotropies lead to stringent limits on the amplitude of magnetic perturbations in high-$\beta$ plasmas, above which a fluctuation can destabilize itself through the parallel firehose instability. This causes the wave frequency to approach zero, ``interrupting'' the wave and stopping its oscillation. These effects are explored in detail in the collisionless and weakly collisional ``Braginskii'' regime, for both standing and traveling waves. The focus is on simplified models in one dimension, on scales much larger than the ion gyroradius. The effect has 
interesting implications for the physics of magnetized turbulence in the 
high-$\beta$ conditions that are prevalent in many astrophysical plasmas.

\end{abstract}

\vspace{2pc}

\submitto{\NJP}

\section{Introduction}\label{sec: intro}

In this paper, we derive and discuss stringent nonlinear  limits on the amplitude of shear-Alfv\'en (SA)
fluctuations
  in weakly collisional plasmas.  The result, which was first presented in \citet{Squire:2016aw}, is that collisionless  linearly polarized SA waves -- that is, low-frequency incompressible oscillations 
  of magnetic field ($\delta B_{\perp}$) and velocity ($ u_{\perp}$)  perpendicular to a background field $B_{0}$ -- 
cannot oscillate when
\begin{equation}
\frac{\delta B_{\perp}}{B_{0}} \gtrsim \beta^{\,-1/2},\label{eq: first basic limit}\end{equation}
where $\beta \equiv 8\pi p_{0}/B^{2}$ is the ratio of thermal to magnetic pressure.
Above this limit (or a related limit \eqref{eq: brag basic limit}
in the weakly collisional regime), standing-wave fluctuations are ``interrupted'' before 
even a quarter oscillation, while traveling waves are heavily nonlinearly damped. 
In both cases, the magnetic field rapidly forms a sequence of  zig-zags -- piecewise straight field line segments
 with zero magnetic tension -- and evolves at later times with the magnetic energy far in excess of the kinetic energy (i.e., effectively in a near-force-free state).

What is the cause of such dramatic nonlinear behavior, even in regimes ($\delta B_{\perp}/B_{0}\ll1$ for $\beta\gg 1$) where linear physics might appear to be applicable?
As we now explain, the effect depends on the development of \emph{pressure anisotropy} -- i.e., a pressure
tensor that differs in the directions perpendicular and parallel to the magnetic field. In a magnetized plasma in which the ion gyro-frequency $\Omega_{i}$ is much larger than the collision frequency $\nu_{c}$, a decreasing (in time) magnetic field leads
to a decreasing pressure perpendicular to the magnetic field ($p_{\perp}$), while the parallel pressure ($p_{\parallel}$) increases. Such behavior originates in part from conservation of the particle's first magnetic moment
$\mu = m v_{\perp}^{2}/2B$, which suggests that $p_{\perp}/B$ should be conserved as $B$ changes in a collisionless plasma. This anisotropy, $\Delta p \equiv p_{\perp}-p_{\parallel}<0$, provides an additional stress in the momentum equation that can neutralize the restoring effects of magnetic tension, even destabilizing the  SA wave and triggering the \emph{parallel firehose instability} if $\Delta p  < - {B^{2}}/{4\pi}$
\citep{Rosenbluth:1956,Chandrasekhar:1958,Parker:1958,Schekochihin:2010bv}. 

Consider the ensuing dynamics if 
we start with $\Delta p =0$, but with a field that, in the process
of decreasing due to the Lorentz force,  \emph{generates a pressure anisotropy that would be sufficient to 
destabilize the wave}. This is a nonlinear effect not captured in linear models of SA waves.
As $\Delta p$ approaches the firehose limit, the magnetic tension disappears and the Alfv\'en frequency approaches 
zero, ``interrupting'' the development of the wave.  Because the wave perturbs the field magnitude 
by  $\delta B_{\perp}^{2}$, an amplitude 
 $\delta B_{\perp} /B_{0} \gtrsim \beta^{\,-1/2}$ is sufficient
to generate such a $\Delta p$ in a collisionless plasma. 
As the field decrease is interrupted at the firehose stability boundary, the plasma self-organizes to 
prevent  further changes in field strength, leading to the nullification of the Lorentz force through
the development of piecewise-straight (and therefore, tension-less) field-line structures.
In addition, as this process proceeds, there is a net transfer of the mechanical energy  of the wave to the plasma thermal energy due to ``pressure-anisotropy heating,''  which   occurs because of spatial  correlations between the wave's self-generated pressure anisotropy and $dB/dt$.

A similar effect also occurs in the weakly collisional ``Braginskii'' regime  \citep{Braginskii:1965vl}. Here, 
collisions act to balance the anisotropy generation and SA waves cannot oscillate if [cf. Eq.~\eqref{eq: first basic limit}] 
\begin{equation}
\frac{\delta B_{\perp}}{B_{0}} \gtrsim \sqrt{\frac{\nu_{c}}{\omega_{A}}}\beta^{\,-1/2},\label{eq: brag basic limit}\end{equation}
where $\omega_{A}$ is the wave frequency and $\nu_{c}$ the ion collision frequency (with $\omega_{A}\ll \nu_{c}$ required for the Braginskii equations to be valid).
In addition, because a \emph{changing} magnetic field is required to balance the collisional relaxation of $\Delta p $, an ``interrupted'' 
wave slowly decays in time until its amplitude is below the limit \eqref{eq: brag basic limit}, at which
point it can oscillate. 
Although the details of the nonlinear dynamics differ from the collisionless regime, the dynamics in both regimes share some generic features, in particular the strong dominance of magnetic energy over kinetic energy after a wave is  interrupted.

The results described in the previous paragraphs are of  interest 
because  the low-frequency shear-Alfv\'en wave has historically been the most robust 
plasma oscillation \citep{Cramer:2011uc}. In particular, unlike its cousins, the fast and slow waves, it is 
linearly unaltered by  kinetic  physics (except at very high $\beta \gtrsim \Omega_{i}/\omega_{A}$; \citealp{Foote:1979ee,Achterberg:1981}),
and it survives unmodified in even the simplest plasma models (e.g., incompressible magnetohydrodynamics; MHD). This includes kinetic models of plasma turbulence involving low-frequency, low-amplitude, but fully nonlinear fluctuations \citep{Schekochihin:2009eu,Kunz:2015gh}.
For these reasons, SA waves play a key theoretical role in most sub-disciplines and applications of plasma physics:  magnetized turbulence 
phenomenologies
 \citep{Goldreich:1995hq,Ng:1996ik,Boldyrev:2006ta},
 the solar wind and its interaction with  Earth \citep{Eastwood2005,Ofman:2010jg,Bruno:2013hk},
 the solar corona \citep{Marsch:2006vp}, solar and stellar interiors \citep{Gizon:2008vi}, cosmic-ray transport \citep{CosmicRays:2015}, astrophysical disks \citep{Quataert:1999gn}, and
magnetic fusion \citep{Heidbrink:2008eg}, to name a few. 

This myriad of applications has in turn led to 
intense study of the SA wave's basic properties across many plasma regimes \citep{Cramer:2011uc}. 
The most relevant to our study here are several papers noting that linearly polarized SA waves are  Landau damped nonlinearly
at the rate $\sim \! \omega_{A}\beta^{1/2}(\delta B_{\perp}/B_{0})^{2}$ at high $\beta$  \citep{Hollweg:1971kq,Lee:1973ky,Stoneham:1981,Fla:1989},
although this rate is reduced by particle trapping effects at high wave amplitudes \citep{Kulsrud:1978cr,Cesarsky:1981wt,Volk:1982vr}. This Landau damping has a similar form to the collisionless ``pressure-anisotropy damping'' that plays a key role in  some of the effects described in this work.
There have also been a wide variety of studies  
considering nonlinear effects due to parametric instabilities and compressibility (e.g., \citealp{Galeev_SOPH_63,Hollweg_JGR_71,Goldstein_APJ_78,Derby_APJ_78,Medvedev:1996ex,Medvedev:1997dk,Delzanna:2001,Matteini_GRL_10,Tenerani_JGR_13}), which have generally found large-amplitude
SA waves to be unstable to parametric decay at low $\beta$, but
with  stability increasing as $\beta $ approaches $\sim 1$ \citep{Bruno:2013hk}. 
Our study here 
 complements these previous works by showing that in  the limit $\beta\gg 1$, linearly polarized finite-amplitude SA waves in weakly  collisional plasmas can
be nonlinearly modified so strongly that they are unable to oscillate at all. Note, however, that circularly polarized 
SA fluctuations are unmodified by these effects  because their magnetic field strength is constant in time.

The role of shear-Alfv\'en waves in magnetized turbulence deserves special emphasis:
turbulence is fundamental to 
 many areas of astrophysics and geophysics and may be significantly modified by the nonlinear amplitude limit.
The  well-accepted phenomenology of  \citet{Goldreich:1995hq} suggests that strong magnetized turbulence should be understood 
in terms of nonlinear  interactions between SA wave packets, which cascade in such a way that their linear physics is of comparable 
importance to their nonlinear interactions (this is known as ``critical balance''). 
Because of the resilience of SA waves to kinetic physics, it is often assumed -- and patently true in some cases, e.g., the solar wind at $\beta \lesssim 1$ -- that Alfv\'enic cascades  survive in
collisionless plasmas \citep{Schekochihin:2009eu} even though naive estimates suggest the plasma viscosity is very large.\footnote{As recently argued by \citet{2016ApJ...831..128V} for the solar wind, large-amplitude compressive fluctuations may also play an important role in high-$\beta$ turbulence, aiding in the isotropization of the distribution function.}
The nonlinear interruption of Alfv\'enic fluctuations above the amplitude $\delta B_{\perp}/B_{0} \sim \beta^{\,-1/2}$
may thus significantly alter our understanding of turbulence in weakly collisional plasmas at high $\beta$ -- conditions
that occur, for example,  in regions of the solar wind \citep{Bale:2009de,Bruno:2013hk}, the intracluster medium (ICM)\footnote{This is the hot plasma 
that fills the space between galaxies in clusters.} \citep{Rosin:2011er,Zhuravleva:2014}, and hot astrophysical disks \citep{Balbus:1998tw,Quataert:2001vz}.
The picture described above and in what follows suggests a limit on the amplitude (in comparison to a background field)
of such turbulence, above which motions are quickly damped, leaving longer-lived magnetic
perturbations in their wake.

This paper, which extends the results of \citet{Squire:2016aw},
is organized as follows. In Sec.~\ref{sec:equations}, we present the  Landau-fluid model \citep{Snyder:1997fs} used 
throughout this work to analyze nonlinear SA wave dynamics. This model is chosen as 
the simplest extension of MHD to weakly collisional plasmas with motions 
on scales that are large  compared to the ion gyroradius. Given the model's relative simplicity in comparison
to full  Vlasov-Maxwell equations, particular focus is given to gaining
qualitative understanding of various physical effects:  the pressure anisotropy, collisions, and  heat fluxes. 
Section~\ref{sec:general description} then contains a very brief description and definition of the two main physical effects -- termed \emph{interruption} and \emph{nonlinear damping} --
that form the basis for our results. 
We then treat Braginskii ($\Omega_{i}\gg\nu_{c} \gg |\nabla \bm{u}|$)
and collisionless ($\Omega_{i}\gg |\nabla \bm{u}|\gg \nu_{c} $) SA wave dynamics  in Secs.~\ref{sec:Braginskii} and \ref{sec:collisionless}, respectively. 
Because standing waves are primarily affected by the interruption effect, whereas traveling-wave dynamics are more naturally thought of in terms of nonlinear damping, we split each of these sections and separately discuss standing and traveling waves in each case. For all cases, we derive various scalings, amplitude limits, and damping rates, and describe the physics 
qualitatively with the aid of numerical examples. In Sec.~\ref{sec:kinetic stuff}, we 
discuss the importance of kinetic physics that is not included in our model, both due to the  limitations of a 1-D domain of the Landau-fluid prescription for the heat fluxes. These considerations underscore
 the importance of future  two- and three-dimensional kinetic simulations for further study of the effect.
 For an impatient reader, the summary of key results   in Sec.~\ref{sec:conclusion} should be (mostly) comprehensible
 without reference to  the main text. 

Finally, the appendices deserve some mention here, being somewhat
separate in character and content than the main text of the paper. In these, we  
derive the nonlinear wave equations asymptotically,   both in the collisionless limit (\ref{app:asymptotics}; we also 
consider the zero-heat-flux double-adiabatic equations there), and in the Braginskii regime (\ref{app:asymptotics Brag}). 
These calculations serve two main purposes. The first is to  justify more formally many of the approximations
in the main text. In this capacity, they may help comfort a reader who is skeptical of our arguments relating, e.g., 
to heat fluxes in collisionless waves. The second purpose is to  derive explicitly various  effects that are only 
heuristically derived in the main text, e.g., the  damping rate for traveling waves. These 
calculations also provide a useful reference point for  future fully kinetic studies that could  account more formally for various 
effects not included in the Landau-fluid model.

\section{Macroscopic equations for a weakly collisional plasma}\label{sec:equations}

Throughout this work, our philosophy is to consider the simplest modifications to macroscopic\footnote{Here ``macroscopic'' refers
to scales that are large compared to the plasma microscales, i.e., to those scales that relate to the 
gyrofrequency, particle Larmor radius, plasma frequency, and Debye length.} plasma dynamics due to kinetic physics. 
We thus consider a two-species, fully ionized plasma, and assume that the pressure tensor is \emph{gyrotropic} -- i.e., invariant under rotations about the field lines --
but can develop an anisotropy, viz., a different pressure parallel and perpendicular to the magnetic field lines.  This approximation
is generally valid for motions on spatiotemporal scales much larger than those relating to ion gyromotion.  It leads to the following 
equations for the magnetic field and the first three moments of the plasma distribution function \citep{CGL:1956,Kulsrud:1980tm,Schekochihin:2010bv}:
\begin{gather} 
 \partial_{t}\rho +\nabla \cdot (\rho \bm{u} ) = 0,\label{eq:KMHD rho} \\[2ex]
\rho \left(\partial_{t}\bm{u} +  \bm{u}\cdot \nabla \bm{u} \right)= -  \nabla\left( p_{\perp} + \frac{B^{2}}{8\pi}\right)+ \nabla \cdot \left[\hat{\bm{b}} \hat{\bm{b}}\left( \Delta p + \frac{B^{2}}{4\pi}\right)\right],\label{eq:KMHD u}\\[2ex]
 \partial_{t} \bm{B} = \nabla \times (\bm{u} \times \bm{B} ) ,\label{eq:KMHD B} \\[2ex]
\partial_{t}p_{\perp} + \nabla \cdot (p_{\perp}  \bm{u}) + p_{\perp} \nabla \cdot \bm{u} + \nabla \cdot (q_{\perp}\hat{\bm{b}}) +q_{\perp} \nabla \cdot \hat{\bm{b}}    = p_{\perp} \hat{\bm{b}}\hat{\bm{b}}: \nabla \bm{u}  -  \nu_{c}\Delta p,\label{eq:KMHD pp}
 \\[2ex]
\partial_{t}p_{\parallel} + \nabla \cdot (p_{\parallel}  \bm{u}) + \nabla \cdot (q_{\parallel}\hat{\bm{b}}) -2q_{\perp} \nabla \cdot \hat{\bm{b}}     = -2 p_{\parallel} \hat{\bm{b}}\hat{\bm{b}}: \nabla \bm{u}  +2\nu_{c}\Delta p.\label{eq:KMHD pl}
\end{gather}
Here Gauss units are used, $\bm{u}$ and $\bm{B}$ are the ion flow velocity and magnetic field, $B\equiv \left| \bm{B} \right|$ and $\hat{\bm{b}}=\bm{B}/B$  denote the field strength and direction, $\rho$ is the mass density, 
$\nu_{c}$ is the ion collision frequency, $p_{\perp}$ and $p_{\parallel}$ are the components of the pressure tensor perpendicular and parallel to the magnetic field, and $q_{\perp}$ and $q_{\parallel}$ are fluxes of perpendicular and 
parallel heat in the direction parallel to the magnetic field.  
Note that  $p_{\perp}$ and $p_{\parallel}$ in Eq.~\eqref{eq:KMHD u} are summed over both particle species, while $\rho$ and $\bm{u}$ in Eqs.~\eqref{eq:KMHD rho}--\eqref{eq:KMHD B} 
are the ion density and flow velocity (although for $k \rho_{i}\ll 1$ and $m_{e}/m_{i}\ll 1$, they may equivalently be viewed as the total density and flow velocity).
The pressure equations \eqref{eq:KMHD pp} and  \eqref{eq:KMHD pl} should in principle be solved separately for each species; however, in this work we consider only the ion pressures, 
an approximation that may be formally justified by an expansion in the electron-ion mass ratio when the electrons are moderately collisional (see, e.g., Appendix~A of \citealp{Rosin:2011er}).
The double-dot notation used in Eqs.~\eqref{eq:KMHD pp}--\eqref{eq:KMHD pl} means $\hat{\bm{b}}\hat{\bm{b}}: \nabla \bm{u}\equiv \hat{b}_{i}\hat{b}_{j}\nabla_{i}u_{j}= \hat{\bm{b}}\cdot (\hat{\bm{b}}\cdot \! \nabla \bm{u} )$. Note that nonideal 
corrections to the magnetic-field evolution, which are important for motions at scales approaching $\rho_{i}$, are not included in Eq.~\eqref{eq:KMHD B} and
will be ignored throughout this work (see, e.g., \citealp{Schekochihin:2010bv}).  We also define $\Delta \equiv \Delta p/p_{0}$ with $p_{0} = 2p_{\perp}/3+ p_{\parallel}/3$ (note that $\Delta p \ll p_{0}$ for $\beta =8\pi p_{0}/B^{2}\gg 1$),  the Alfv\'en speed $v_{A} = B_{0}/\!\sqrt{4\pi\rho}$ (with $B_{0}$ a constant background field), the sound speed $c_{s}=\sqrt{p_{0}/\rho}$, parallel sound speed $c_{s\parallel}=\sqrt{p_{\parallel}/\rho}$, and denote the 
ion gyroradius and gyrofrequency $\rho_{i}$ and $\Omega_{i}$, respectively. 
Although Eqs.~\eqref{eq:KMHD rho}-\eqref{eq:KMHD pl} are derived directly from the 
Vlasov equation assuming $k \rho_{i} \ll 1$ and $\omega/\Omega_{i} \ll 1$ (where $k$ and $\omega$ are characteristic wavenumbers
and frequencies of the system), the heat fluxes $q_{\perp,\parallel}$ remain unspecified and  must be solved
for using some closure scheme  (or the full kinetic equation) as discussed below. 

\subsection{The importance of pressure anisotropy at high $\beta$}

In a changing magnetic field, the terms 
\begin{equation}
\hat{\bm{b}}\hat{\bm{b}}: \nabla \bm{u} =\frac{1}{B}\frac{dB}{dt}+\nabla \cdot \bm{u}\end{equation}
(where $d/dt$ is the Lagrangian derivative) in Eqs.~\eqref{eq:KMHD pp} and \eqref{eq:KMHD pl} locally force 
a pressure anisotropy $\Delta = \Delta p/p_{0}$. Importantly, because 
this anisotropy generation depends on $\hat{\bm{b}}$ rather than $\bm{B}$, its
dynamical influence increases  as $\beta$ increases (aside from the limiting effects of microsinstabilities; see below). Namely, the final term of Eq.~\eqref{eq:KMHD u} shows that $\Delta p$
has a strong dynamical influence (i.e., is comparable to the Lorentz force) when $\Delta p \sim B^{2}$; i.e., when $\Delta\sim \beta^{-1}$.  For $\beta>1$, the pressure anisotropy  generated by changing $B$ will generally cause a  stress that is \emph{stronger} than the Lorentz force. It is also worth noting the importance of the spatial form of $\Delta p$, which, as we shall show, can strongly influence the nonlinear dynamics. As will become clear below (Secs.~\ref{subsub: Braginskii} and \ref{subsub: noC}), this spatial variation in $\Delta p $ depends on the balance between the driving $\hat{\bm{b}}\hat{\bm{b}}: \nabla \bm{u}$ and the other terms in Eqs.~\eqref{eq:KMHD pp} and \eqref{eq:KMHD pl} (e.g., the heat fluxes or collisionality), so we should expect nonlinear wave dynamics to depend significantly on a particular physical regime.

In this work, we focus on  two such regimes for the evolution of $\Delta p$, neglecting compressibility for simplicity
in both cases [this neglect  is valid at $\beta \gg 1$, $\delta B_{\perp}/B_{0}\ll 1$; see \ref{app:initial wave} around Eq.~\eqref{eq: pressure 2 is smooth} and \ref{app:asymptotics Brag} around Eq.~\eqref{eq:Brag high coll}]. The first approximation is Braginskii MHD, which is valid in weakly collisional plasmas when $\Omega_{i}\gg\nu_{c} \gg |\nabla \bm{u}|$; the second is collisionless ($\nu_{c}=0$, or equivalently $\Omega_{i}\gg|\nabla \bm{u}| \gg \nu_{c}$), which we model using a simple 
Landau fluid (LF) closure for the heat flux.

\subsubsection{Braginskii MHD. } \label{subsub: Braginskii}When collisions dominate ($| \nabla \bm{u} |\ll \nu_{c}$), we may neglect 
$\partial_{t}p_{\perp}$ and $\partial_{t}p_{\parallel}$ in comparison 
to $\nu_{c} \Delta p$ in Eqs.~\eqref{eq:KMHD pp} and \eqref{eq:KMHD pl}. For $\beta\gtrsim 1$, 
these approximations also imply $\Delta p \ll p_{0}$, leading to
\begin{equation}
\Delta p \approx \frac{p_{0}}{\nu_{c}}\left(\hat{\bm{b}}\hat{\bm{b}}:\nabla \bm{u} - \frac{1}{3}\nabla \cdot \bm{u} \right) \approx 
\frac{p_{0}}{\nu_{c}} \frac{1}{B}\frac{dB}{dt}.\label{eq:Brag closure}
\end{equation}
We have  neglected  $q_{\perp,\parallel}$ for simplicity in deriving Eq.~\eqref{eq:Brag closure}, although 
this is only valid in the limit $\delta p_{\perp,\parallel} /p_{\perp,\parallel} \ll |\bm{u}|/c_{s}$ (where $ \delta p_{\perp,\parallel}$ 
denotes the spatial variation in $p_{\perp,\parallel}$; see \citealp{Mikhailovskii:1971,Rosin:2011er}).\footnote{For $\Delta p \ll p_{0}$, this condition is approximately equivalent to $\nu_{c}\sim c_{s}/\lambda_{\mathrm{mfp}}\gg k_{\parallel} c_{s}$ (where $\lambda_{\mathrm{mfp}}$ is the ion mean-free path).}
An expression for $\Delta p$ with heat fluxes included is derived in \ref{app:asymptotics Brag} [Eqs.~\eqref{eq: brag D4 av} and \eqref{eq: Brag D4}], where
we also
briefly discuss how the  nonlinear SA wave dynamics are modified by the resulting different spatial form of  $\Delta p$.
However, given the extra complexity of including this effect, we ignore the heat fluxes 
in the discussion of Braginskii  dynamics in Sec.~\ref{sec:Braginskii}.

\setcounter{footnote}{0}

\subsubsection{Collisionless plasma. } \label{subsub: noC}
The evolution of $\Delta$ is strongly influenced by heat fluxes when $\nu_{c} \ll | \nabla \bm{u} |$ and $\beta \gtrsim 1$. As a simple
prescription, we employ a \emph{Landau fluid} (LF) closure \citep{Snyder:1997fs,Hammett:1990,Hammett:1992,Passot:2012go},
which has been extensively used in the fusion community, and to a lesser degree for astrophysical applications \citep{Sharma:2006dh,Sharma:2007cr}.
The heat fluxes are chosen to reproduce linear Landau damping rates, namely,
\begin{gather}
q_{\perp} = -\frac{2 c_{s\parallel}^{2}}{\sqrt{2 \pi }c_{s\parallel} |k_{\parallel}|+\nu_{c}} \left[ \rho  \nabla_{\parallel} \left(\frac{p_{\perp}}{\rho}\right)  - p_{\perp}\left(1-\frac{p_{\perp}}{p_{\parallel}} \right)\frac{\nabla_{\parallel} B}{B}  \right], \label{eq:GL heat fluxes p}\\ 
q_{\parallel} = - \rho\frac{8 c_{s\parallel}^{2}}{\sqrt{8 \pi }c_{s\parallel} |k_{\parallel}|+(3\pi-8)\nu_{c}} \nabla_{\parallel} \left(\frac{p_{\parallel}}{\rho}\right),\label{eq:GL heat fluxes l}
\end{gather} 
where $\nabla_{\parallel}$ is the parallel gradient operator, while the parallel wavenumber $|k_{\parallel}|$ must
 be considered as an operator.
In the regime of interest, $\Delta p \ll p_{0}$ and $\nu_{c}=0$, with small perturbations
to the magnetic field, the dynamical effect of $q_{\perp,\parallel}$ can be easily understood. Equations~\eqref{eq:GL heat fluxes p} and \eqref{eq:GL heat fluxes l} are
\begin{equation}
q_{\parallel} \approx - \sqrt{\frac{8 }{\pi }} \rho c_{s}\frac{\nabla_{\parallel}}{| k_{\parallel}|} \left(\frac{p_{\parallel}}{\rho}\right), \: q_{\perp} \approx - \sqrt{\frac{2 }{\pi }} \rho c_{s}\frac{\nabla_{\parallel}}{| k_{\parallel}|} \left(\frac{p_{\perp}}{\rho}\right).\label{eq:simp heat fluxes}
\end{equation}
These, combined with $\hat{\bm{b}}\cdot \nabla q_{\perp,\parallel} \gg q_{\perp,\parallel} \nabla \cdot \hat{\bm{b}} $ (valid 
for small perturbations to the background field), imply that the heat-flux contributions to the pressure equations \eqref{eq:KMHD pp} and \eqref{eq:KMHD pl} simplify to 
\begin{equation}
\partial_{t} p_{\perp} \sim -\rho c_{s}|k_{\parallel}| (p_{\perp}/\rho), \quad \partial_{t} p_{\parallel}\sim -\rho c_{s}|k_{\parallel}| (p_{\parallel}/\rho).\label{eq:q effect on p} \end{equation}
These terms, which model the Landau damping of temperature perturbations,\footnote{Since  the  spatial variation in 
$\rho$ will generally be similar to that of $p_{\perp,\parallel}$, the effective damping 
is less than what it would be if the variation in $\rho$ were ignored in Eq.~\eqref{eq:q effect on p}. 
However, because the
 spatial variation in $T_{\perp,\parallel}$ is of the same order as that of $p_{\perp,\parallel}$, a  damping of the pressure alone $ - c_{s}|k_{\parallel}| p_{\perp}$ may be used for heuristic estimates. 
 A full asymptotic calculation of the relative contributions of $\rho$ and $p_{\perp,\parallel}$ is 
 given in \ref{app:asymptotics} [see Eqs.~\eqref{app: eq: p rho pert 1 1}--\eqref{app: eq: p rho pert 1 2} and \eqref{app: eq: p rho q pert 2}--\eqref{app: eq: p rho q pert 3}].}
 suppress spatial variation in $p_{\perp,\parallel}$ over the particle crossing
time $\tau_{\mathrm{damp}}\sim (| k_{\parallel}| \,c_{s})^{-1}$. 
This damping implies that if $\tau_{\mathrm{damp}}\ll |\nabla \bm{u}|^{-1}\sim \omega_{A}$, the $k_{\parallel}\neq 0$ part of $\Delta$ is suppressed 
by a factor of $\sim v_{A}/c_{s}\sim \!\beta^{\,-1/2}$ compared to its mean.\footnote{This estimate arises from the balance between the driving, on timescale $\omega_{A}^{-1}\sim (k_{\parallel}v_{A})^{-1}$, and the damping, on timescale $\tau_{\mathrm{damp}}\sim (| k_{\parallel}| \,c_{s})^{-1}$. It is derived in detail in \ref{app:asymptotics}; see Eq.~\eqref{eq: D3 explicit}.} This leads us to the simple interpretation that 
the heat fluxes \emph{spatially average} $\Delta p$, by damping $k_{\parallel}\neq 0$ components of the pressure perturbations, 
giving 
\begin{equation}
\Delta = 3\!\int \!\left< \hat{\bm{b}}  \hat{\bm{b}} :\nabla \bm{u}\right>dt \left[1  + \mathcal{O}(\beta^{\,-1/2})(\bm{x}) \right]\approx  3 \left< \ln \frac{B(t)}{B(0)} \right>,\label{eq:noC Delta}
\end{equation}
where $\langle\cdot \rangle$ denotes the spatial average.
The spatial form of the $\mathcal{O}(\beta^{\,-1/2})(\bm{x})$ term
generally follows the spatial variation  of $\hat{\bm{b}}  \hat{\bm{b}} :\nabla \bm{u}$, 
and is calculated by asymptotic expansion in various regimes in \ref{app:initial wave} and \ref{app:near interruption}
[see Eqs.~\eqref{eq: D3 explicit} and \eqref{eq: final D4}; these calculations also   justify more formally the spatial-averaging action of the heat fluxes derived heuristically 
above].

\subsection{Microinstabilities}\label{sub:microinstabilities equations}
An important limitation of Eqs.~\eqref{eq:KMHD rho}-\eqref{eq:KMHD pl}, which 
exists for both the Braginskii and LF closures, 
is their inability to  capture correctly certain plasma microinstabilities. For our purposes, 
at high $\beta$, the most important of these are the firehose and mirror instabilities.
Both of these grow fastest  on scales approaching the Larmor radius, which are explicitly 
outside the validity of Eqs.~\eqref{eq:KMHD rho}--\eqref{eq:KMHD pl}. Assuming $\Delta p \ll p_{0}$, the firehose is 
unstable if $\Delta < -2/\beta$
and comes in two flavors: the parallel firehose, which is present in fluid models 
and is the cause of SA wave interruption, and the oblique firehose  \citep{Yoon:1993of,Hellinger:2008ob}, 
which grows fastest at $k_{\perp}\neq 0$, and is not correctly captured  by  Eqs.~\eqref{eq:KMHD rho}-\eqref{eq:KMHD pl}. 
The mirror instability is unstable if $\Delta > 1/\beta$ and  grows with $k_{\perp}\gg k_{\parallel}$.
Although the linear mirror instability is contained in the LF model \citep{Snyder:1997fs}, its nonlinear evolution, which involves trapped particle dynamics \citep{Rincon:2015mi}, presumably requires a fully kinetic model.
It is worth noting that 1-D fully kinetic simulations would also not 
correctly include either the oblique firehose or mirror instabilities.

It has been common in previous literature (e.g., \citealp{Sharma:2006dh,Kunz:2012bi,SantosLima:2014cn})
to model the effect of these instabilities in fluid simulations
by applying ``hard-wall'' boundaries on $\Delta$, limiting its value by the appropriate microinstability threshold. This is motivated by the fact that  both in kinetic simulations and, it appears, in the observed solar-wind, microinstabilities act to limit the pressure anisotropy at its marginal values (see, for example, \citealp{Hellinger:2006ge,Bale:2009de,Kunz:2014kt,Servidio_Valentini_Perrone_Greco_Califano_Matthaeus_Veltri_2015}). In addition, the enormous scale separation between 
the micro- and macroscales in many astrophysical plasmas 
implies that the effect  of microscale instabilities on large-scale dynamics should be effectively instantaneous \citep{Melville:2015tt}.
Motivated by the fact that the parallel firehose instability \emph{is} 
contained in fluid models, most of the numerical results in this 
work will (where appropriate) apply a limit on  positive anisotropies (to model the action of the mirror instability),
but not on negative anisotropies. 

A more thorough discussion of microinstabities is given in Sec.~\ref{sec:kinetic stuff}, focusing in particular on the implications 
of previous kinetic results for  SA wave dynamics and 
the possible changes that might result from  a multi-dimensional fully kinetic treatment.

\subsection{Energy conservation}\label{sub:energy equations}

Energy conservation arguments are used heavily throughout the paper, forming
the basis for our estimates of traveling-wave damping rates in Secs.~\ref{sub:Brag traveling} and \ref{sub:noC traveling}. With the kinetic, magnetic, and thermal energies  defined as
 \begin{equation}
E_{K} = \int d\bm{x}  \frac{\rho u^{2}}{2},\quad E_{M} = \int d\bm{x} \frac{B^{2}}{8\pi}\,\quad E_{\mathrm{th}}=\int d\bm{x} \left( p_{\perp}+\frac{p_{\parallel}}{2} \right), \label{eq:energies}\end{equation}
Eqs.~\eqref{eq:KMHD rho}-\eqref{eq:KMHD pl} conserve the total energy:  \begin{equation}
\partial_{t}(E_{K} + E_{M} + E_{\mathrm{th}})=0.\end{equation}
A key difference compared to standard MHD arises in the evolution equation for the mechanical energy $E_{\mathrm{mech}}=E_{K}+E_{M}$:
\begin{equation}\partial_{t}E_{\mathrm{mech}}= - \partial_{t}E_{\mathrm{th}}=\partial_{t} (E_{K} + E_{M}) =  \int d\bm{x}\,p_{\parallel}\nabla \cdot \bm{u} - \int d\bm{x}\,\Delta p\frac{1}{B}\frac{dB}{dt} .\label{eq:kinetic energy}\end{equation}
The final term in this equation describes the transfer of mechanical to thermal energy due to the presence of a 
spatial correlation between $\Delta p$ and $B^{-1} dB/dt$. In the (high-$\beta$) Braginskii limit, where 
$\Delta p \propto B^{-1} dB/dt$ [see Eq.~\eqref{eq:Brag closure}],
 this term is always positive and represents a parallel viscous heating. In the collisionless case, 
 it can in principle have either sign, although we shall see that for SA waves, there is
a positive correlation between $\Delta p$ and $ B^{-1} dB/dt$ that leads to net damping of the waves.

For later reference,
 the mean pressure anisotropy evolves according to
\begin{equation}
\partial_{t }\int d\bm{x}\,\Delta p  = 2\int d\bm{x}\,p_{\parallel}\nabla \cdot \bm{u} -3 \int d\bm{x}\,q_{\perp}\nabla \cdot \hat{\bm{b}} 
+  \int d\bm{x}\,(p_{\perp}+2p_{\parallel})\frac{1}{B}\frac{dB}{dt} - 3\nu_{c}\int d\bm{x}\,\Delta p.\label{eq:mean delta p evolution}\end{equation}

\subsection{Shear-Alfv\'en wave dynamics}\label{sub:wave eq}

It is helpful 
to derive a simple wave equation that isolates the
key features of linearly polarized shear-Alfv\'en waves and the influence of the pressure 
anisotropy.
Although here the derivation is heuristic, with the aim of highlighting  the key features 
of high-$\beta$ SA dynamics, similar equations are derived 
asymptotically from the full LF system \eqref{eq:KMHD rho}--\eqref{eq:KMHD pl} in the Appendices, for 
a variety of different regimes [see Eqs.~\eqref{eq: final CGL wave eq}, \eqref{eq: final Landau1 wave eq}, \eqref{eq:Landau2 final equation}, and \eqref{eq:Brag mod coll}].

Our geometry is that of a
background field $B_{0}\hat{\bm{z}}$, with perturbations perpendicular to $\hat{\bm{z}}$ and the wavevector $\bm{k} = k_{z} \hat{\bm{z}}+ \bm{k}_{\perp}$. Since SA waves are  unmodified by $\bm{k}_{\perp} \neq 0$  (the envelope is simply modulated in the perpendicular direction) and we analyze only linear polarizations,  we   assume  $x$-directed
perturbations that depend only on $z$ and $t$, viz.,
\begin{equation}
\bm{B}=B_{0}\,\hat{\bm{z}} + \delta {B}_{\perp}(z,t)\, \hat{\bm{x}},\;  \bm{u} = \delta {u}_{\perp}(z,t)\, \hat{\bm{x}}.\end{equation}
Note that circularly polarized fluctuations are unaffected by the pressure-anisotropic physics because the field
strength remains constant in time.
Combining Eqs.~\eqref{eq:KMHD u} and \eqref{eq:KMHD B} and neglecting compressibility, the  field perturbation $\delta b = \delta B_{\perp}/B_{0}$ satisfies
\begin{equation}
\frac{\partial^{2}}{\partial t^{2}}\delta b = v_{A}^{2}\frac{\partial^{2}}{\partial z^{2}}\left[  \delta b +  \frac{\delta b}{1+\delta b^{2}}\frac{\beta  \Delta(z)}{2} \right], \label{eq:waves}
\end{equation}
where $\Delta$ is given by Eq.~\eqref{eq:noC Delta} (collisionless closure) or Eq.~\eqref{eq:Brag closure} (Braginskii closure).
In the absence of a background pressure anisotropy, Eq.~\eqref{eq:waves} illustrates that linear long-wavelength
SA fluctuations are unmodified by kinetic effects. Similarly, fixing $\Delta$ and linearizing in $\delta b$,
the parallel firehose instability 
emerges because the coefficient of $\partial_{z}^{2}\delta b $ is 
negative for $\beta \Delta/2 < -1$.

In the following sections, we shall  treat standing and traveling waves separately. While these differ only in their 
initial conditions, they can display rather different nonlinear dynamics.
In the context of Eq.~\eqref{eq:waves}, a standing wave has initial conditions in
either $\delta b$ or $u_{x}/v_{A}$, viz., \begin{equation}
\delta b(t=0)=-\delta b_{0}\cos(k_{z} z),\:u_{x}(t=0)=0,\label{eq: standing b ICs}\end{equation}
or \begin{equation}
\delta b(t=0)=0,\:u_{x}(t=0)/v_{A}=\delta b_{0}\sin(k_{z} z);\label{eq: standing u ICs}\end{equation}
a traveling wave involves initial conditions in both $\delta b$ and $u_{x}/v_{A}$, viz., 
\begin{equation}
\delta b(t=0)=-\delta b_{0}\cos(k_{z} z),\:u_{x}(t=0)/v_{A}=\delta b_{0}\cos(k_{z} z)\label{eq: traveling ICs}\end{equation}
 (for a wave traveling from left to  right).

\subsection{Numerical method}
For all numerical examples, both of the Landau fluid equations and of various reduced equations (in the Appendices), 
we use a simple Fourier pseudospectral numerical method on a periodic domain. Standard  $3/2$ dealiasing is used, along with a $k^{6}$
hyperviscous diffusion operator in all variables, which is tuned so as to damp fluctuations at scales just above the grid scale. {This 
is necessary with Fourier methods because there is little energy dissipation otherwise, and the energy can be spuriously reflected back from high-$k$ into lower-$k$ modes.}
The only further approximation used in solving  Eqs.~\eqref{eq:KMHD rho}--\eqref{eq:GL heat fluxes l}
is the identification of $|k_{\parallel}|$ in Eqs.~\eqref{eq:GL heat fluxes p}--\eqref{eq:GL heat fluxes l} with $|k_{z}|$ (the $|k_{\parallel}|$ operator is nondiagonal in both Fourier and real space and thus somewhat expensive to evaluate).
While this approximation is truly valid only for $\delta b \ll 1$,  various tests have shown that the exact form of the heat fluxes 
makes little difference; for example, the method of \citet{Sharma:2006dh}, which sets $|k_{\parallel}|=k_{L}$ with $k_{L }$ a parameter, 
does not qualitatively modify the solutions presented here.
Results shown in the figures throughout the text were obtained at a resolution $N_{z} =512$, but we see little modification of results at higher or lower resolutions.

\section{Wave interruption and damping through pressure anisotropy}\label{sec:general description}

In this section, we  explain the two 
key mechanisms that can lead to  strongly  nonlinear
behavior of SA waves in high-$\beta$ regimes. These are: (1) the nullification of the 
wave's restoring force (the Lorentz force) through the self-generated pressure
anisotropy, which we term \emph{interruption}, and (2) the channeling of wave energy into thermal 
energy due to spatial correlation of $\Delta$ and $dB/dt$, which we term \emph{nonlinear damping}.

\subsection{Interruption}
It is immediately clear from Eq.~\eqref{eq:waves} that any time $\Delta(z)$ approaches $-2/\beta$,
the solutions to Eq.~\eqref{eq:waves} are fundamentally altered because the  restoring force of the SA wave 
 disappears (i.e., the coefficient of $\partial_{z}^{2}$  approaches zero). We 
 term this effect ``wave interruption,'' because the oscillation halts when this occurs.

 In the Braginskii  limit, with $\Delta$ given by Eq.~\eqref{eq:Brag closure}, the wave is thus interrupted when 
 \begin{equation}
\nu_{c}^{-1} \frac{1}{B}\frac{dB}{dt} \sim -\frac{2}{\beta}.\label{eq: brag interruption condition}\end{equation}
Here $B = B_{0}(1+\delta b^{2})^{1/2}$, which depends
 only on  the current value of the field and its rate of change. 
 
 By contrast, in  the collisionless
 limit, with $\Delta$ given by Eq.~\eqref{eq:noC Delta}, the interruption occurs if a wave evolves 
 so that  \begin{equation}
3 \left<\ln \frac{B(t)}{B(0)} \right>  = -\frac{2}{\beta},\label{eq: noC interruption condition} \end{equation}
 which is interesting for its explicit dependence
 on the initial conditions. Our derivations of amplitude limits in the following sections are 
 simply applications of Eqs.~\eqref{eq: brag interruption condition} and \eqref{eq: noC interruption condition}.
 
 \subsection{Nonlinear damping}
 In the presence of a positive correlation between $\Delta p $ and $B^{-1}dB/dt$, the mechanical energy of the wave 
 is converted to thermal energy at the rate [see Eq.~\eqref{eq:kinetic energy}]
 \begin{equation}
\partial_{t} E_{\mathrm{mech}} = -\int d\bm{x}\,\Delta p\frac{1}{B}\frac{dB}{dt} = -\partial_{t} E_{\mathrm{th}}. \end{equation}
We term this effect ``nonlinear damping'' because the fact that the $B$ perturbation is proportional to $\delta b^{2}$ in a SA wave implies that
the damping rate also scales with $\delta b^{2}$.

In the Braginskii limit [Eq.~\eqref{eq:Brag closure}], $\Delta \propto \nu_{c}^{-1}B^{-1}dB/dt$ and the energy damping rate is, therefore, 
\begin{equation}
\partial_{t} E_{\mathrm{mech}} \sim -\nu_{c}^{-1}\int d\bm{x}\,p_{0}\left(\frac{1}{B}\frac{dB}{dt}\right)^{2}, \end{equation}
which is simply the parallel viscous damping. 

In the collisionless case, there is no fundamental requirement that $\Delta p $ and 
$B^{-1}dB/dt$ have a positive spatial correlation. Nonetheless, given that  $\Delta p $ is driven by $B^{-1}dB/dt$ [see Eqs.~\eqref{eq:KMHD pp} and \eqref{eq:KMHD pl}], one might intuitively expect such a correlation for SA waves, and 
the calculations in Sec.~\ref{sub:noC traveling} and \ref{app:initial wave} show that this is indeed the case.  Note, however, that its numerical value, and thus the wave damping rate, depends on the
effect of the heat fluxes in smoothing $\Delta p$ [this is the $\mathcal{O}(\beta^{-1/2})(\bm{x})$ part in Eq.~\eqref{eq:noC Delta}].
In the collisionless limit, there is also Landau damping of a nonlinear SA wave due to the spatiotemporal variation of the magnetic pressure \citep{Hollweg:1971kq,Lee:1973ky}, which 
turns out to  cause wave damping at a rate similar to the pressure anisotropy damping (neglecting particle trapping effects;  \citealp{Kulsrud:1978cr}).


\section{Braginskii MHD -- the weakly collisional regime}\label{sec:Braginskii}

In this section we work out the behavior of SA waves in the Braginskii limit. As discussed
in Sec.~\ref{sec:equations}, Braginskii dynamics  differ
significantly from fully  collisionless dynamics because the pressure anisotropy is 
determined by the current value of $\partial_{t} B$, rather than the time history of the magnetic field.

\subsection{Standing waves}\label{sub:Brag standing}

\begin{figure}
\begin{center}
\includegraphics[width=0.5\columnwidth]{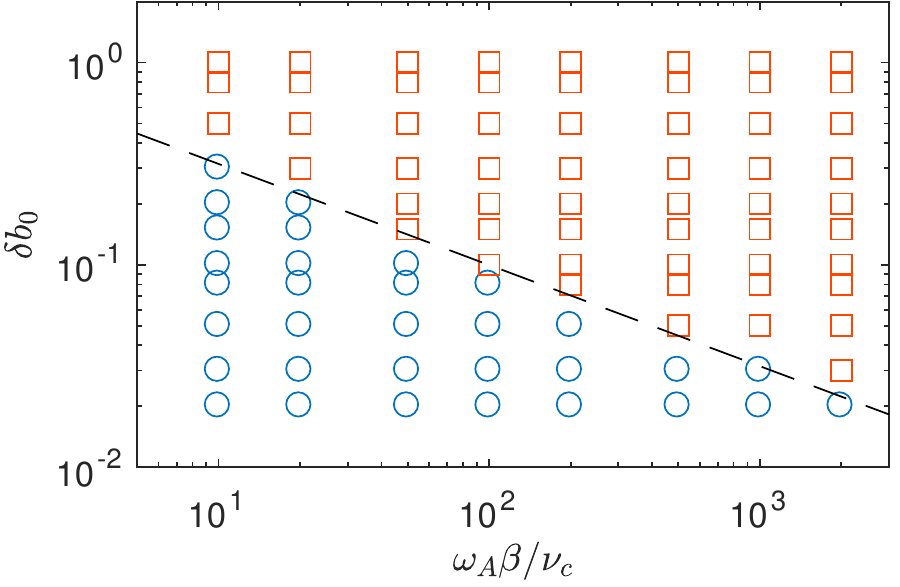}
\caption{Numerical confirmation of the scaling \eqref{eq:Brag standing cond}. Each square represents a numerical solution of the SA wave equation \eqref{eq:waves} with the Braginskii closure \eqref{eq:Brag closure}, starting from a sinusoidal magnetic perturbation [initial conditions \eqref{eq: standing b ICs}], with amplitude $\delta b_{0}$ and some chosen $\omega_{A} \beta / \nu_{c}$ (see Fig.~\ref{fig:Brag standing}). 
A red square indicates that an initial  perturbation was interrupted before a half cycle (as in Fig.~\ref{fig:Brag standing}), while a blue circle indicates that the perturbation flipped polarity without interruption. The dashed line 
is $\delta b_{0} = 2.5(\omega_{A}\beta/\nu_{c})^{\,-1/2}$.
Note that in the incompressible limit, SA wave dynamics are determined entirely by $\delta b_{0}$ and the ratio $\omega_{A} \beta / \nu_{c}$, because the $\nu_{c}^{-1}$ factor in $\Delta$ in Eq.~\eqref{eq:Brag closure} multiplies $\beta/2$ in Eq.~\eqref{eq:waves}.}
\label{fig: Brag on off}
\end{center}
\end{figure}

Starting from a finite-amplitude magnetic perturbation, a Braginskii standing SA fluctuation  will be significantly modified 
(interrupted)
if Eq.~\eqref{eq: brag interruption condition} is satisfied at some point during its decay. If we
consider an unmodified standing wave \begin{equation}
\delta b(z,t) = \delta b_{0} \cos(k_{z}z) \cos(\omega_{A} t)\end{equation}
with $\delta b_{0}\ll 1$, then  \begin{equation}
\Delta = \frac{1}{\nu_{c}}\frac{1}{B}\frac{dB}{dt} \approx -\frac{1}{2}\frac{\omega_{A}}{\nu_{c}} \delta b_{0}^{2} \sin (2\omega_{A} t) \cos^{2}(k_{z} z).\end{equation}
The wave will thus be significantly modified -- i.e., interrupted -- if $\Delta \lesssim -2/\beta$ at some point in space, which occurs if
\begin{equation}
\delta b_{0} \gtrsim 2 \sqrt{\frac{\nu_{c}}{\omega_{A}}} \beta^{-1/2}\equiv \delta b_{\mathrm{max}}. \label{eq:Brag standing cond}\end{equation}
Above this limit, $\Delta p$ can remove the restoring force of the wave in regions where $\delta b(z) \neq 0$ (i.e., around 
the antinodes of the wave).  As we show in Fig.~\ref{fig: Brag on off}, the limit~\eqref{eq:Brag standing cond} is well matched by numerical solutions. 

\begin{figure}
\begin{center}
\includegraphics[width=0.7\columnwidth]{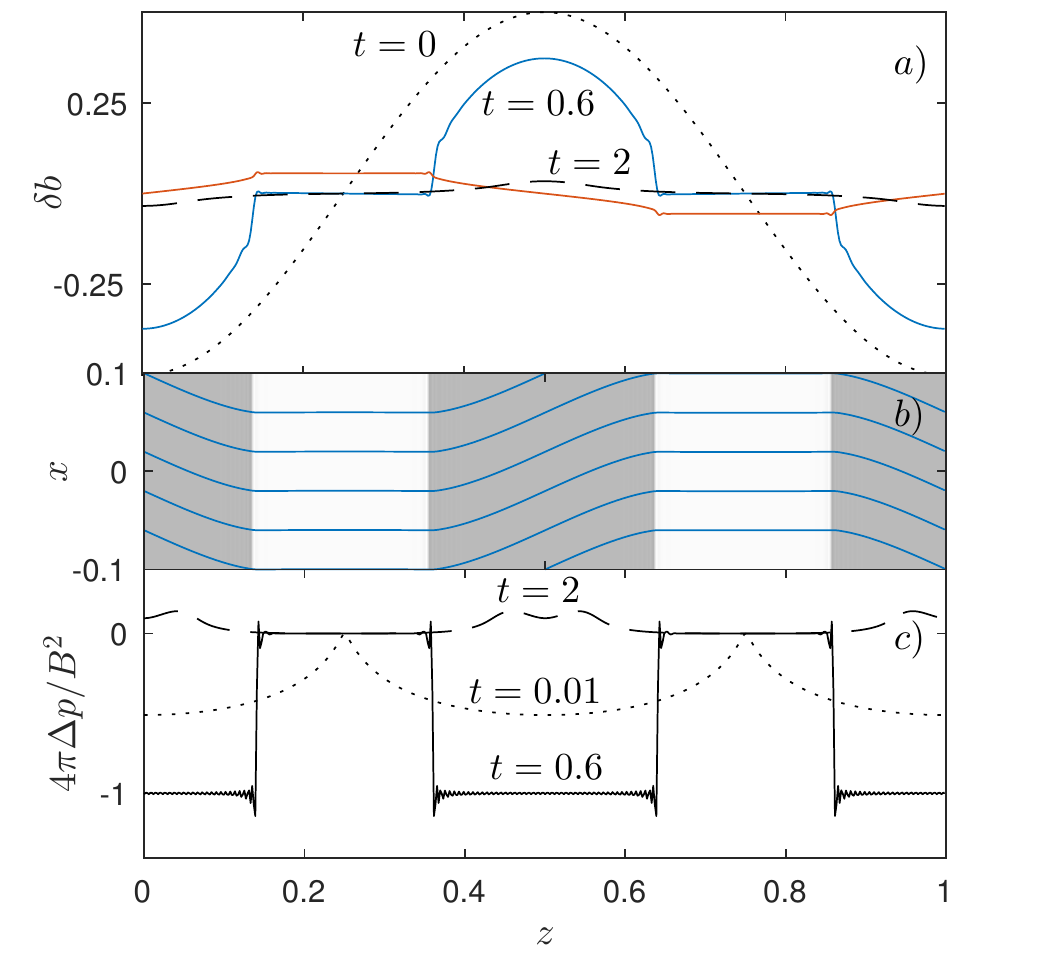}
\caption{Evolution of an initial magnetic perturbation $\delta b = -0.5 \cos(2\pi z)$ within the
Braginskii model at $\sqrt{\nu_{c}/\omega_{A}}\beta^{-1/2}=0.2$. Panel (a) shows $\delta b$ at $t=0$ (black
dotted line), $\delta b$ at $t=0.6\tau_{A}$ (blue solid line), $u_{\perp}/v_{A}$ at $t=0.6\tau_{A}$ (red solid line), 
and $\delta b$ at $t=2\tau_{A}$ (black dashed line), which is after the amplitude has decreased below the interruption limit. Panel (b) illustrates the shape of  the magnetic field lines in space at $t=0.6\tau_{A}$ (blue lines), with the shading showing
where $4\pi \Delta p /B^{2}=0$ (white) or $-1$ (gray). Panel (c) 
shows the anisotropy parameter, $4\pi \Delta p /B^{2}$, which is $-1$ at the firehose limit,  at the same times as in panel (a) (the black dotted line shows $t=0.01\tau_{A}$ to illustrate the initial evolution). 
 Note the velocity is much smaller than the magnetic field during the decay, and the lack of
magnetic tension everywhere in the decaying  wave.}
\label{fig:Brag standing}
\end{center}
\end{figure}

\begin{figure}
\begin{center}
\includegraphics[width=0.7\columnwidth]{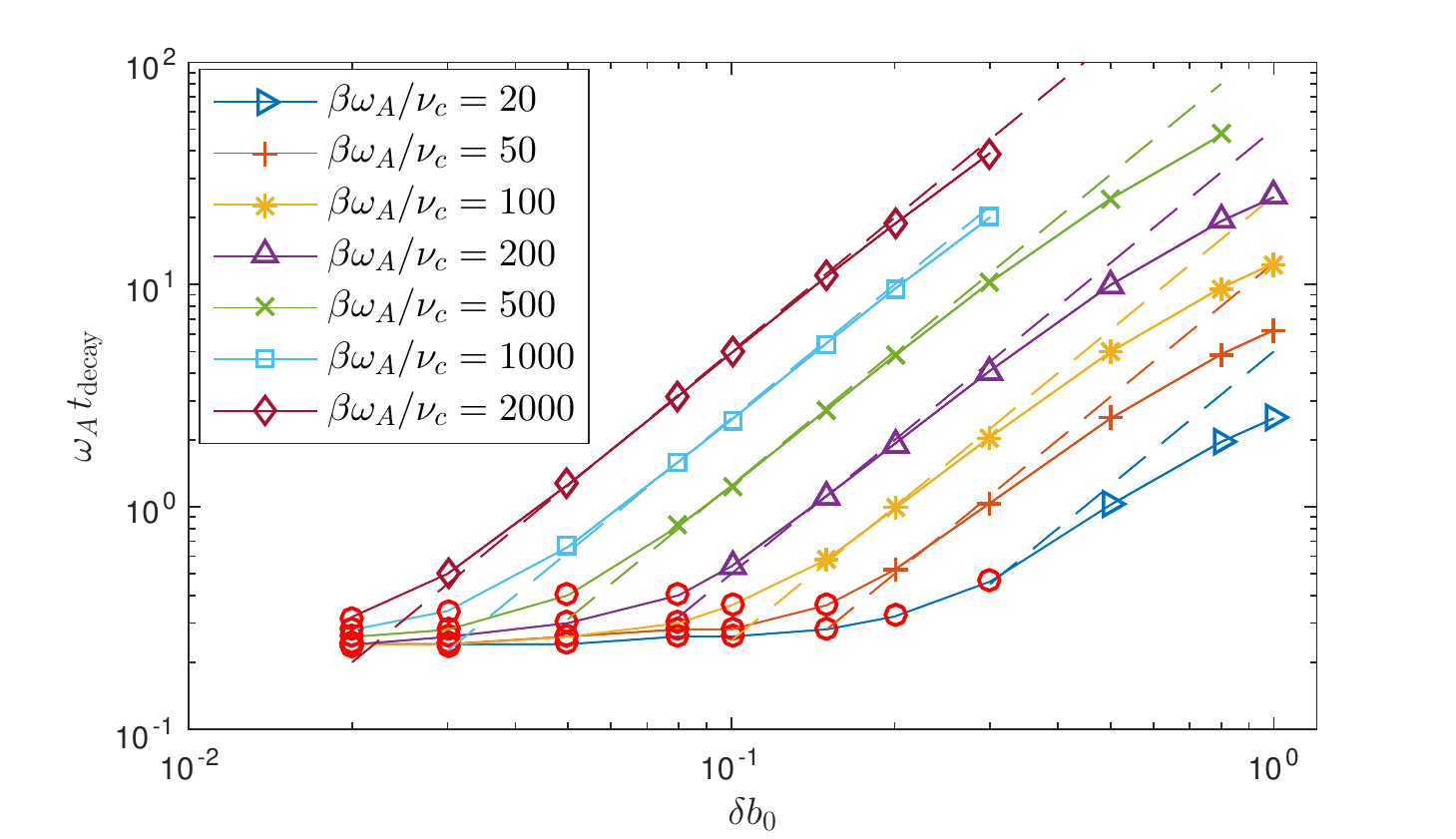}
\caption{Time $t_{\mathrm{decay}}$ for an initial magnetic perturbation of the form $\delta b(z) = \delta b_{0} \cos(k_{z}z) $ to 
decay to an amplitude that is small enough that it can oscillate. Solid lines and symbols show the numerically measured
$t_{\mathrm{decay}}$ (normalized by $\omega_{A}^{-1}$), while
dashed lines of matching color show the theoretical prediction \eqref{eq:Brag decay time estimate}. 
The bright
red circles are those points for which $\delta b_{0} $ is below the limit \eqref{eq:Brag standing cond}, meaning the wave is able to oscillate. The match 
with the theoretical prediction for $t_{\mathrm{decay}}$ is surprisingly accurate, illustrating the usefulness of the simple arguments outlined in Sec.~\ref{sub:Brag standing}. }
\label{fig:Brag standing decay time}
\end{center}
\end{figure}

Figure \ref{fig:Brag standing} illustrates the dynamics of a standing wave above the limit \eqref{eq:Brag standing cond}.
We solve the SA wave equation \eqref{eq:waves}, using the Braginskii closure \eqref{eq:Brag closure}, which 
assumes only 1-D dynamics and incompressibility of the wave.\footnote{Note that for the condition
\eqref{eq:Brag standing cond} to be met at the same time as the 
condition $\nu_{c}\gg \omega_{A}$, required for the validity of the Braginskii equations, the system must be at very high $\beta$. 
Further, since $u \sim v_{A} \sim \beta^{-1/2}c_{s}$, the motions are very subsonic. 
It thus makes sense to assume incompressibility when studying 
 Braginskii waves, and if one wished to study the lower-$\beta$, larger-$\delta b_{0}$, limit between
 the collisionless and Braginskii regimes,
 it would be most sensible to solve the full LF equations \eqref{eq:KMHD rho}--\eqref{eq:KMHD pl} with
 the collisional relaxation terms included.
More discussion, including the effects of heat fluxes, is given in \ref{app:Brag general considerations}.}
Note that within the incompressible limit, the dynamics are entirely determined by $\delta b_{0}$ and the
ratio $\beta^{-1} \nu_{c}/\omega_{A} $, because the $\nu_{c}^{-1}$ factor in $\Delta$ [Eq.~\eqref{eq:Brag closure}] multiplies $\beta/2$ in Eq.~\eqref{eq:waves}.

Although the nonlinear wave dynamics shown in Fig.~\ref{fig:Brag standing} may appear quite bizarre, with angular field structures
and sharp discontinuities in $\Delta p$, many  features can be straightforwardly understood by noting that if
the field is to decrease significantly more slowly than in a linear SA wave,
it must nullify the magnetic tension. This can be achieved: (1) by keeping a pressure anisotropy at the firehose limit, which occurs 
in the $\delta b(z)$ ``humps''  where the field is curved; or (2) by having straight field 
lines, which occurs where $\delta b(z)$ is zero. Then, because the field must keep 
decreasing in order to maintain $\Delta =-2/\beta$ (since $\Delta \sim \nu_{c}^{-1}dB/dt$), it slowly decays in time, with the 
regions where the field is small reaching $\delta b(z)=0$ first.
Once the amplitude of the wave decays  below the level at which  it can sustain $\Delta = -2/\beta$ (i.e., when the 
field at the wave antinode is $\delta b^{2} \sim \nu_{c}/\omega_{A}\beta^{-1}$), the wave can 
oscillate freely again with an amplitude below the interruption limit (although $\delta b$  is
not sinusoidal because the final stages of the interrupted decay are nonsinusoidal). Note 
that throughout this decay process, the perturbation's magnetic energy dominates over the
kinetic energy. This is because the pressure anisotropy stress cancels out the Lorentz force in the momentum equation, leading to 
a magnetic field that changes more slowly than that in a similar-amplitude linear SA wave.

We can use these ideas to calculate the decay time of the field,  $t_{\mathrm{decay}}$, as a function of $\beta$ and the initial amplitude $\delta b_{0}$. 
The idea is simply  to ignore the spatial dependence of the solution, focusing 
on the antinode of the wave, where $\delta b$ is maximal. The condition $\Delta=-2/\beta$ is
then
\begin{equation}
-\frac{2}{\beta}=\Delta=\frac{1}{\nu_{c}}\frac{1}{B}\frac{dB}{dt} \sim \frac{1}{\nu_{c}} \delta b \frac{\partial \delta b}{\partial t}, \end{equation}
which has  the solution 
\begin{equation}
\delta b^{2}  = \delta b_{0}^{2} - 4 \frac{\nu_{c}}{\beta} t.\label{eq:Brag sol}\end{equation}
By  solving for $\delta b^{2}=0$, we arrive at a prediction for the time for the interrupted field to decay to an amplitude at which 
it can oscillate:
\begin{equation}
t_{\mathrm{decay}}=\frac{\beta}{4 \nu_{c}} \delta b_{0}^{2}.\label{eq:Brag decay time estimate}\end{equation}
As shown in Fig.~\ref{fig:Brag standing decay time}, this estimate agrees very 
well with numerically computed decay  times (taken from calculations like that in Fig.~\ref{fig:Brag standing}) even quantitatively, 
illustrating the effectiveness of the simple dynamical model proposed above.


\subsubsection{Standing waves with an initial velocity perturbation.}
It is worth briefly describing also the dynamics that one observes after initializing with a velocity rather than a magnetic perturbation.\footnote{This arguably represents a more natural situation  physically, since it 
is hard to envisage how a static  magnetic perturbation might arise.} Because in such a situation the field initially grows rather than decays, the
ensuing dynamics depend on what occurs at positive pressure anisotropies, when $\Delta  > 1/\beta$.
Specifically, one expects growing mirror fluctuations (which are not captured in 1-D models) 
to act to limit $\Delta$ at $1/\beta$, and
that this limiting action will be fast compared to $\omega_{A}$, so long as there is significant scale 
separation with the gyroscale
 (see Sec.~\ref{sub:microinstabilities equations}; \citealp{Kunz:2014kt,Melville:2015tt}).
 If this limit on $\Delta$ does not exist -- i.e., if $\Delta$ can grow without bound as the magnetic field 
 grows in the standing wave -- the extra magnetic tension arising from $\Delta>0$ acts to reverse the fluctuation of 
 the wave at low magnetic-field amplitudes, while  strong nonlinear damping  causes the wave to damp to an amplitude below the interruption limit \eqref{eq:Brag standing cond} in less than a wave period  (see Sec.~\ref{sub:Brag traveling}). However, if the anisotropy is limited at positive values, 
this allows the field to grow to much higher amplitudes, viz., $\delta b \approx u_{\perp 0}/v_{A}$ (for  $u_{\perp 0}/v_{A} \gg \beta^{-1/2}$ and $\beta \gg 1$). When the field then starts decreasing
again (once $u_{\perp}\approx 0$) it does so from an amplitude that is above the interruption limit, and 
thus behaves in effectively the same way as an initial purely magnetic perturbation. Thus, the dynamics -- as long
as growing mirror fluctuations  act to limit positive pressure anisotropies -- are similar to those for an initial static magnetic perturbation,
and the  limit on the amplitude of an initial velocity perturbation $u_{\perp 0}/v_{A}$ is similar to Eq.~\eqref{eq:Brag standing cond}. In  Sec.~\ref{sec:kinetic stuff}, we give a  more detailed 
discussion of this physics.\footnote{The
 reader may be puzzled that we are not applying such a ``$\Delta$-limit'' argument also to the negative anisotropies. 
 The salient point is that  $\Delta$ can only be limited by firehose microinstabilities once $\Delta < -2/\beta$, by which 
 point the magnetic tension has already been removed.  More extensive discussion of this and related issues is given in  Sec.~\ref{sec:kinetic stuff}. }

\subsection{Traveling waves}\label{sub:Brag traveling}

With the Braginskii model, because
$\Delta \approx \nu_{c}^{-1} B^{-1}dB/dt$, the anisotropy and the rate of change of the magnetic field $B^{-1}dB/dt$ are always strongly correlated.
From 
Eq.~\eqref{eq:kinetic energy}, this implies the wave energy $E_{\mathrm{wave}}=E_{\mathrm{mech}}=E_{K}+E_{M}$ is nonlinearly damped at
the rate 
\begin{equation}
\partial_{t}E_{\mathrm{wave}} = -\frac{1}{\nu_{c}}\int d \bm{x}\, p_{0} \left( \frac{1}{B}\frac{dB}{dt} \right)^{2}.\label{eq: brag wave damping 1}\end{equation}
A sinusoidal traveling SA wave, 
$\bm{B} = B_{0} (\delta b \sin(k_{z} z -\omega_{A} t),0,1),$
creates a changing magnetic field 
\begin{equation}
\frac{1}{B}\frac{dB}{dt} = -\frac{1}{2}\delta b^{2}\omega_{A}\sin(2 k_{z}z - 2\omega_{A}t),\end{equation}
which, from Eq.~\eqref{eq: brag wave damping 1}, causes the wave to damp at the rate 
\begin{equation}
\partial_{t}E_{\mathrm{wave}} \approx -\frac{1}{8}p_{0} \frac{\omega_{A}^{2}}{\nu_{c}} \delta b^{4}\label{eq:Brag wave energy damping}.\end{equation}
This is effectively a parallel viscous damping, which occurs because  there is a
component of $\bm{u}$ in the field-parallel direction due to the finite amplitude of the wave.
Noting that $E_{\mathrm{wave}} = (8\pi)^{-1} \delta b^{2}B_{0}^{2}$, we conclude that Eq.~\eqref{eq:Brag wave energy damping} implies a wave damping rate
\begin{equation}
\frac{1}{2E_{\mathrm{wave}}}\frac{dE_{\mathrm{wave}}}{dt} \approx -\frac{1}{4} \omega_{A} \frac{\beta \omega_{A}}{\nu_{c}} \delta b^{2}  = -\omega_{A} \frac{\delta b^{2}}{\delta b_{\mathrm{max}}^{2}},\label{eq:brag wave damping}\end{equation}
where $\delta b_{\mathrm{max}}$ is the interruption limit, given by Eq.~\eqref{eq:Brag standing cond}.

\begin{figure}
\begin{center}
\includegraphics[width=0.5\columnwidth]{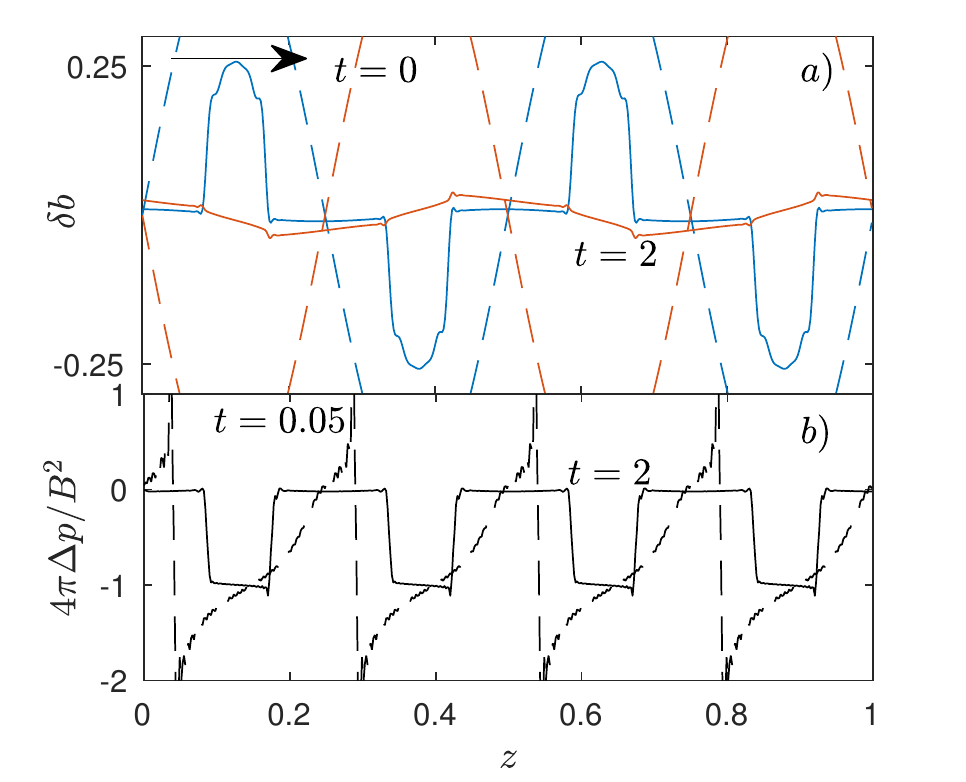}\includegraphics[width=0.5\columnwidth]{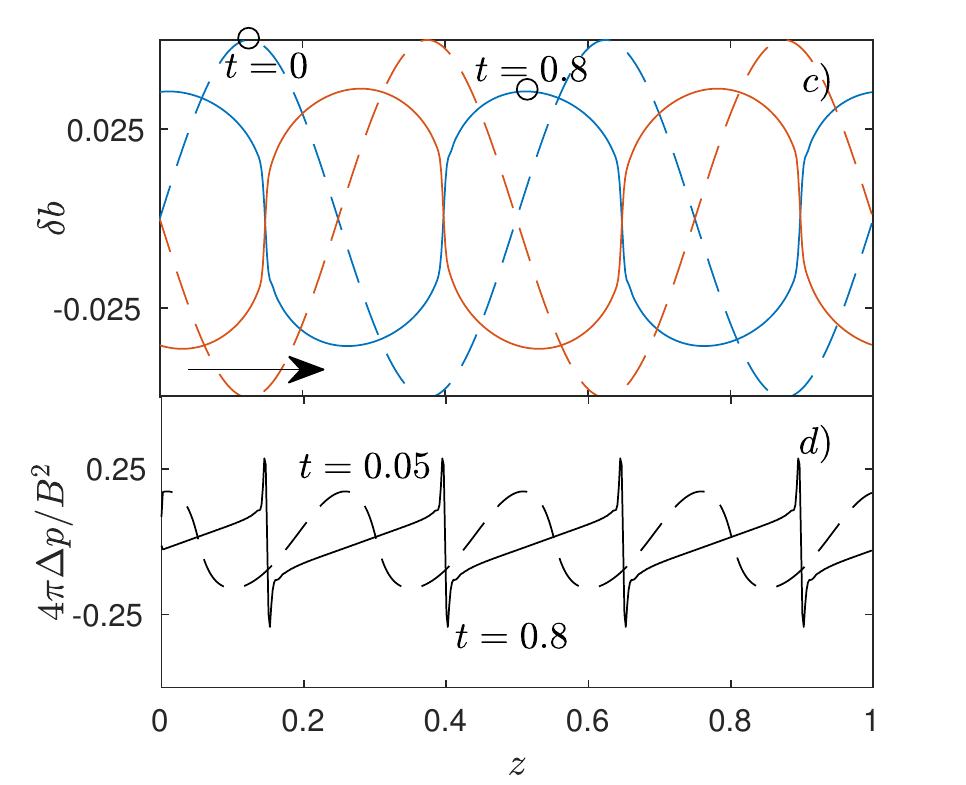}
\caption{Evolution of a traveling waves with the initial condition $\delta b =-u_{\perp}/v_{A}= -\delta b_{0} \cos(4\pi z)$ [Eq.~\eqref{eq: traveling ICs}] within the
Braginskii model at $\sqrt{\nu_{c}/\omega_{A}}\beta^{-1/2}=0.2$. In the top panels, blue lines illustrate $\delta b$ and red lines show $u_{\perp}/v_{A}$, while the bottom panels show the anisotropy 
parameter $4\pi \Delta p /B^{2}$ (this is $-1$ at the firehose limit). The left panels (a)-(b) show a wave with $\delta b_{0} = 0.5$ -- i.e., above the interruption limit -- at $t=0$ (dashed lines; we show $t=0.05$ instead for the anisotropy parameter to illustrate the initial evolution) and $t=2 \tau_{A}$ (solid lines) [the initial condition in panel (a) is sinusoidal; a reduced plot range is used to better see the later times].  The right panels (c)-(d) show a wave with $\delta b_{0} = 0.05$ -- i.e., below the interruption limit -- at $t=0$ [dashed lines; $t=0.05$ in panel (b)], $t=0.8 \tau_{A}$ (solid lines) (the earlier time is  shown so as to fit the full evolution into one panel). In panel (c) hollow circles mark the same position on the wave front as it propagates. Above the interruption limit, there is such a strong
nonlinear modification of the wave that it is effectively interrupted and stopped before it can propagate. In contrast, below 
the limit, the wave undergoes minor shape changes and slow nonlinear decay.  }
\label{fig:Brag traveling}
\end{center}
\end{figure}

\begin{figure}
\begin{center}
\includegraphics[width=0.6\columnwidth]{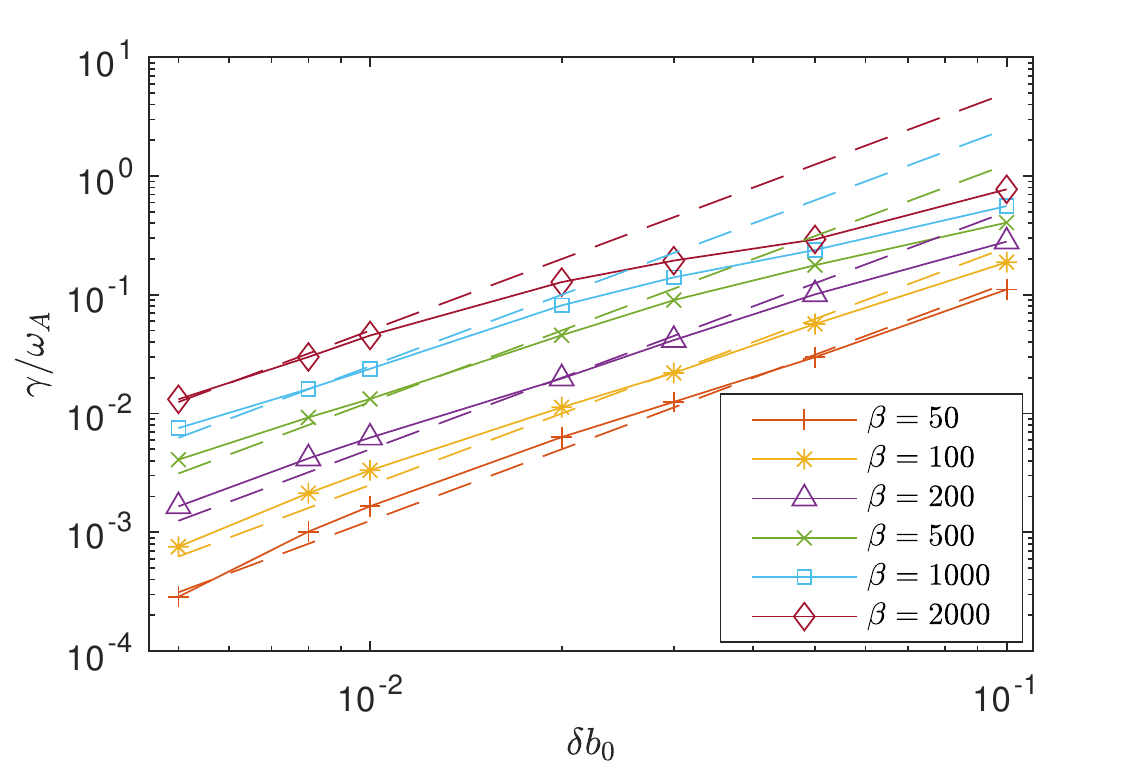}
\caption{Initial decay rate $ \gamma/\omega_{A} \equiv 0.5\, E_{\mathrm{wave}}^{-1} {dE_{\mathrm{wave}}}/{dt}$ of Braginskii traveling waves. Solid lines and symbols show the numerical measurements, while dashed lines are the theoretical prediction \eqref{eq:brag wave damping}. The agreement is reasonably good, although at larger decay rates, where the nonlinear effects are stronger, the wave reduces
its damping rate by becoming more square.}
\label{default}
\end{center}
\end{figure}

For amplitudes above the interruption limit \eqref{eq:Brag standing cond}, Eq.~\eqref{eq:brag wave damping} implies 
a damping rate greater than the frequency of the wave itself. In this case, the damping will cause such a strong nonlinear modification
of the wave that it might be considered more accurately as an interruption, effectively stopping the wave. Indeed,
the local pressure anisotropy will reach the firehose limit in regions where $B^{-1}dB/dt<0$ (and the mirror limit where $B^{-1}dB/dt>0$), so we should expect some of the arguments of Sec.~\ref{sub:Brag standing} to apply here. 
A traveling wave (with the same parameters as the standing wave in Fig.~\ref{fig:Brag standing}) is illustrated in Figs.~\ref{fig:Brag traveling}(a)-(b), showing how the wave is virtually stopped with a larger magnetic
than kinetic energy, and an anisotropy that is similar to the standing wave  [cf. Fig.~\ref{fig:Brag standing}(c)]. Thus, there is effectively an ``interruption'' of the same kind as for a standing wave.

Consider now a traveling wave that is well below  the interruption limit, illustrated in Fig.~\ref{fig:Brag traveling}(c)-(d).
Such a  wave exhibits much slower damping and moderate nonlinear modification to the wave shape, which should be
expected because the mechanism causing the wave damping has nonlinear spatial variation in space. The more angular 
structures that the wave develops act to reduce $B^{-1}dB/dt$ over much of the wave, and thus reduce the damping rate 
somewhat. 

The damping rate is measured quantitatively in the right-hand panel of Fig.~\ref{fig:Brag traveling}, where  the theoretical 
prediction \eqref{eq:brag wave damping} is compared to the rates  measured in simulations. While our prediction agrees very 
well in the low-decay-rate limit, where the nonlinear modifications to the wave shape are small, it deviates as the 
wave damping increases and the pressure anisotropy causes more significant changes to the shape of the wave. 

Finally, note that although, for consistency with the upcoming analysis of collisionless plasmas (Sec.~\ref{sec:collisionless}), we have   discussed  standing
waves  and traveling waves separately,  this
distinction is less important for Braginskii dynamics. Indeed, we have seen that a Braginskii 
traveling wave above the amplitude limit \eqref{eq:Brag standing cond} is effectively interrupted, because the anisotropy is so strong 
that it stops the wave. Analogously,  a standing wave below the interruption limit will oscillate but will be nonlinearly damped at the rate
\eqref{eq:brag wave damping}, because there 
is still parallel variation in $\nabla \bm{u}$ that is damped by the Braginskii viscosity.
In the next section,  we shall see that there is 
a stronger distinction between standing and traveling waves 
 (and between interruption and nonlinear damping) for collisionless wave  dynamics, because of the smoothing effect of the heat 
fluxes.

\section{Collisionless waves}\label{sec:collisionless}

We now consider the dynamics of SA waves in a collisionless plasma at high $\beta$.
Collisionless dynamics differ significantly from the Braginskii limit discussed in the 
previous section because, in a collisionless plasma,  the pressure anisotropy {remembers} the \emph{time history} 
of $B^{2}$, rather than being set by the instantaneous value of $dB/dt$.
This implies that, once the magnetic tension is removed when the anisotropy reaches
the firehose limit, the field is not able to decrease;  any further decrease in $B$ would drive the plasma unstable. In contrast, in the Braginskii limit, maintaining $\Delta p$ at the firehose limit \emph{requires} $dB/dt<0$.
In addition, the heat fluxes always play a significant dynamical role in collisionless 
plasmas at high $\beta$, acting to smooth $\Delta p$. This leads to near perfect
zig-zag magnetic field lines that minimize the spatial variation in $B^{2}$.

Compared to the Braginskii model, which may be rigorously derived 
from the kinetic equations via a perturbative expansion  \citep{Braginskii:1965vl}, 
 Landau-fluid closures are  only  heuristically motivated (see Sec.~\ref{sec:equations}). Nonetheless,
for the clarity of presentation throughout this section, we shall primarily 
focus our discussion on physics contained within the LF model, viz., large heat fluxes that  result from particles streaming along
field lines, with no particle scattering.  A variety of other 
physical  effects that may be important (e.g., particle trapping, or particle scattering by magnetic fluctuations due to  microinstabilities)
are discussed in Sec.~\ref{sec:kinetic stuff}.

\subsection{Standing waves}\label{sub:noC standing}

As discussed in Sec.~\ref{sec:equations} and more formally justified in \ref{app:asymptotics} [see Eq.~\eqref{eq: pressure 2 is smooth} and related discussion], 
the primary effect of the heat fluxes is to damp all  $k_{\parallel}\neq 0$ components of $\Delta p$, giving
\begin{equation}
\Delta =   3 \left< \ln \frac{B(t)}{B(0)} \right>\left[ 1 + \mathcal{O}(\beta^{-1/2})(\bm{x})+ \cdots\right].\label{eq:noC anisotropy standing}
\end{equation}
Since $\langle  {B(t)}/{B(0)} \rangle$ decreases in time as  a standing wave evolves, a wave 
 will  reach $\Delta = -2/\beta$ if 
\begin{equation}
 \frac {3}{2}\left< \ln \frac{1}{1+\delta b_{0}(\bm{x})^{2}} \right> \approx -\frac{3}{2}\langle \delta b_{0}(\bm{x})^{2} \rangle  < -\frac{2}{\beta}.\end{equation}
Assuming a sinusoidal initial perturbation $\delta b_{0}(\bm{x})= \delta b_{0} \cos(k_{z} z) $,  a SA wave  is 
interrupted if
\begin{equation}
\delta b_{0}\gtrsim   \sqrt{\frac{8}{3}}\beta^{-1/2}\equiv \delta b_{\mathrm{max}}.\label{eq:noC standing wave cond}\end{equation}
This limit, including the $\sqrt{8/3}$ numerical coefficient, matches numerical simulations using the full LF model nearly perfectly \citep{Squire:2016aw}.

\begin{figure}
\begin{center}
\includegraphics[width=0.75\columnwidth]{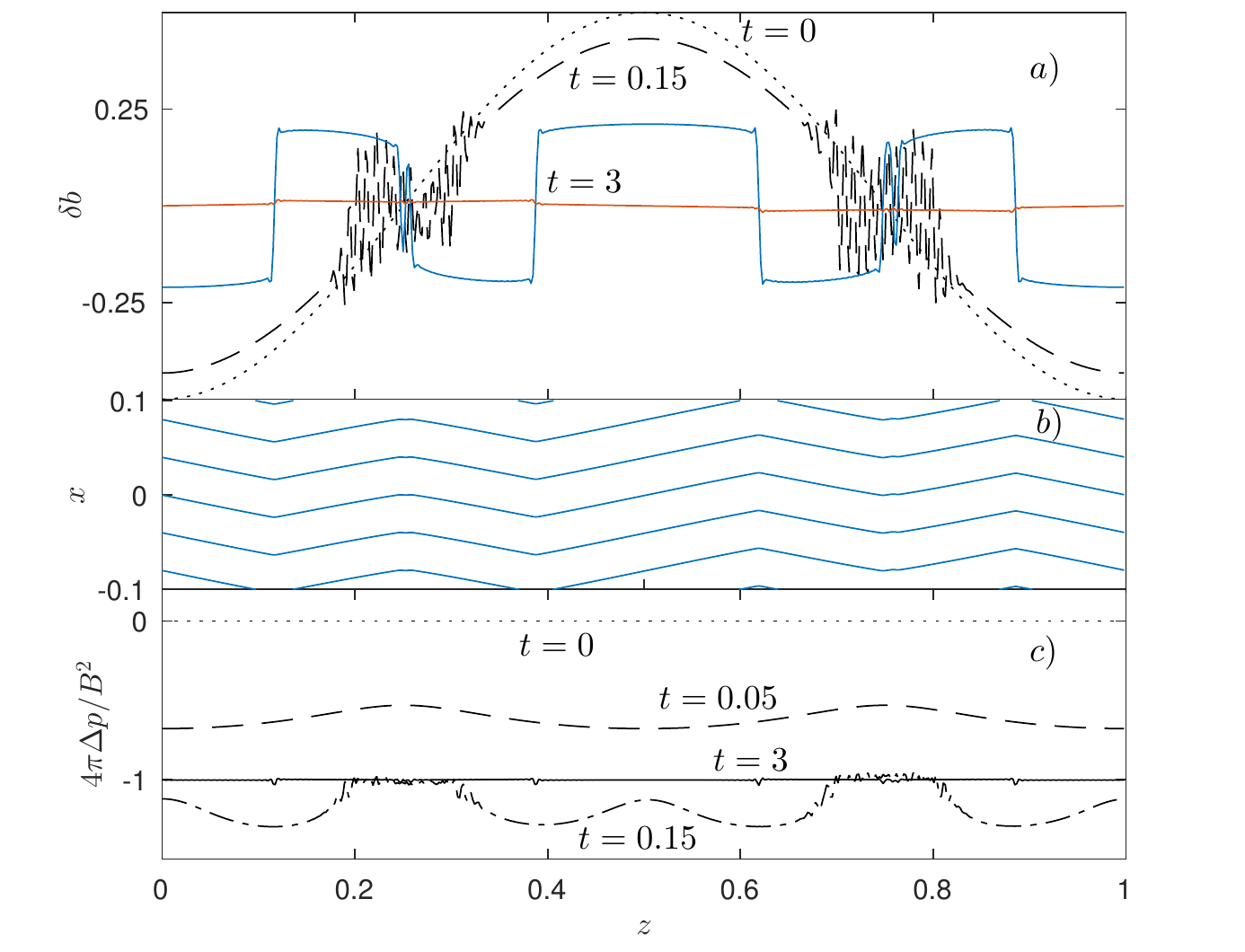}
\caption{Evolution of a shear-Alfv\'en standing wave in the collisionless LF model at $\beta=100$, at an initial amplitude above the interruption limit \eqref{eq:noC standing wave cond}, with initial condition $\delta b=-0.5 \cos(k_{z}z)$. Panel (a) shows $\delta b$ at $t=0$, $t=0.15\tau_{A}$, and $t=3 \tau_{A}$ (dotted black, dashed black, and blue lines, respectively), and $u_{\perp}/v_{A}$ at $t=3 \tau_{A}$ (red line). Panel (b) shows the magnetic-field lines of the perturbation shown in  (a) at $t=3\tau_{A}$.  Panel (c) shows the anisotropy $4\pi \Delta p/B^{2} = \Delta \beta/2$ associated with the evolution shown in (a), at $t=0$, $t=0.05\tau_{A}$, $t=0.15\tau_{A}$ and $t=3 \tau_{A}$ (dotted, dashed, dot-dashed, and solid lines, respectively). 
In stark contrast to the highly nonlinear behavior of collisionless waves shown here, an MHD perturbation at these parameters is almost perfectly linear.  A detailed description of each phase of evolution is given in Sec.~\ref{sub:noC standing}.}
\label{fig:noC standing}
\end{center}
\end{figure}

The dynamics of a perturbation that starts \emph{above} the limit \eqref{eq:noC standing wave cond} are 
illustrated in Fig.~\ref{fig:noC standing}, which shows a solution of the LF equations \eqref{eq:KMHD rho}--\eqref{eq:GL heat fluxes l}.
Despite the  bizarre appearance of the highly angular, zig-zag structures that develop here [see Fig.~\ref{fig:noC standing}(c)], the main features 
can be relatively easily understood. Let us consider qualitatively  the wave evolution in three phases:
\begin{description}
\item[Approach to interruption.]{During the initial evolution of the wave, before the anisotropy reaches $\Delta = -2/\beta$, 
the spatial shape of the wave is largely unaffected by the developing anisotropy. This is because  
spatial variation of $\beta \Delta  $ is $\mathcal{O}(\beta^{-1/2})$ [see Eq.~\eqref{eq:noC anisotropy standing}] even though $\beta \Delta/2 $ itself 
is $\mathcal{O}(1)$ during this phase. The nonlinearity due to the pressure anisotropy $\beta \partial_{z}^{2}(\delta b \Delta)$ 
[see Eq.~\eqref{eq:waves}] is thus $\sim \beta \Delta \partial_{z}^{2}\delta b$, which 
simply slows down the wave without modifying its spatial structure. The 
pressure anisotropy during this phase is shown as a  dashed line in Fig.~\ref{fig:noC standing}(b), while  $\delta b (z)$ looks 
very similar to the initial condition [dotted line in Fig.~\ref{fig:noC standing}(b)] at these parameters.}
\item[Early nonlinear evolution.]{As the pressure anisotropy approaches $\Delta = -2/\beta$, the linear term in the 
wave evolution equation \eqref{eq:waves}, $1+\beta \Delta/2$, becomes very small and then turns negative when the
anisotropy overshoots the firehose limit.  This overshoot has two effects: the first is to reverse the decrease in the magnetic field of
 the largest-scale mode (since the linear term has changed sign); the second is to
 cause small-scale firehose fluctuations
to grow rapidly in the regions of low $\delta b(z) $ (i.e., in the neighborhood of the wave nodes; the overshoot of the firehose
limit is greatest at low fields because $\Delta$ does not vary significantly in space). 
These growing small-scale modes act very quickly to return the anisotropy back to its marginal level. 
This process can be seen in the $t=0.15\tau_{A}$ curves in Fig.~\ref{fig:noC standing}(a) and (c), which show the 
field and the pressure anisotropy just after the small-scale firehose modes have grown at the antinodes and returned
the anisotropy to the marginal level. 
During this phase, the presence of a spatially varying nonlinearity -- i.e., the $\mathcal{O}(\beta^{-1/2})$ spatial variation in $\Delta$ [Eq.~\eqref{eq:noC anisotropy standing}] and
the $(1+\delta b^{2})^{-1}$ nonlinearity arising from field strength variation [see Eq.~\eqref{eq:waves}] -- is crucial
to the dynamics, because the linear term $1+\beta \Delta/2$ is small. Without these nonlinearities, there is no preferential location 
for the growth of firehose instabilities and the entire wave erupts in a sea of small-scale fluctuations. 
It is critical (but nontrivial) to account for such a
nonlinearity in a reduced equation that describes interruption dynamics [see \ref{app:near interruption},
 Eq.~\eqref{eq:Landau2 final equation} and Fig.~\ref{app:fig:LF asym}, for such an equation]. }
 \item[Late-time evolution.]{As the firehose modes push the anisotropy back to its marginal level, the smallest-scale fluctuations decay rapidly \citep{Melville:2015tt}. Following a transient
 period during which the pressure anisotropy slowly oscillates around (and  decays towards) $\Delta=-2/\beta$ \citep{Schekochihin:2008en}, the system 
 relaxes into a final state with regions of straight fields separated by sudden corners. 
Despite this state's bizarre appearance,  the basic cause of the plasma's preference for such structures may be inferred from
a rather simple physical argument (within the LF model).
 This argument follows from  three important properties of the collisionless dynamics: (i) without 
particle scattering any decrease in the magnetic field will lead to a decrease in the pressure anisotropy towards more negative values\footnote{See Eq.~\eqref{eq:mean delta p evolution}
for the evolution of the spatially averaged anisotropy.
The asymptotic scalings discussed in \ref{app:initial wave} show that the compressional term is small at high $\beta$, while the heat-flux term, 
$3\langle q_{\perp} \nabla \cdot \hat{\bm{b}}\rangle$, relies on magnetic curvature and so will decrease with $\delta b$. Thus, the only effect
able to cancel the creation of anisotropy through $3p_{0}\langle B^{-1}dB/dt \rangle$ is the collisional damping $-3\nu_{c}\langle \Delta p\rangle$.
}; (ii)
the only way in which the magnetic-field strength an be constant in time is either for  the anisotropy to be at the firehose limit
or for the field lines to be straight (or both);
 (iii) the heat fluxes continue to remove 
 the spatial variation in $\Delta$  during the slow transient phase following the initial interruption.
Property (i) tells us that the magnetic field cannot continue decreasing, as it did in the Braginskii model, 
without creating small-scale firehose fluctuations everywhere in the plasma.
Then, if we assume that the plasma has reached some quasi-steady state with nonzero $B$, the plasma cannot 
be everywhere at the firehose limit and also have $\partial_{z}B \neq 0$, because the heat fluxes continue to 
flatten $\Delta p$ [see Fig.~\ref{fig:noC standing}(b)]. Thus, in the absence of  oscillatory 
behavior due to the Lorentz force, properties (ii)-(iii) together imply that  that  
$B$ is effectively also flattened by the heat fluxes, which in turn suggests $\delta b(z)$ must be piecewise constant if it is nonzero.
The result is the zig-zag field lines, shown in Fig.~\ref{fig:noC standing}(c).
Note that the double-adiabatic model, which neglects the heat fluxes and so lacks property (iii), does not produce  
constant $B$ fields (see Fig.~\ref{fig:app: CGL evolution}); instead, fields with curvature may be tensionless by being everywhere 
at the firehose limit but with a spatially varying $\Delta p$, as in the case of Braginskii interruption (see Sec.~\ref{sub:Brag standing}).  
  }
\end{description}

\begin{figure}
\begin{center}
\includegraphics[width=0.65\columnwidth]{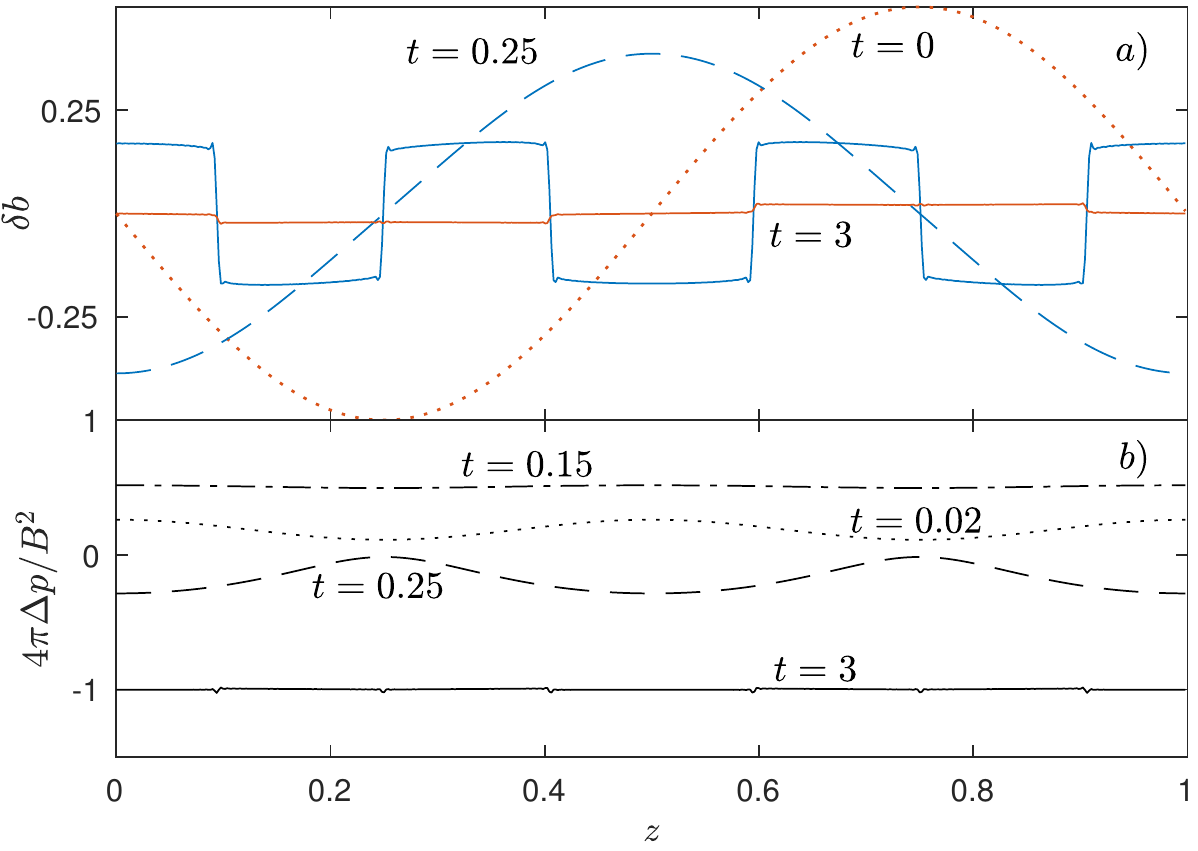}
\caption{Evolution of a SA standing wave in the collisionless LF model at $\beta=100$ (same parameters as Fig.~\ref{fig:noC standing}), with initial conditions in the velocity above the interruption limit $u_{\perp}/v_{A}=-0.5 \sin(k_{z}z)$ [i.e., initial conditions \eqref{eq: standing u ICs}]. Panel (a) shows $u_{\perp}/v_{A}$ at $t=0$ and $t=3\tau_{A}$ (dotted and solid red lines respectively), and $\delta b $ at $t=0.25\tau_{A}$ (when $\delta b$ is at its maximum) and $t=3\tau_{A}$ (dashed and solid blue lines respectively).   We limit $\Delta \leq 1/\beta $
to  capture heuristically the anisotropy-limiting behavior of the mirror instability; this enables $\delta b$ to reach amplitudes approaching that of the initial $u_{\perp}/v_{A}$.
Panel (b) shows the anisotropy parameter $4\pi \Delta p/B^{2}$ at $t=0.02$ (dotted line; this shows the early time increase in $\Delta p$), at $t=0.15$ (dot-dashed line; when the anisotropy is limited at $\Delta = 1/\beta$), $t=0.25$ (dashed line), and $t=3$ (solid line).
Following the decrease in the magnetic field from the profile shown at $t=0.25\tau_{A}$, the evolution
is relatively similar to that shown in Fig.~\ref{fig:noC standing} (we do not illustrate  intermediate times
to avoid clutter).}
\label{fig:noC standing-U0}
\end{center}
\end{figure}

It is worth noting that, within the LF model that we use [Eqs.~\eqref{eq:KMHD rho}--\eqref{eq:KMHD pl} with Eqs.~\eqref{eq:GL heat fluxes p}--\eqref{eq:GL heat fluxes l}],
the firehose fluctuations grow fastest at the smallest scale accessible in the simulation, which is set by an artificial hyperdiffusion operator.
In reality, this scale is set by the gyroradius, where the parallel firehose growth rate decreases due to 
finite-Larmor radius (FLR) effects \citep{1968PhFl...11.2259D,Schekochihin:2010bv}. A more detailed study of FLR and other kinetic effects
will be the subject of future work (see Sec.~\ref{sec:kinetic stuff}), but it is worth noting that we see very similar 
macroscopic dynamics independently of the numerical resolution (for resolutions $N_{z}\gtrsim 128$). This suggests that the 
exact scale separation between the SA wave and the firehose fluctuations that erupt in the ``early nonlinear evolution'' phase is 
not important for the  late-time large-scale evolution (so long as the scale separation is sufficiently large).

\subsubsection{Standing waves with an initial velocity perturbation.}

The discussion above  concerned the evolution of a wave starting from a magnetic perturbation. For an initial 
perturbation in the velocity, the anisotropy initially grows in the positive direction, effectively increasing the restoring force of the
wave. Starting from
$\Delta p=0$ but without a mechanism to limit $\Delta p$ at positive values,
 this increase of $\Delta p $ as $B$ grows is exactly the same as its decrease after $B$ has reached its maximum,
so the system never reaches $\Delta <0$. This results in nonlinear standing-wave oscillations with a frequency $\omega>\omega_{A}$ and $ u_{\perp}/v_{A} >  B_{\perp}/B_{0}$,\footnote{Using Eq.~\eqref{eq:noC anisotropy standing} and assuming
that $\partial_{z}\Delta = 0$, it is straightforward to derive an equation for the amplitude $\delta b$ of a (sinusoidal) perturbation. After normalizing time by $\omega_{A}$, this is \begin{equation}
\partial_{t}^{2} \delta b(t)= - \delta b - \frac{3\beta}{8}\delta b^{3} ,\label{eq: footnote u ev} 
\end{equation}
with the initial condition $\delta b(0) = 0$, $ \partial_{t} \delta b(0) = \delta u_{0} = u_{\perp}(0)/v_{A}$. Equation~\eqref{eq: footnote u ev} is an undamped Duffing equation, and for $\beta^{1/2} \delta u_{0} \gg 1$ has the approximate solution
\begin{equation}
\delta b(t) \approx \left( \frac{16 \delta u_{0}^{2}}{3 \beta}\right)^{1/4} \mathrm{sn} \left[ \left(\frac{3 \beta \delta u_{0}^{2}}{16}\right)^{1/4} t,-1\right],\label{eq: u init jacobi sol}\end{equation}
where $\mathrm{sn}$ denotes the Jacobi elliptic function. These solutions oscillate in time, with a maximum amplitude $\delta b_{\mathrm{max}}<\delta u_{0}$ and a frequency larger then $\omega_{A}$ (this is $1$ in these time units). There is no  damping of the wave, which arises from the neglected spatial variation in the pressure anisotropy.}  which  decay in time because of pressure-anisotropy damping arising from the (small) spatial variation in $\Delta p$ (see Sec.~\ref{sub:noC traveling} for details).
However, the
mirror instability (which is excited when $\Delta>\beta^{-1}$)  breaks this symmetry, allowing the magnetic field to grow in time  while $\Delta $ is fixed at $\Delta \approx \beta^{-1} $. 
As the magnetic field starts decreasing again, the mirror modes that sustained
$\Delta \approx \beta^{-1} $ presumably decay quickly (\citealp{Melville:2015tt}; see also our Sec.~\ref{sec:kinetic stuff}), implying that
the anisotropy starts decreasing from $\Delta=\beta^{-1}$ and can  reach 
negative values. Thus, the limit on $u_{\perp}/v_{A}$ will be similar to Eq.~\eqref{eq:noC standing wave cond},
with perhaps a larger numerical prefactor  to account for the fact that the magnetic-field decrease starts
from a positive pressure anisotropy. 
The process is illustrated in Fig.~\ref{fig:noC standing-U0}, in which we artificially limit the
anisotropy to $\Delta \leq \beta^{-1}$.  Although  $\Delta=\beta^{-1}>0$ when the magnetic perturbation reaches its maximum (at $t\approx \tau_{A}/4$),\footnote{The magnetic perturbation $\delta B_{\perp}/B_{0}$ that results from $u_{\perp}$ is also
 slightly smaller than the initial $u_{\perp}/v_{A}$ because of the larger restoring force when $\Delta>0$.}
 the resulting final state is  similar to Fig.~\ref{fig:noC standing}.

\subsection{Traveling waves}\label{sub:noC traveling}

A traveling wave, with $\delta b \propto \cos(k_{z}z - \omega_{A}t)$, does not change $\langle B(t) \rangle$ as
it evolves. The arguments developed for standing waves in the previous section thus no longer apply, 
since the $\mathcal{O}(1)$ part of Eq.~\eqref{eq:noC anisotropy standing} is zero. 
However, even though $\langle B^{-1}dB/dt\rangle = 0$, $B^{-1}dB/dt$ itself is large. The
resulting $\Delta p$ -- which is reduced by a factor $\sim \beta^{1/2}$ by the heat fluxes  -- is correlated in space with $B^{-1}dB/dt$
and thus damps wave energy into thermal energy at the rate \eqref{eq:kinetic energy}, viz.,
\begin{equation}
\partial_{t} E_{\mathrm{wave}} = -\int d\bm{x}\,\Delta p\frac{1}{B}\frac{dB}{dt}.\label{eq:wave damping noC sec}\end{equation} 
As the wave is damped, the resulting decrease in $\langle B(t) \rangle$ causes $\langle \Delta p \rangle$ to decrease
also [according to Eq.~\eqref{eq:noC anisotropy standing}], slowing down the wave and eventually causing it to stop (interrupt) if 
the initial $B$ is sufficiently large for $\langle\Delta \rangle$ to reach $-2/\beta$. The key point here is that during the process of wave decay, there is 
no mechanism to isotropize the $k_{\parallel}=0$ component of the pressure, implying any decrease in $B$ must also 
be accompanied by a decrease in $\Delta p$.

\begin{figure}
\begin{center}
\includegraphics[width=0.85\columnwidth]{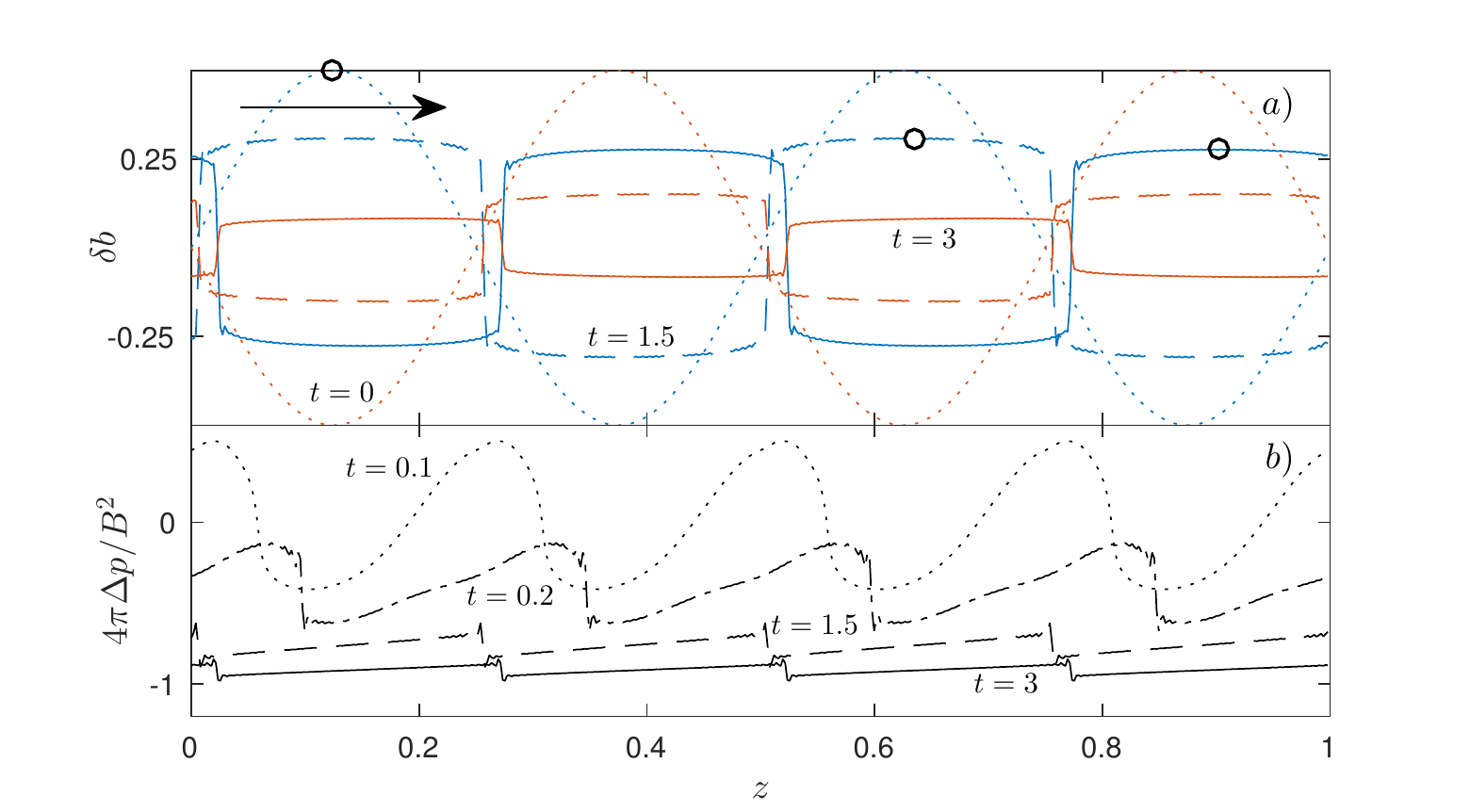}
\caption{Evolution of a shear-Alfv\'en traveling wave in the collisionless LF model at $\beta=100$, with an
initial amplitude above the interruption limit, $\delta b = -u_{\perp}/v_{A} = 0.5 \sin(4\pi z)$. Panel (a) 
illustrates $\delta b$ (blue) and $u_{\perp}/v_{A}$ (red) at $t=0$ (dotted lines), $t=1.5\tau_{A}$ (dashed lines), and $t=3 \tau_{A}$ (solid lines). Panel (b) illustrates the anisotropy, $4\pi \Delta p/B^{2} = \Delta \beta/2$, associated with the evolution shown in (a), at $t=0.1\tau_{A}$, $t=0.2\tau_{A}$, $t=1.5\tau_{A}$ and $t=3 \tau_{A}$ (dotted, dashed, dot-dashed, and solid lines respectively). Note the slow decrease in mean anisotropy, which forces the magnetic field to dominate over the velocity 
and slows the wave. At later times  (not shown due to clutter in the figure), the velocity continues to damp and the wave eventually comes to a standstill, with a similar final state to that of the standing waves shown in Figs.~\ref{fig:noC standing} and \ref{fig:noC standing-U0}.}
\label{fig:noC traveling}
\end{center}
\end{figure}

The process described above is illustrated in Fig.~\ref{fig:noC traveling}, which shows the 
solution of the LF equations at the same
parameters as the standing-wave examples (Figs.~\ref{fig:noC standing} and \ref{fig:noC standing-U0}). 
At early times, the pressure anisotropy [the dotted line in Fig.~\ref{fig:noC traveling}(b)] is a strong 
function of space\footnote{Note that the smoothing effect of the heat
fluxes is very strong here. Without it, the early time anisotropy shown in Fig.~\ref{fig:noC traveling}(b)
would be much larger. In the example shown in Fig.~\ref{fig:noC traveling}, the early time anisotropy is 
below the mirror and firehose thresholds ($|\Delta|\lesssim \beta^{-1}$); however, for even larger amplitudes, $\delta b \gtrsim \beta^{-1/4}$  the wave can cause $|\Delta|\gtrsim \beta^{-1}$ in the regions where the field is changing fastest [this estimate results from $|\Delta |_{\mathrm{max}}\sim \beta^{-1/2}\delta b^{2}$; see Eq.~\eqref{eq: D3 explicit}], causing even stronger nonlinear modifications to the wave.}  with $\langle \Delta p \rangle=0$.
This spatially periodic $\Delta p$ then damps the wave, as well as causing significant nonlinear modifications to its shape (which becomes
more angular, reducing $B^{-1}dB/dt$). As is also clear in Fig.~\ref{fig:noC traveling}(b), this damping drives the mean anisotropy to negative values. This   effectively reduces the Alfv\'en speed to $\tilde{v}_{A}=v_{A}(1+\beta \Delta/2)^{1/2}$, which causes the 
velocity to decay faster than the magnetic field [compare red and blue dashed and solid lines in Fig.~\ref{fig:noC traveling}(a)] because $\delta u_{\perp}/\tilde{v}_{A} =  \delta B_{\perp}/B_{0} $ in a traveling wave and $\tilde{v}_{A}<v_{A}$.
Although not shown in Fig.~\ref{fig:noC traveling} to avoid clutter, the velocity continues to be damped faster than the field as $\Delta$ approaches $-2/\beta$, and we are left with an angular, magnetically dominated final state that is  similar to the final 
state of   
the standing-wave evolution (Figs.~\ref{fig:noC standing} and \ref{fig:noC standing-U0}).

\setcounter{footnote}{0}

\subsubsection{Landau damping versus pressure-anisotropy damping.}
It is worth noting that the Landau damping rate of a linearly polarized SA wave  due to the (nonlinear) spatiotemporal
variation of the magnetic pressure is similar to the pressure-anisotropy damping, causing the wave to be damped at the 
rate $\gamma/\omega_{A }\sim \! \beta^{1/2}\delta b^{2}$ [\citealp{Hollweg:1971kq,Lee:1973ky}; see discussion around Eq.~\eqref{eq:Ewave noC} below].\footnote{This estimate neglects particle trapping effects, 
which reduce the damping rate \citep{Kulsrud:1978cr,Cesarsky:1981wt,Volk:1982vr}; i.e., the damping-rate estimate used here results from 
an application of linear Landau damping to a nonlinear wave.
Such particle trapping effects are  not included in our model, because our Landau-fluid closure uses linear 
Landau-damping rates to calculate the heat fluxes. } 
This effect is different  from pressure-anisotropy damping (but is still captured by the LF model),  and can in fact also be included in models of wave propagation that do not include a pressure anisotropy (see, e.g., 
\citealp{Medvedev:1996ex,Medvedev:1997dk}). The difference between the pressure-anisotropy damping and Landau damping  can be explained as follows. Heat fluxes are necessary for the Landau damping of a nonlinear wave, because they directly damp out the pressure perturbation that arises from the magnetic-field-strength variation. In contrast, heat fluxes act to \emph{reduce} the pressure-anisotropy damping, by  smoothing the spatial variation in the pressure anisotropy (in other words, the Landau damping damps  the pressure-anisotropy damping!).
In the discussion below, our estimate of the damping rate is heuristic,  so there is no need to work out these two effects separately; however, the Landau-damping effects
are  included in the calculation of the wave-decay rate given in \ref{app:initial wave}.

\subsubsection{Semi-quantitative description of traveling wave evolution}

\begin{figure}
\begin{center}
\includegraphics[width=0.48\columnwidth]{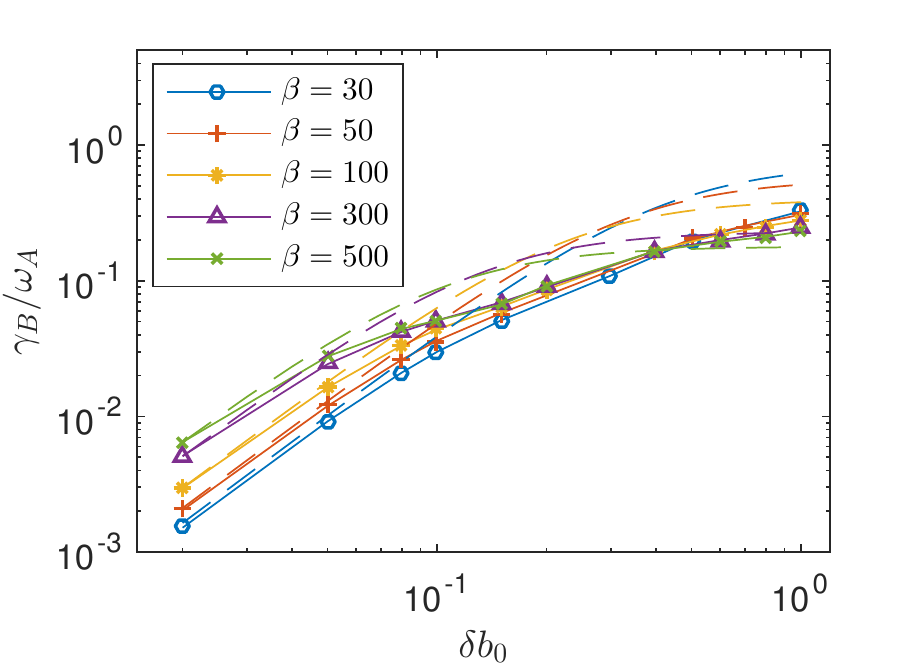}\includegraphics[width=0.48\columnwidth]{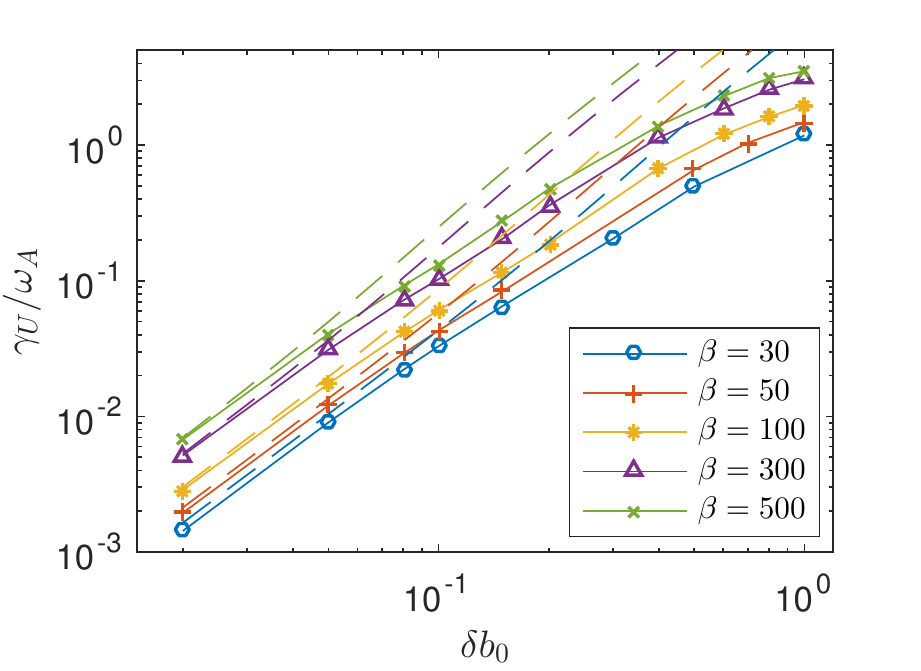}
\caption{Initial traveling-wave decay rates. 
The solid lines and symbols show  decay rates from the first $2\tau_{A}$ of the wave's evolution, measured in numerical simulation
of the LF equations with initial conditions $\delta b(z,0)=-\delta b_{0}\cos(k_{z} z),\:u_{x}(z,0)/v_{A}=\delta b_{0}\cos(k_{z} z)$. 
The dashed lines show the theoretical predictions \eqref{eq:traveling u b decay rates}. The left panel shows the magnetic-field decay rate $\gamma_{B}/\omega_{A}$ while the 
right panel shows the velocity decay rate $\gamma_{U}/\omega_{A}$ [see Eq.~\eqref{eq:decay rate defs}]. Although the agreement is not perfect, the theoretical predictions do capture
the  qualitative trends rather well, considering the crudeness of the sinusoidal approximation used.}
\label{fig:noC decay traveling}
\end{center}
\end{figure}

We now analyze the traveling-wave decay process in more detail, 
deriving a simple ordinary differential equation (ODE) to describe
the process of decay and interruption.
We assume that the wave remains sinusoidal throughout its 
evolution, which, although far from quantitatively justified (see Fig.~\ref{fig:noC traveling}), allows
one to construct an ODE that describes qualitatively how the nonlinear damping 
and mean anisotropy affect the magnetic-energy decay. This is then used 
to derive the decay rates of kinetic and magnetic energy,
\begin{equation}
\gamma_{U} \equiv \frac{1}{2E_{K}}\frac{dE_{K}}{dt},\qquad\gamma_{B} \equiv \frac{1}{2E_{M}}\frac{dE_{M}}{dt},\label{eq:decay rate defs}\end{equation}
 as a function of $\beta $ and $\delta b$. These turn out to match reasonably well the numerical LF solutions.

Our first step is to work out the decay rate of the wave due to pressure-anisotropy and Landau damping. Because this relies on the LF 
prescription for the heat fluxes, which in turn depends on
the compressible response of the plasma [because $q_{\perp,\parallel}\sim \partial_{z}(p_{\perp,\parallel}/\rho)$], a formal calculation of this damping is somewhat involved and
 is worked out in detail in \ref{app:initial wave} [see Eqs.~\eqref{eq:third order wave damping}--\eqref{eq:quant traveling damping}].
Here we give a heuristic derivation so as to present a relatively simple description of the important physics.
Defining $\Delta = \bar{ \Delta}+ \Delta_{k}$, where $\bar{\Delta}=\langle\Delta\rangle$ and $\Delta_{k}$ is 
the spatially varying part of $\Delta$ that arises from the perturbation $\delta b(z)$ with wavenumber $k$,
the important terms in the equation for $\Delta_{k}$ are [see Eq.~\eqref{eq:q effect on p}]
\begin{equation}
\partial_{t}\Delta_{k} \approx 3\frac{1}{B}\frac{dB}{dt} - a_{1}\sqrt{\frac{p_{0}}{\rho}}|k_{\parallel}|\Delta_{k},\label{eq:toy Dk}\end{equation}
where $a_{1}$ is an $\mathcal{O}(1)$ dimensionless coefficient, which depends on the details of a closure for the heat fluxes. 
We assume a monochromatic traveling wave of amplitude $\delta b$, giving 
\begin{equation}
2\frac{1}{B}\frac{dB}{dt}\approx \frac{3}{2}\delta b^{2}\omega_{A}(\bar{\Delta}) \sin[2 k z-2\omega_{A}(\bar{\Delta})t],\end{equation}
where $\omega_{A}(\bar{\Delta}) = v_{A}k (1+\beta \bar{\Delta}/2)^{1/2}$ accounts
for the slowing down of the wave as $\bar{\Delta}$ becomes negative. 
 Since $k_{\parallel}c_{s} \sim \beta^{1/2} \omega_{A}$, we may neglect the time derivative\footnote{The heat fluxes suppress spatial variation in $\Delta_{k}$ on the time scale $\tau_{\mathrm{damp}}\sim (k_{\parallel}c_{s})^{-1}\ll |\nabla \bm{u}|^{-1}\sim \omega_{A}$, so $\partial_{t}\Delta \sim \omega_{A}\Delta$ is small compared to $|k_{\parallel}|c_{s} \Delta$.} on the left-hand side of Eq.~\eqref{eq:toy Dk}, leading to
\begin{equation}
\Delta_{k}\sim {\beta^{-1/2}} \delta b^{2} \left(1+\beta \frac{\bar{\Delta}}{2}\right)^{1/2} \sin[2 k z - 2\omega_{A}(\bar{\Delta})t].\end{equation}
Evaluating   the integral \eqref{eq:wave damping noC sec} one finds,
\begin{equation}
\partial_{t}E_{\mathrm{wave}} \approx a_{2} \beta^{1/2} B_{0}^{2} \delta b^{4} \omega_{A}\left( 1+ \frac{\beta \bar{\Delta}}{2}\right),\label{eq:Ewave noC}\end{equation}
where $a_{2} \approx \sqrt{\pi}/8$ is calculated in \ref{app:initial wave}  [Eq.~\eqref{eq:quant traveling damping}]. This expression is similar in form to the  damping rate \eqref{eq:Brag wave energy damping} in the Braginskii regime, but
is reduced by $\beta^{1/2}$ due to the smoothing effect of the heat fluxes. 

Armed with the energy damping rate \eqref{eq:Ewave noC}, we now formulate an equation for the slow (compared to the sound propagation, $\omega\sim k_{\parallel}c_{s}$) dynamics
of the
wave amplitude $\delta b$.  For a sinusoidal wave, the magnetic 
energy is $E_{M} = \delta b^{2}B_{0}^{2}/16\pi$, while the assumption that the  wave remains traveling rather than standing
(i.e., that it does not generate a global oscillation in time) gives
 $E_{K} = (1+\beta \bar{\Delta}/2)E_{M}$. Noting also that $\bar{\Delta} = 3/4(\delta b^{2} - \delta b_{0}^{2})$,
 Eq.~\eqref{eq:Ewave noC} becomes
 \begin{equation}
\partial_{t}\left[ 2 \delta b^{2} + \frac{3\beta}{8}\delta b^{2}(\delta b^{2} - \delta b_{0}^{2})\right] = 
-a_{2} \beta^{1/2}\omega_{A}\delta b^{4} \left[ 1+ \frac{3\beta}{8}(\delta b^{2} - \delta b_{0}^{2})\right].
\end{equation}
This can be reformulated in the variables $\zeta = \beta \delta b^{2}$ and $\bar{t} = \omega_{A}\beta^{1/2} t$ as 
\begin{equation}
\partial_{t}\left[ 2 \zeta + \frac{3}{8}\zeta(\zeta-\zeta_{0})\right] = 
-a_{2} \zeta^{2} \left[ 1+ \frac{3}{8}(\zeta^{2}-
\zeta_{0}^{2})\right], \label{eq:noC wave damping equation}
\end{equation}
which has the benefit of being controlled by just one parameter $\zeta_{0} = \beta \delta b_{0}^{2}$.

A full  analytic solution to Eq.~\eqref{eq:noC wave damping equation} is intractable, but numerical solutions (not shown) match our expectations based on the qualitative discussion in Sec.~\ref{sub:noC traveling} above. Specifically, above
the interruption limit ($\zeta_{0}\gtrsim1$), $ u_{\perp}$ decays much faster than $ \delta B_{\perp}$ in time and $\delta B_{\perp}$ asymptotes to a constant nonzero value at late times, whereas below the interruption limit ($\zeta_{0}\lesssim 1$) the nonlinear damping more equally affects $u_{\perp}$ and $\delta B_{\perp}$  and there is simply a 
slow decay of both.

We now use Eq.~\eqref{eq:noC wave damping equation} 
to derive the initial decay rates of  $u_{\perp}$ and $\delta B_{\perp}$. This is most easily done by 
linearizing Eq.~\eqref{eq:noC wave damping equation} about $\zeta = \zeta_{0}$, viz., letting 
$\zeta =\zeta_{0}(1+\delta \zeta)$ and expanding in $\delta \zeta$. This gives 
\begin{equation}
\delta \zeta = -\frac{a_{2}\zeta_{0}}{2+3\zeta_{0}/8} \bar{t}.\end{equation}
Rewriting $\zeta $ in terms of $\delta b$, substituting $\bar{t} = \omega_{A}\beta^{1/2} t$, then calculating the decay rates \eqref{eq:decay rate defs} gives
\begin{equation}
\gamma_{B} \approx \frac{1}{2}a_{2}\omega_{A}\beta^{1/2}\delta b_{0}^{2}\frac{8}{16+3\beta\delta b_{0}^{2}},\qquad
\gamma_{U} \approx \frac{1}{2}a_{2}\omega_{A}\beta^{1/2}\delta b_{0}^{2}\frac{8+3\beta \delta b_{0}^{2}}{16+3\beta\delta b_{0}^{2}}.\label{eq:traveling u b decay rates}
\end{equation}
Although these expressions appear rather complicated, they agree nicely with our intuitive picture 
described earlier. In particular, the decay transitions from a regime where $\gamma_{B}\approx \gamma_{U}$ below 
the interruption limit $\beta \delta b_{0}^{2}\ll 1$, to one where $\gamma_{B}\ll \gamma_{U}$ (with $\gamma_{B}$ independent
of $\delta b_{0}$) when $\beta \delta b_{0}^{2}\gg1$.

A comparison of the damping rates \eqref{eq:traveling u b decay rates}
to the full LF traveling-wave solutions is presented in Fig.~\ref{fig:noC decay traveling}, where we show
 the decay rates  measured numerically for solutions starting with a sinusoidal traveling wave.
We find good agreement with the damping rates at low $\delta b_{0}$, when they are small,
and qualitative agreement with the trends predicted by Eq.~\eqref{eq:traveling u b decay rates} 
at larger $\delta b_{0}$. Note in particular that the decay rate of $\delta B_{\perp}$ 
changes from increasing to decreasing with $\beta $ at high $\delta b_{0}$, whereas 
the decay rate of $u_{\perp}$ does not. 
The quantitative agreement at high $\beta \delta b_{0}^{2}$ is lacking, and there are clear reasons for this discrepancy. First, there is our assumption
that the wave remains sinusoidal, which is patently not true when $\beta \delta b_{0}^{2}>1$ (see Fig.~\ref{fig:noC traveling}). The strong nonlinear shape modifications that do occur early in the evolution presumably 
involve some exchange of energy between $u_{\perp}$ and $B_{\perp}$ in ways that are not included in
our model. Secondly, the measurement of a decay rate is ambiguous for the strongly nonlinear $\beta \delta b_{0}^{2}>1$ 
solutions. For simplicity, we have fit the amplitude evolution from $t=0$ to $t=2\tau_{A}$ to a decaying exponential function, but the decay rate can vary
significantly over this  range at $\beta \delta b_{0}^{2}>1$. We have explored a variety of methods
for determining this initial decay rate of the wave, and although the quantitative results vary with 
method, the general properties and qualitative agreement with the predictions \eqref{eq:traveling u b decay rates} are robust.

\section{Fully kinetic and multi-dimensional effects}\label{sec:kinetic stuff}

Throughout the preceding sections, we have primarily focused on physical effects contained within the simplest
1-D Landau fluid equations \eqref{eq:KMHD rho}--\eqref{eq:GL heat fluxes l}.
Importantly, the mirror and oblique firehose instabilities are not included in this model,\footnote{The linear mirror instability can be  captured relatively accurately by the LF model that we use (if the equations are solved in 2 or 3 dimensions; see Sec.~8 of \citealp{Snyder:1997fs}). However, given the importance of trapped particles in the nonlinear mirror evolution \citep{Schekochihin:2008en,Kunz:2014kt,Rincon:2015mi,Melville:2015tt}, it seems quite unlikely that a LF model could  correctly reproduce the pressure-anisotropy-limiting behavior of the mirror instability,  although we know of no relevant study that tests this (see Sec.~\ref{sub:kinetic stuff, mirror} for discussion). } because
these grow at $k_{\perp}\sim k_{\parallel}$ on the Larmor scale and thus require 2-D or 3-D kinetic simulations
to be resolved correctly. 
In this section, we discuss -- based on previous fully kinetic theory and simulations -- some
possible effects of these microinstabilities on the global wave evolution, focusing on which aspects of the simple 1-D picture
described above are robust, and which may be modified by the inclusion of this physics. 
We also discuss other  kinetic effects that could modify our results, including FLR effects (these
were neglected by assuming  $k_{\parallel}\rho_{i}\ll 1$), particle trapping (this is not contained with the
LF prescription for the heat fluxes), heat-flux limits from the gyrothermal instability, and other scattering effects.
Of course, this discussion is in no way intended to be a replacement for future fully kinetic theory and simulations in two or three dimensions; rather, 
its purpose is to motivate the design of such studies and provide some guidance for interpreting their results. 

\subsection{Mirror instability}\label{sub:kinetic stuff, mirror}

In our discussion of standing waves (Sec.~\ref{sub:noC standing}),  the mirror instability was invoked 
to justify a limit on positive pressure anisotropies when starting from a velocity perturbation. This in
turn allowed the magnetic field to grow, reach its maximum, and then be interrupted in a similar way to an initial  purely
magnetic perturbation. Without this limiting effect, an initial velocity perturbation will create an oscillating 
wave [albeit not a linear SA wave because the restoring force is enhanced by the positive anisotropy; see Eqs.~\eqref{eq: footnote u ev} and \eqref{eq: u init jacobi sol}].
Thus, although the mirror instability is not crucial for the interruption effect itself, its presence does imply that
the effect cannot be significantly modified based on the initial conditions. We now discuss  in  more detail
why it is reasonable to assume that the mirror instability should have this effect. 

A variety of recent kinetic results \citep{Kunz:2014kt,Rincon:2015mi,Hellinger:2015mi,Melville:2015tt}  show that mirror fluctuations, which are unstable when $\Delta\gtrsim \beta^{-1}$ and cause perturbations in the field strength $\delta B$,
limit $\Delta p$ by trapping particles. Namely, as the macroscopic field grows and attempts to raise the pressure anisotropy, a larger and larger fraction of particles becomes trapped in the magnetic wells
and  ``sees'' a lower field. Thus, even though the volume-averaged field continues to increase, $\Delta $ is maintained 
at the marginal level $\beta^{-1}$ because a larger proportion of particles is trapped in the ever-deepening mirror wells. 
During this phase, the magnetic mirrors grow in time as $|\delta B/B|\sim (|\nabla \bm{u}|t)^{2/3}$ and there is very little particle scattering because their parallel scale is significantly larger than the Larmor radius. Further, since the mirrors only saturate and start scattering particles when $|\delta B/B|\sim 1$,  they  should {never} saturate and cause significant particle scattering for any SA wave initial
conditions with
$u_{x}(0)/v_{A} \sim |\nabla \bm{u}| \tau_{A}<1$. 
The numerical  experiments and arguments of \citet{Melville:2015tt} are particularly relevant to what happens 
as the magnetic field reaches its maximum and starts to decrease. For $\beta \ll \Omega_{i} /|\nabla \bm{u}|$ -- the ``moderate-$\beta$'', or large-scale-separation, regime most relevant to our results -- the mirrors should freely decay on time scales much shorter  than $\tau_{A }$ (the decay time is $ \sim \beta/\Omega_{i}$), releasing their trapped particles and allowing the anisotropy
to decrease towards the firehose limit. Although less well understood, it seems that in the opposite limit, $\beta \gg \Omega_{i} /|\nabla \bm{u}|$, the firehose limit is \emph{also} quickly reached (see \citealp{Melville:2015tt}, Sec.~3.2), probably because the  smaller-scale firehose fluctuations are able to grow on top of the larger-scale decaying mirrors.
Overall, it is thus reasonable to surmise that the mirror instability will effectively act as a passive limiter, ensuring 
$\Delta \lesssim \beta^{-1}$ but not strongly affecting large-scale wave dynamics.\footnote{As the scale separation $(k_{\parallel} \rho_{i})^{-1}$ is reduced, the mirrors will presumably become less effective (see \citealp{Kunz:2014kt}), allowing $\Delta$ to overshoot $\beta^{-1}$ before acting to limit the anisotropy. Thus very large  domains (compared to $\rho_{i}$) are likely
essential to see these effects in fully kinetic simulations.}

It is worth reiterating  a fundamental difference between the mirror and firehose limits for SA waves. At the firehose limit, 
the anisotropic stress nullifies the wave restoring force (i.e., the magnetic tension). In contrast, 
a plasma at the  mirror limit merely feels a modestly stronger (factor $3/2$)  restoring force. This difference explains
why the firehose limit is of much greater importance than the mirror limit for SA wave dynamics.

\subsection{Oblique firehose instability}\label{sub:kinetic stuff, firehose}

The oblique firehose instability is not included in our model, both  because it operates
at  $k_{\perp}\sim k_{\parallel}$ and because kinetic theory is required for its correct description. Linearly, oblique firehose fluctuations grow faster than parallel firehose fluctuations because of their smaller scale
\citep{Yoon:1993of,Hellinger:2008ob}, and are also seen clearly at larger
amplitudes in nonlinear regimes \citep{Kunz:2014kt,Melville:2015tt}.
Further, unlike mirror fluctuations, the firehose fluctuations will saturate and start scattering
particles after $t_{\mathrm{sat}}\sim \beta^{1/2}(|\nabla \bm{u}|\Omega)^{-1/2}$ (for $\beta \ll \Omega /|\nabla \bm{u}|$; see \citealp{Kunz:2014kt,Melville:2015tt}); 
i.e., after a very short time set by the microphysics. 

The most obvious question that arises is then whether the differences between  oblique  and parallel firehose dynamics will cause 
significant differences in the nonlinear interruption of SA waves, compared to 1-D models where only the parallel firehose exists.
This remains unclear, and understanding such issues will require fully kinetic simulations in 2-D or 3-D with $k_{z}\rho_{i}\ll 1$. Either scenario -- that the oblique firehose does or does not significantly  modify the SA wave dynamics -- can be plausible.  On the one hand, the oblique firehose
may behave similarly to the parallel firehose in 1-D simulations: be strongly excited during the early phases of wave interruption,  but then die away at later times because the pressure anisotropy is pushed back above the firehose limit. 
On the other hand, 
the enhanced particle scattering in kinetic oblique  firehose fluctuations
could possibly continue until the magnetic field decays completely, potentially leading to collisionless SA dynamics that more closely resemble a Alfv\'en wave in the Braginskii regime\footnote{In support of this idea, including an artificial hard-wall firehose limit
 at $\Delta = -2/\beta$ in a standing-wave LF simulation leads to the wave being strongly nonlinearly modified and then rather 
 quickly decaying to oscillate with an amplitude below the interruption limit. The effect on a 
 traveling wave is less severe, because the wave remains above the firehose limit for much of its decay.} (see Sec.~\ref{sec:Braginskii}). However, 
in either case, the presence of oblique firehose fluctuations cannot circumvent the interruption limit itself -- 
they have the same instability threshold as the parallel firehose, so  are not active until the wave restoring force has already disappeared.

\subsection{Other kinetic effects}\label{sub:kinetic stuff, other}

Here we outline several other possible kinetic effects. Unlike the mirror and oblique firehose modes discussed above, most
of these effects could be studied using 1-D, but fully kinetic simulations.
 
\paragraph{Finite-Larmor radius effects.}
Although various FLR effects can be included in Landau fluid models \citep{Goswami:2005cl,Ramos:2005ci,Passot:2012go,2015JPlPh..81a3203S}, the simple closure that 
we used here does not include these corrections. Assuming
 large scale separation, the most obvious effect from such corrections in 1-D 
is  the regularization of the small scales for the parallel firehose, which 
has its peak growth rate at $k_{\parallel} \rho_{i}\sim |\Delta+2/\beta|^{1/2}$ \citep{1968PhFl...11.2259D,Schekochihin:2010bv}.
Since we found that collisionless wave-interruption dynamics did not depend significantly on numerical resolution (which effectively sets the fastest-growing firehose mode in our fluid model), it 
seems unlikely that the direct effect of this regularization will be particularly important to the large-scale interruption
(but note that the requirement $k_{\parallel}\rho_{i} \ll 1$ could be quite severe, because $|\Delta+2/\beta|^{1/2}\ll 1$ and
we need significant separation between the firehose modes and the wave). 
There could, however, be other effects that  
are of some significance. For example, FLR effects enable a new instability --
 the ``gyrothermal instability'' \citep{Schekochihin:2010bv} -- which may act to limit the heat fluxes  before the SA wave hits the interruption limit,  in 
 a similar way to how the firehose instability limits the pressure anisotropy \citep{Rosin:2011er}.
Through the gyro-viscous terms (the off-diagonal elements of the pressure tensor) and the Hall effect,\footnote{The Hall effect  becomes important when $k_{\parallel}d_{i}\sim 1$, where $d_{i}\sim \sqrt{\beta}\rho_{i}$ is the 
 ion skin depth.} FLR effects can also act to circularly polarize the wave, creating a $B_{y}$
 perturbation from a spatially varying $B_{x}$.   However, this is  presumably only directly important for the macroscopic wave when the scale separation is modest, or in regions with large gradients that form during
 nonlinear evolution. Simple extensions of the LF model (solved numerically throughout Sec.~\ref{sec:collisionless}) to include gyro-viscous effects and/or the Hall effect (not shown) have illustrated that these terms cause only minor changes to the SA wave evolution, so long as $k_{\parallel}\rho_{i}$ is sufficiently small. 
 
\paragraph{Particle trapping.}
Since LF closures prescribe the heat fluxes based on linear 
Landau-damping rates, effects of particle trapping are not included in these closures
and  may provide an order-unity correction to the heat fluxes. In particular,  trapping can be important whenever 
the bounce frequency $\omega_{b}$ of particles approaches the frequency of large-scale motions (this is $\sim\!\omega_{A}$ for a SA wave). Given that particles with velocity $\bm{v}$ and parallel velocity $v_{\parallel} $ are trapped 
if $\xi = v_{\parallel}/|\bm{v}| <\xi_{\mathrm{tr}} \sim |\delta B/B_{0}|^{1/2}$, while $|\delta B/B_{0}|^{1/2}\sim |\delta B_{\perp}/B_{0}|$, a simple estimate for the bounce frequency is $\omega_{b}\sim \beta^{1/2}\delta b\, \omega_{A}$. Thus, trapping can be important if $\delta b >\beta^{-1/2}$; i.e., for a wave above the interruption limit.
Trapping has the effect of reducing the Landau damping rate of nonlinear SA traveling waves \citep{Oneil:1965hd,Kulsrud:1978cr} and presumably also modifies  pressure-anisotropy damping.  These effects  will be considered in detail in future work.

\paragraph{Other scattering effects.}
If the  ``corners'' that develop in the magnetic-field lines (e.g., Figs.~\ref{fig:noC standing}--\ref{fig:noC traveling}) are
on the Larmor scale, unresolved by our LF closure, these may scatter particles. 
This would provide an interesting case where a  plasma could set its own mean free path  $\lambda_{\mathrm{mfp}}\sim k_{\parallel}^{-1}$
based on the large-scale driving. If real, this effect would most significantly modify traveling-wave dynamics, because
the square structures that develop \emph{before} interruption (see Fig.~\ref{fig:noC traveling}) could possibly cause
sufficient scattering to damp the global pressure anisotropy fast enough so that the wave decayed before reaching the interruption limit.

\vspace{4mm}
Overall, we would like to stress that although the details of wave interruption may be modified by the addition of other kinetic
physics, our basic result -- that weakly collisional SA waves cannot 
exist in their linear  form above the limits \eqref{eq:Brag standing cond} and \eqref{eq:noC standing wave cond} -- 
is robust. The dominance of magnetic energy over kinetic energy is also 
 a generic consequence of interruption, because in the approach to the firehose
 limit, the equipartition of energy in an Alfv\'en wave is modified by the decrease of magnetic tension. 
We find  generic agreement on these points between 
the Landau fluid, Braginskii, and double-adiabatic models ($q_{\perp}=q_{\parallel}=0$; see \ref{app:CGL}). Many of our key 
 results are thus quite insensitive to the form of the heat fluxes or particle scattering,
relying purely on the physics of pressure-anisotropy generation in a changing magnetic field.

\section{Conclusion}\label{sec:conclusion}

In this paper, we have explored the nonlinear   ``interruption'' and damping
of linearly polarized shear-Alfv\'en (SA) waves in weakly collisional plasmas. 
These effects, which arise due to the pressure anisotropy that is generated
in the changing magnetic field of the wave, lead to a limit on the amplitude 
of propagating/oscillating SA waves  in the collisionless regime:
\begin{equation}
\frac{\delta B_{\perp}}{B_{0}}\lesssim \beta^{\,-1/2}.\label{eq:conclusion amp limit}\end{equation}
In the weakly collisional Braginskii limit, which applies when $\nu_{c}\gg \omega_{A}$, propagating/oscillating SA waves are also limited in amplitude, to 
\begin{equation}
\frac{\delta B_{\perp}}{B_{0}}\lesssim \sqrt{\frac{\nu_{c}}{\omega_{A}}}\beta^{\,-1/2}.\label{eq:conclusion amp limit B}
\end{equation}
We summarize our main findings as follows:
\begin{itemize}
\item {Above the limit \eqref{eq:conclusion amp limit}, collisionless SA waves are ``interrupted'' when their self-generated pressure anisotropy
reaches the firehose boundary $\Delta p \approx -B^{2}/4\pi$.
At this boundary, the wave's restoring force (the Lorentz force) is cancelled by the anisotropy, and
the magnetic energy dominates the kinetic energy because 
the effective Alfv\'en speed  goes to zero.}
\item{Due to the  correlation between $\Delta p $ and  $B^{-1}dB/dt$, there is a net transfer of 
wave energy to thermal energy of the plasma through ``pressure-anisotropy heating'' at the rate $\int d\bm{x}\, \Delta p B^{-1}dB/dt$. This results in a nonlinear damping of the
wave, even below the limit \eqref{eq:conclusion amp limit}.}
\item{Heat fluxes are always  important in the high-$\beta$ collisionless limit (because the thermal 
velocity is larger than $v_{A}$) and act to  smooth the spatial dependence of the pressure anisotropy. }
\item{In the collisionless limit,  standing and traveling SA waves behave in qualitatively different ways
because the spatial average of $B$ decreases during a standing wave's evolution, whereas it does not for a traveling
wave. Thus, while a standing wave above the limit  \eqref{eq:conclusion amp limit} is interrupted within  half a wave period,  
a traveling wave is first nonlinearly damped [at the rate $\sim \omega_{A}  (\delta B_{\perp}/B_{0})^{2}\beta^{\,1/2}$], leading to a decreasing  $B$ and eventual interruption of  the wave. }
\item{The kinetic energy in a collisionless traveling wave is damped significantly faster than the magnetic energy
 for amplitudes approaching (or exceeding) the limit \eqref{eq:conclusion amp limit}, and the  magnetic energy can be a large fraction of 
 its initial value
 when the wave interrupts.
This occurs because, as the wave decays, the global decrease in $\Delta p $ reduces $v_{A}$, which changes the ratio of $u_{\perp}$ and $\delta B_{\perp}$ (this also slows down the wave; see Fig.~\ref{fig:noC traveling}).
}
\item{Barring additional kinetic and higher-dimensional effects not contained within our Landau-fluid model (see Sec.~\ref{sec:kinetic stuff}), 
the outcome of wave interruption is the creation of a magnetically dominated state of nearly perfect zig-zag magnetic field lines (see Figs.~\ref{fig:noC standing} and \ref{fig:noC traveling}) -- 
i.e., a quasi-periodic pattern with spatially constant magnetic field strength. The emergence of this state may be understood by noting that it is 
the only state that has both zero magnetic tension and a spatially smooth pressure anisotropy along the field lines (because spatial variation in $\Delta p $ is damped by the 
heat fluxes).}
\item{Wave interruption in the Braginskii limit involves a slow decay 
of the wave over the timescale $t_{\mathrm{decay}}\sim {\beta}/\nu_{c} (\delta B_{\perp}/B_{0})^{2}$, which occurs 
because a slowly changing $B$ is necessary to maintain the anisotropy at the firehose limit. The characteristic 
field-line structures (Fig.~\ref{fig:Brag standing}) differ from  collisionless waves because
the magnetic tension is zero if the anisotropy is at the firehose limit, even if there is 
spatial variation in $B^{2}$. }
\item{The pressure-anisotropy damping of SA waves is  large in a Braginskii plasma because the spatially varying part of $\Delta p$ is comparable to its mean. For waves below the limit \eqref{eq:conclusion amp limit B}, this leads to  the wave energy being damped at the rate  $\sim \omega_{A}^{2}/\nu_{c}  (\delta B_{\perp}/B_{0})^{2}\beta$.\footnote{This result is valid 
in the regime where  the 
heat fluxes are unimportant; see \ref{app:asymptotics Brag}.}  }
\item{The amplitude limits do not apply to circularly polarized SA waves because for these, $dB/dt = 0$.}
\end{itemize}

\subsection{Implications and applications}\label{sub:implications and applications}

Given the ubiquity of shear-Alfv\'en waves in plasma physics (see Sec.~\ref{sec: intro}),  the stringent limits on their 
amplitude derived here may have interesting implications in a variety of hot, low-density (and therefore, 
weakly collisional) astrophysical plasmas. Although a detailed study of all these implications is beyond the scope of this work,   it is worth commenting
on magnetized turbulence in particular,  given its
importance in many subdisciplines of astrophysics.
The salient point is that in well-accepted phenomenologies 
of strong magnetized turbulence (in particular, \citealp{Goldreich:1995hq}, and extensions, e.g., \citealp{Boldyrev:2006ta}), the physics of 
shear-Alfv\'en waves is critical at all scales in the turbulence.\footnote{A more common way to say this is that the cascade is in \emph{critical balance} \citep{Goldreich:1995hq}, which states that the linear  
(Alfv\'en) time is equal to the nonlinear turnover time at all scales in a strong MHD cascade.}
A strong modification to SA wave physics would thus be expected to significantly modify
the turbulent cascade. 
One might expect such modifications to be even stronger in the weak turbulence regime \citep{Ng:1996ik,2000JPlPh..63..447G,2012PhRvE..85c6406S} given the relative weakness of nonlinear 
interactions in comparison to the SA wave physics in such turbulence.

\setcounter{footnote}{0}

More explicitly,  turbulence in a weakly collisional high-$\beta$ plasma may depend on the 
amplitude of its forcing. Since velocities are strongly damped 
when a wave is interrupted, it may behave as a fluid with Reynolds number $\lesssim 1$ when  $ u_{\perp}/v_{A}\gtrsim \beta^{-1/2}$ (and $ u_{\perp}\lesssim v_{A}$; otherwise the waves are of such
large amplitude that the turbulence would likely be in a dynamo regime, which is not particularly well understood even in MHD, but is less obviously Alfv\'enic).
In contrast, for perturbations of amplitude $ u_{\perp}/v_{A}\lesssim \beta^{-1/2}$, the linear SA wave physics
is mostly unaffected, and a standard Alfv\'enic cascade should develop.
Since pressure-anisotropy heating is able to dissipate large-scale wave energy when wave interruption is important, a turbulent 
cascade may not be necessary for the plasma to absorb the energy injected by a continuous mechanical forcing \citep{Kunz:2010gv}.  
While  further study is necessary to  understand this physics better, it is at least clear that the immediate 
interruption of SA fluctuations with amplitudes exceeding $\delta B_{\perp}/B_{0}\sim \beta^{\,-1/2}$  should significantly limit
the application of MHD-based turbulence phenomenologies to high-$\beta$ weakly collisional plasmas. 

\subsection{Future work}
A first priority for future studies of wave interruption is the inclusion of the  kinetic effects discussed in 
Sec.~\ref{sec:kinetic stuff}. Unfortunately (from a computational standpoint), due to the 2-D kinetic nature of the mirror and oblique firehose
instabilities, significant progress in this endeavor requires kinetic simulations in two spatial dimensions and three velocity-space dimensions.
Since  firehose instabilities grow on scales somewhat above the gyroscale (see Sec.~\ref{sec:kinetic stuff}), 
if one hopes to study the
 asymptotic regime $k_{\parallel}\rho_{i}\ll 1$,
the required scale separation between the gyroscale and the SA wave is likely quite large. We thus expect a detailed kinetic 
study to be rather computationally expensive,  although certainly feasible.
That said, there will also be a variety of interesting insights to be gained 
from purely 1-D kinetics:  for example, the role of the gyrothermal instability, particle scattering off magnetic discontinuities,    and particle-trapping effects. 
Further, the behavior of SA waves with limited scale separation between $k_{\parallel}^{-1}$ and $\rho_{i}$ is also 
of interest physically, in particular for the solar wind, where observations easily probe turbulent fluctuations down to the ion gyroscales and below.
A separate line of investigation for future work involves applications of the amplitude limit, in particular to turbulence, as discussed in Sec.~\ref{sub:implications and applications}.
This may be productively pursued using a 3-D Landau fluid code (as in \citealp{Sharma:2007cr,Sharma:2006dh}) or 
using Braginskii MHD.

 Overall, although many questions remain, 
 both the limit $\delta {B}_{\perp}/{B}_{0} \lesssim \beta^{\,-1/2}$ itself and the strong 
 dominance of magnetic over kinetic energy are robust, appearing across  a variety of models. Given the stringent nature 
of the  amplitude limit and the interesting implications for high-$\beta$ magnetized turbulence in weakly collisional plasmas, we anticipate a
 range of future applications to heliospheric, astrophysical, and possibly laboratory \citep{Forest:2016,LAPD:2016} plasmas. 

\vspace{4mm}

\section*{Acknowledgements}

It is a pleasure to thank S.~Cowley, M.~Kunz, S.~Bale, C.~H.~K~Chen, S.~Balbus, L.~Sironi, F.~Rincon, and M.~Strumik for useful and enlightening discussions. JS is  endebted to Merton College for supporting a stay in Oxford where some of this work was carried out. JS and AAS would also like to thank the Wolfgang Pauli Institute in Vienna for its hospitality during the $9^{\mathrm{th}}$ Plasma Kinetics Working Group Meeting. JS was funded in part by the Gordon and Betty Moore Foundation
through Grant GBMF5076 to Lars Bildsten, Eliot Quataert and E. Sterl
Phinney. AAS was supported in part by grants from  UK STFC and EPSRC. EQ was supported by Simons Investigator awards from the Simons Foundation and  NSF grant AST 13-33612.

\vspace{4mm}

\appendix

\label{HERE IS THE APPENDIX}
\section{Asymptotic wave equations -- collisionless limit}\label{app:asymptotics}

In this appendix, we derive a variety of wave equations to describe standing and traveling shear-Alfv\'en waves in collisionless
regimes (the Braginskii regime is treated in \ref{app:asymptotics Brag}).
This is carried out by means of  asymptotic expansions of Eqs.~\eqref{eq:KMHD rho}--\eqref{eq:GL heat fluxes l} in $\epsilon \sim \delta b = \delta B_{\perp}/B_{0}$, with $\delta b/\delta b_{\mathrm{max}}=\delta b\, \beta^{1/2}\sim \mathcal{O}(1)$. 
These calculations formally justify some of the ideas presented  in the main text; e.g., the flattening effects of the 
heat fluxes and the scaling of traveling wave damping. In addition,
the  theory allows the determination of the numerical value for the initial decay rate of a traveling wave [i.e., $a_{2}$ in Eq.~\eqref{eq:Ewave noC}], and the form of the  nonlinearity that arises from the $\mathcal{O}(\beta^{\,-1/2})$ spatially 
varying part of $\Delta$ near interruption [see Eq.~\eqref{eq:noC Delta}]. In all cases, we consider a strictly one-dimensional wave, as in the main
text. 

Although an asymptotic expansion, as promised above, is in principle straightforward, in practice there are several issues that arise. 
Most importantly, 
our ordering scheme does not allow for  a single wave equation that describes both the early evolution of a wave [i.e., when 
$1+ \beta \Delta/2\sim  \mathcal{O}(1)$]
and  the final approach to the interruption (when $1+\beta \Delta/2\ll 1 $). This problem is related to 
 $\Delta$ being spatially constant at lowest order because of the smoothing effect of the heat fluxes, even though 
the spatially variation in $\Delta$
plays a key role in forming square structures  as the wave approaches the interruption limit. 
This motivates two
separate expansions: the first is valid when $1+\beta \Delta/2 \sim \mathcal{O}(1)$, the second is valid when $1+\beta \Delta/2 \sim \mathcal{O}(\epsilon^{2})$ (i.e., when the wave has already evolved to be close to the interruption limit). 

These difficulties motivate our arrangement of this appendix as follows. We start in \ref{app:CGL} by
considering the double-adiabatic version of Eqs.~\eqref{eq:KMHD rho}--\eqref{eq:KMHD pl}, with 
$q_{\perp}=q_{\parallel}=0$. This leads to a simple nonlinear wave equation that is free from the issues
mentioned in the previous paragraph because there is large spatial variation in $\Delta$. We then consider the initial wave evolution 
using the Landau-fluid closure in \ref{app:initial wave}, ordering $u_{\perp}/v_{A}\sim \delta B_{\perp}/B_{0} \sim \mathcal{O}(\epsilon)$ and $1+\beta \Delta/2\sim \mathcal{O}(1)$,
which can be used to find the initial damping rate of a traveling wave. Finally we 
derive an equation for waves as they get very close to the interruption limit itself in \ref{app:near interruption}.  This
involves a spatially varying nonlinearity arising from both the field curvature and the spatial variation of $\Delta$.
Unfortunately, a closure problem prevents true asymptotic determination of the evolution of
the spatial mean of $\Delta$, although the expansion is still helpful for determining 
the residual spatial variation of $\Delta$ and formulating a simple nonlinear
wave equation that  describes the wave's approach to zig-zag field-line structures.

\subsection{Nondimensionalized equations}\label{app:equations}

For the sake of algebraic simplicity and to emphasize the appearances of $\beta$, throughout
the following sections we work in dimensionless variables. 
These are chosen such that the Alfv\'en frequency and wavenumber are both unity:
\begin{gather}
\bm{x} = k_{\parallel}^{-1}\bar{\bm{x}},\quad t=\omega_{A}^{-1}\bar{t},\quad{\bm{u}}=v_{A0}\bar{\bm{u}},\quad{\bm{B}}=B_{0}\bar{\bm{B}},\quad\rho = \rho_{0}\bar{\rho},\quad p_{\perp}=p_{0}\bar{p}_{\perp},\quad p_{\parallel}=p_{0}\bar{p}_{\parallel}, \nonumber \\  q_{\perp}=c_{s} p_{0} \bar{q}_{\perp},\quad q_{\parallel}=c_{s} p_{0} \bar{q}_{\parallel},\quad \beta_{0} \equiv \frac{8\pi p_{0}}{B_{0}^{2}},\quad v_{A0}\equiv \frac{B_{0}}{\sqrt{4\pi \rho_{0}}}, \nonumber \\
 c_{s}^{2}=2\frac{p_{0}}{\rho_{0}}=\beta_{0}v_{A0}^{2},\quad\omega_{A}\equiv k_{\parallel}v_{A0},\quad\nu_{c}=\omega_{A}\bar{\nu}_{c},\quad \Delta\equiv \bar{p}_{\perp}-\bar{p}_{\parallel}.\label{eq:dimensionless variables}
\end{gather}
Substituting these definitions into Eqs.~\eqref{eq:KMHD rho}--\eqref{eq:KMHD pl} and Eqs.~\eqref{eq:GL heat fluxes p}--\eqref{eq:GL heat fluxes l}, 
one obtains
\begin{gather}
\partial_{t} \rho + \nabla \cdot (\rho \bm{u})=0,\label{eq:NDMHD rho}\\[2ex]
\rho(\partial_{t}\bm{u}+\bm{u}\cdot \nabla \bm{u}) = -\frac{\beta_{0}}{2}\nabla p_{\perp} - \nabla \frac{B^{2}}{2}+\nabla \cdot \left[ \hat{\bm{b}}\hat{\bm{b}}\left(\frac{\beta_{0}}{2}\Delta + B^{2} \right)\right], \label{eq:NDMHD u}\\[2ex]
\partial_{t}\bm{B}+\bm{u}\cdot \nabla \bm{B} = \bm{B}\cdot \nabla \bm{u} - \bm{B}\nabla \cdot \bm{u},\label{eq:NDMHD B}\\[2ex]
\partial_{t}p_{\perp} + \nabla \cdot (p_{\perp}\bm{u})+ p_{\perp}\nabla \cdot \bm{u} + \beta_{0}^{\,1/2} \left[ \nabla \cdot (q_{\perp}\hat{\bm{b}}) + q_{\perp}\nabla \cdot \hat{\bm{b}}\right] = p_{\perp}\hat{\bm{b}}\hat{\bm{b}}:  \nabla \bm{u} -\bar{\nu}_{c}\Delta,\label{eq:NDMHD pp}\\[2ex]
\partial_{t}p_{\parallel} + \nabla \cdot (p_{\parallel}\bm{u}) + \beta_{0}^{\,1/2} \left[ \nabla \cdot (q_{\parallel}\hat{\bm{b}}) -2 q_{\perp}\nabla \cdot \hat{\bm{b}}\right] = -2p_{\parallel}\hat{\bm{b}}\hat{\bm{b}}:  \nabla \bm{u} +2\bar{\nu}_{c}\Delta,\label{eq:NDMHD pl}\\[2ex]
q_{\perp} = -\sqrt{\frac{ p_{\parallel}}{\pi \rho}} \frac{1}{|k_{\parallel}| +\bar{\nu}_{c}  (\beta \pi p_{\parallel}/\rho)^{-1/2}} \left[ \rho  \nabla_{\parallel} \left(\frac{p_{\perp}}{\rho}\right)  - p_{\perp}\left(1-\frac{p_{\perp}}{p_{\parallel}} \right)\frac{\nabla_{\parallel} B}{B}  \right], \label{eq:NDMHD qp}\\[2ex]
q_{\parallel} = -2\sqrt{\frac{ p_{\parallel}}{\pi \rho}} \frac{1}{|k_{\parallel}| +(3\pi/2-4)\bar{\nu}_{c}  (\beta \pi p_{\parallel}/\rho)^{-1/2}} \rho \nabla_{\parallel} \left(\frac{p_{\parallel}}{\rho}\right),\label{eq:NDMHD ql}
\end{gather}
where $B^{2}\equiv \bm{B}\cdot \bm{B}$, and $\hat{\bm{b}}\equiv \bm{B}/B$. The  bars on  variables are henceforth suppressed. 

We shall carry out all calculations in 1-D in $z$ (as in the main text) with
 the imposed background magnetic field $\bm{B}=B_{0}\hat{\bm{z}}$ [this is $\bar{\bm{B}}=1\hat{\bm{z}}$ in the dimensionless variables \eqref{eq:dimensionless variables}].
Magnetic-field perturbations $\delta \bm{B}_{\perp}$ are taken to be in the $\hat{\bm{x}}$ direction, 
which implies that $\hat{\bm{y}}$ directed vector components of $\bm{u}$ and $\bm{B}$ are identically zero. Note 
that because $B_{0}=1$, the $B_{x}$ used throughout the following sections is the same 
as  the $\delta b$ used in the main text [e.g., Eq.~\eqref{eq:waves}]. 
Our expansion is carried out in $\epsilon \sim B_{x}$ with $B_{x} \beta_{0}^{1/2}\sim \epsilon^{0}$, which implies
$\beta_{0}\sim \epsilon^{-2}$.
With such a scaling ($\beta_{0}\gg 1$), it is immediately apparent that the pressure terms [in the 
momentum equation \eqref{eq:NDMHD u}] and the heat-flux terms [in the pressure equations \eqref{eq:NDMHD pp}--\eqref{eq:NDMHD pl}] dominate, since 
all space and time derivatives for an Alfv\'en wave are $\sim \mathcal{O}(1)$. 

Throughout the following sections
we also define the spatial average 
\begin{equation}
\langle f \rangle \equiv \frac{1}{2\pi}\int dz\, f,
\end{equation}
and the spatially varying  part of a quantity
\begin{equation}
\tilde{f}\equiv f-\langle f\rangle.\end{equation}

\subsection{The double-adiabatic limit}\label{app:CGL}
It transpires that an asymptotic expansion of the  double-adiabatic equations  [$q_{\perp}=q_{\parallel}=0$ in Eqs.~\eqref{eq:NDMHD rho}--\eqref{eq:NDMHD pl}]
is significantly simpler 
than that with the Landau closure for the heat fluxes [Eqs.~\eqref{eq:NDMHD qp} and \eqref{eq:NDMHD ql}].
This is because a  nonlinearity with spatial dependence
appears in the lowest-order wave equation, due to the 
spatial variation of $\Delta$ being comparable to its mean. 
For this reason we start by outlining 
the procedure for the double-adiabatic equations, 
even though the neglect of the heat fluxes is not a valid approximation  in the high-$\beta$ limit. 

Our asymptotic ordering, motivated by our interest in solutions near the
interruption limit with $\delta B_{\perp}/B_{0}\sim \beta_0^{\,1/2}$, is 
\begin{gather}
B_{x} = \epsilon B_{x1} + \epsilon^{2}B_{x2} + \mathcal{O}(\epsilon^{3}),\label{eq:Bx expansion 1}\\
u_{x} = \epsilon u_{x1} + \epsilon^{2}u_{x2} + \mathcal{O}(\epsilon^{3}),\\
u_{z} = \epsilon^{2} u_{z2} + \epsilon^{3} u_{z3}+ \mathcal{O}(\epsilon^{4}),\\ 
\rho = 1+ \epsilon^{2}\rho_{2} +\epsilon^{3}\rho_{3} +\mathcal{O}(\epsilon^{4}),\\
p_{\perp} = 1+ \epsilon^{2} p_{\perp 2}+ \epsilon^{4} p_{\perp 4} + \mathcal{O}(\epsilon^{4}),\\ 
p_{\parallel} = 1+ \epsilon^{2} p_{\parallel 2} + \epsilon^{3} p_{\parallel 3}  \mathcal{O}(\epsilon^{4}).\label{eq:pl expansion 1}
\end{gather}
In addition, $B_{z}=1$ due to $\nabla \cdot \bm{B}=0$.
This leads to 
\begin{gather}
B = 1+\frac{1}{2}\epsilon^{2}B_{x1}^{2}+ \mathcal{O}(\epsilon^{3}),\\
\hat{b}_{x} = \epsilon B_{x1} + \epsilon^{2} B_{x2} + \epsilon^{3} \left( B_{x3} - \frac{B_{x1}^{3}}{2} \right)+ \mathcal{O}(\epsilon^{4}),\\
\hat{b}_{z} = 1 - \frac{1}{2} \epsilon^{2} B_{x1}^{2}+ \mathcal{O}(\epsilon^{3}),\\
\Delta = \epsilon^{2}(p_{\perp 2} - p_{\parallel 2}) + \epsilon^{3}(p_{\perp 3} - p_{\parallel 3}) + \mathcal{O}(\epsilon^{4}) = \epsilon^{2}\Delta_{2} + \epsilon^{3}\Delta_{3} + \mathcal{O}(\epsilon^{4}).
\end{gather}
We now insert the expansions \eqref{eq:Bx expansion 1}--\eqref{eq:pl expansion 1} into  Eqs.~\eqref{eq:NDMHD rho}--\eqref{eq:NDMHD pl}  and expand in $\epsilon$
to obtain a wave equation for $B_{x1}$.

\paragraph{Order $\epsilon^{0}$.} 
There is only one contribution at $\mathcal{O}(\epsilon^{0})$, which comes from the $z$ component of  Eq.~\eqref{eq:NDMHD u},
\begin{equation}
-\frac{\beta_0}{2}\partial_{z} p_{\perp2} + \frac{\beta_0 }{2}\partial_{z}(\hat{b}_{z0}\hat{b}_{z0}p_{\perp2}-p_{\parallel2})=-\frac{\beta_0}{2} p_{\parallel 2} = 0.\label{eq: O0 parr mom eq}\end{equation}
This says that $p_{\parallel 2}$ has no spatial variation, $\widetilde{p_{\parallel 2}}=0$, expressing the parallel pressure balance.

\paragraph{Order $\epsilon^{1}$.} 
The perpendicular velocity at $\mathcal{O}(\epsilon$) satisfies
\begin{equation}
\partial_{t} u_{x1} = \partial_{z} \left[ B_{x1 } \left(1+ \frac{\beta_0}{2} \Delta_{2}\right)\right], \label{eq:CGL asym wave}\end{equation}
while the induction equation \eqref{eq:NDMHD B} is simply
\begin{equation}
\partial_{t} B_{x1} = \partial_{z}u_{x1}.  \label{eq:CGL asym induc}\end{equation}
The parallel momentum equation \eqref{eq:NDMHD u} again gives $\partial_{z} p_{\parallel 3} = 0$ and there is no contribution 
from the continuity \eqref{eq:NDMHD rho} or pressure equations \eqref{eq:NDMHD pp}--\eqref{eq:NDMHD pl}. 
From Eq.~\eqref{eq:CGL asym wave}, it is clear that we need an expression for $\Delta_{2}$, and, 
for the system to be closed, this must depend only on $u_{x1}$ and $B_{x1}$.

\begin{figure}
\begin{center}
\includegraphics[width=0.5\columnwidth]{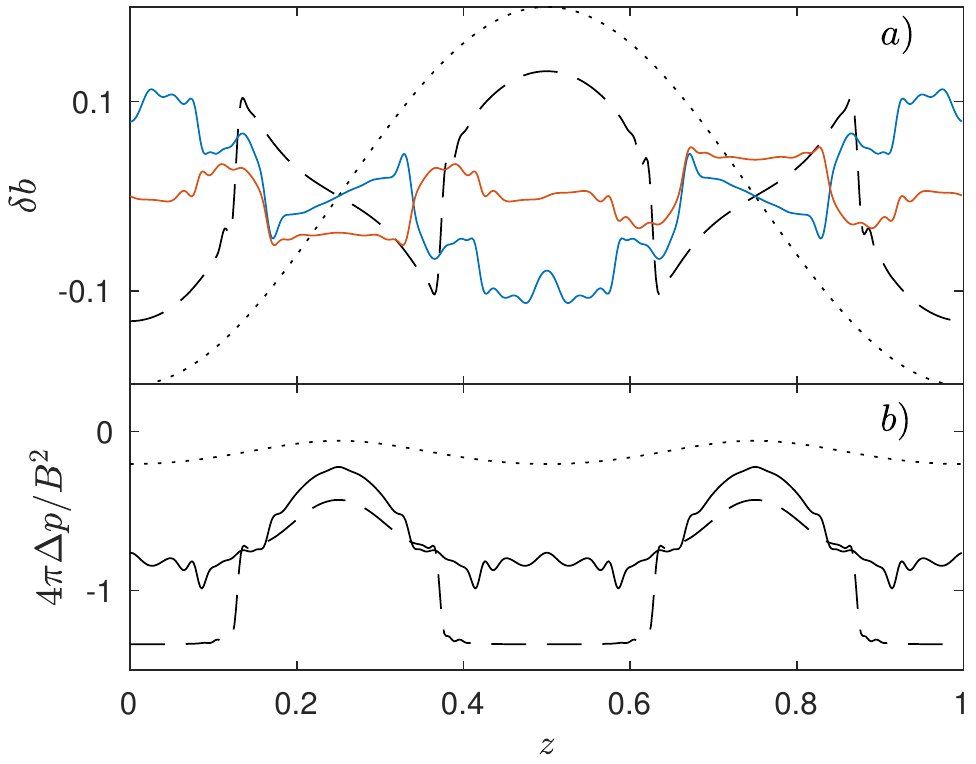}~~\includegraphics[width=0.5\columnwidth]{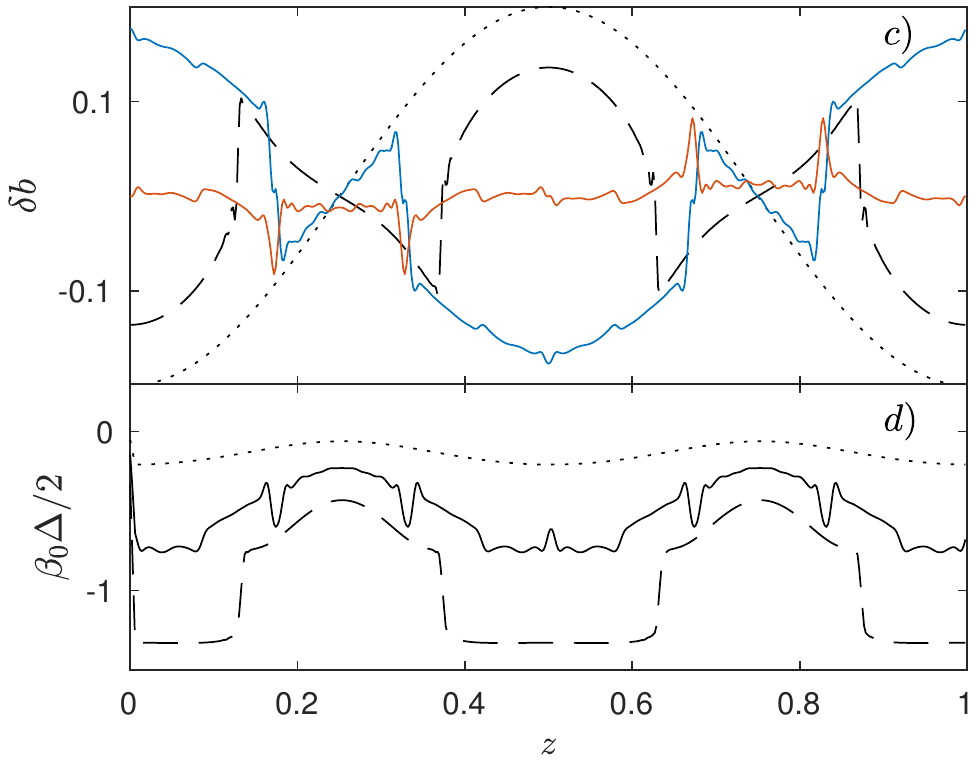}
\caption{Evolution of an initial magnetic perturbation [standing wave; initial conditions \eqref{eq: standing b ICs}] within the CGL model, at $\beta_0=100$ with $\delta b(t=0)=B_{x1}(0) =0.2$. Panels (a)--(b) show the solutions of the full CGL model [Eqs.~\eqref{eq:NDMHD rho}--\eqref{eq:NDMHD pl}], while panels (c)--(d) shows the solution to the asymptotic wave equation \eqref{eq: final CGL wave eq}.
The top panels show $\delta b$ at $t=0$ (black dotted line), $\delta b $ at $t=0.25 \tau_{A}$ (black dashed line), $\delta b$ at $t=\tau_{A}$ (blue solid line), and $u_{x}/v_{A}$ at $t=\tau_{A}$ (red solid line). 
The bottom panels show the pressure anisotropy parameter, $4\pi \Delta p/B^{2}$ in panel (b) and $\beta_{0}\Delta/2$ in panel (d) (these are $-1$ at the firehose limit in each case), at $t=0.05\tau_{A}$ (dotted line), $t=0.25\tau_{A}$ (dashed line), and $t=\tau_{A}$ (solid line).
Note the differences between these solutions and   the evolution of a wave with the Landau fluid prescription for the heat fluxes (e.g., Fig.~\ref{fig:noC standing}), which should be expected due to the different form of 
the spatially varying nonlinearity. 
   }
\label{fig:app: CGL evolution}
\end{center}
\end{figure}

\paragraph{Order $\epsilon^{2}$.} 

To calculate $\Delta_{2}$, we require only $p_{\perp 2}$ and $p_{\parallel 2}$, so may ignore the momentum \eqref{eq:NDMHD u} and induction \eqref{eq:NDMHD B} equations at this 
order. 
Noting that $\hat{\bm{b}}\hat{\bm{b}}:  \nabla \bm{u} = \hat{{b}}_{z}\hat{{b}}_{z} \partial_{z}u_{z} +  \hat{{b}}_{x}\hat{{b}}_{z} \partial_{z}u_{x} $,
the perpendicular and parallel pressure equations \eqref{eq:NDMHD pp}--\eqref{eq:NDMHD pl} become
\begin{gather}
\partial_{t}p_{\perp 2} + \partial_{z} u_{z2} = B_{x1}\partial_{z}u_{x1},\label{eq:pp CGL expansion}\\
\partial_{t}p_{\parallel 2} + 3 \partial_{z} u_{z2} = -2 B_{x1}\partial_{z}u_{x1}.
\end{gather}
From the $\mathcal{O}(\epsilon^{0})$ parallel momentum equation \eqref{eq: O0 parr mom eq}, we know that $\widetilde{p_{\parallel 2}}=0$,
so, using $\langle{\partial_{z} u_{z2}}\rangle =0$, we obtain 
\begin{equation}
\partial_{z} u_{z2} = -\frac{2}{3}\widetilde{B_{x1}\partial_{z}u_{x1}}.\end{equation}
Combining this with the perpendicular pressure equation \eqref{eq:pp CGL expansion} and assuming  $\Delta_{2}(t=0)=0$, we find 
\begin{equation}
\Delta_{2} = \frac{5}{6}(B_{x1}(t)^{2} - B_{x1}(0)^{2}) + \frac{2}{3}\langle B_{x1}(t)^{2} - B_{x1}(0)^{2}\rangle. \label{eq: CGL D2}\end{equation}
Inserting Eq.~\eqref{eq: CGL D2} into Eq.~\eqref{eq:CGL asym wave} and using Eq.~\eqref{eq:CGL asym induc}, we obtain a closed wave equation for $B_{x1}$:
\begin{equation}
\partial_{t}^{2} B_{x1} = \partial_{z}^{2} \left[ B_{x1}\left( 1+ \frac{5}{12} \beta_0[B_{x1}^{2}]_{0}^{t}+ \frac{1}{3}\beta_0\langle [ B_{x1}^{2} ]_{0}^{t}\rangle  \right) \right], \label{eq: final CGL wave eq}\end{equation}
where $[f]_{0}^{t} \equiv f(t)-f(0)$. 

Because there is a strong spatially varying nonlinearity arising from the $[B_{x1}^{2}]_{0}^{t}$ term, Eq.~\eqref{eq: final CGL wave eq} can  represent solutions of the full double-adiabatic equations \eqref{eq:NDMHD rho}--\eqref{eq:NDMHD pl}
both initially and near the interruption limit.
As illustrated in Fig.~\ref{fig:app: CGL evolution}, where we compare
solutions of the full double-adiabatic model with those of the wave equation \eqref{eq: final CGL wave eq}, this nonlinearity causes a significant change in the shape of the wave as it evolves. As should be expected because
of the different spatial form of the nonlinearity, the nonlinear evolution at the  interruption limit is quite different from
the evolution
when heat fluxes are included (cf. Fig.~\ref{fig:noC standing}). 
 In particular, since the heat fluxes no longer act to spatially 
smooth the pressure anisotropy, $B^{2}$ can vary in space even with the anisotropy everywhere 
close to the firehose limit. The model equation does a reasonably good job at capturing the main qualitative features of interruption and 
is nearly perfect for the initial wave interruption (compare dashed lines in each panel).
It is worth noting that the relative spatial variation of $p_{\perp}$ compared to its mean (and the lack of variation in 
$p_{\parallel}$) can also be obtained by considering the compressible 
part of the CGL equations as a forced oscillator system (see \citealp{Squire:2016mri}), 
and this agrees with numerical solutions of the full equations \eqref{eq:NDMHD rho}--\eqref{eq:NDMHD pl}.

\subsection{Landau-fluid closure: Initial evolution}\label{app:initial wave}

Here, we repeat the calculation of the previous 
section but  include the Landau-fluid prescription \eqref{eq:NDMHD qp}--\eqref{eq:NDMHD ql} for the heat fluxes. These act to smooth
 pressure perturbations on the sound-crossing timescale,
leading to a pressure perturbation that is constant in space to lowest order, i.e., $\widetilde{\Delta_{2}}=0$. 
The resulting equation for $B_{x}$ is thus not accurate 
when $1+\beta_0\Delta /2 \ll 1$ (i.e., in the approach to the interruption limit),
because there is no spatially local nonlinearity to steepen the wave 
into zig-zag structures during the slow dynamics when $\beta_0\Delta /2 \approx -1$. 
As well as motivating the use of a second expansion with  $1+\beta_0\Delta /2$ ordered
small (this is done in the \ref{app:near interruption}), the expansion presented here 
is used to calculate the damping rate of a traveling wave due to 
the spatial correlation of $B^{-1}dB/dt$ and $\Delta$, as used in the arguments in Sec.~\ref{sub:noC traveling}. This is nonzero only at order $\mathcal{O}(\epsilon^{3})$, 
because $\widetilde{\Delta_{2}}=0$.

We use the same ordering as the previous section, Eqs.~\eqref{eq:Bx expansion 1}--\eqref{eq:pl expansion 1}. 
To lowest order, the heat fluxes \eqref{eq:NDMHD qp}--\eqref{eq:NDMHD ql} simplify significantly to 
\begin{gather}
q_{\perp} = \epsilon^{2}\, \pi^{-1/2}\frac{\partial_{z}}{|k_{z}|}( p_{\perp 2} - \rho_{2} ) + \epsilon^{3} \,\pi^{-1/2}\frac{\partial_{z}}{|k_{z}|}( p_{\perp 3} - \rho_{3} ) + \mathcal{O}(\epsilon^{4}),\\
q_{\parallel} = \epsilon^{2} \,2\pi^{-1/2}\frac{\partial_{z}}{|k_{z}|}( p_{\parallel 2} - \rho_{2} ) + \epsilon^{3} \,2\pi^{-1/2}\frac{\partial_{z}}{|k_{z}|}( p_{\parallel 3} - \rho_{3} ) + \mathcal{O}(\epsilon^{4}),
\end{gather}
while $\nabla \cdot (q_{\perp} \hat{\bm{b}} ) = \epsilon^{2}\partial_{z} q_{\perp,2} + \epsilon^{3}\partial_{z} q_{\perp,3} + \mathcal{O}(\epsilon^{4})$ (and similarly for $q_{\parallel}$).
These simplifications are tantamount to stating that the heat  flows along the mean field (i.e., along $\hat{\bm{z}}$) at 
the lowest two orders.  As we did for the double-adiabatic calculation in \ref{app:CGL}, let us go through each order of the
expanded equations. 

\paragraph{Order $\epsilon^{0}$.} 
This is unchanged from the double-adiabatic calculation, giving $\widetilde{p_{\parallel 2}}=0$ due to parallel pressure balance [see Eq.~\eqref{eq: O0 parr mom eq}].

\paragraph{Order $\epsilon^{1}$.} 
The perpendicular momentum equation \eqref{eq:NDMHD u} and induction equation \eqref{eq:NDMHD B}  at this order remain unchanged compared to the double-adiabatic result  [Eqs.~\eqref{eq:CGL asym wave} and \eqref{eq:CGL asym induc}]. The parallel momentum equation \eqref{eq:NDMHD u} is also unchanged, giving \begin{equation}
\widetilde{ p_{\parallel 3}}=0.\label{eq: pl3 equals 0}\end{equation}
However, due to the $\beta_0^{1/2}$ terms in Eqs.~\eqref{eq:NDMHD pp} and \eqref{eq:NDMHD pl}, there
is now a contribution at $\mathcal{O}(\epsilon)$ in the pressure equations,
\begin{gather}
\pi^{-1/2} \beta_0^{1/2}|k_{z}| (p_{\perp 2}-\rho_{2}) =0,\label{eq:lowest order p2 condition 1} \\ 2\pi^{-1/2} \beta_0^{1/2} |k_{z}| (p_{\parallel 2}-\rho_{2}) =0,\label{eq:lowest order p2 condition 2}\end{gather}
where we have used $\partial_{z}^{2}/|k_{z}| = -|k_{z}|$ to simplify the nonlocal diffusion operators. 
Combined with $\partial_{z} p_{\parallel 2}=0$ and with the $\mathcal{O}(\epsilon^{2})$ continuity equation $\partial_{t}\rho_{2}+\partial_{z}u_{z2}=0$, Eqs.~\eqref{eq:lowest order p2 condition 1}--\eqref{eq:lowest order p2 condition 2} 
imply
\begin{equation}
\widetilde{p_{\perp 2}}  =  \widetilde{\rho_{2}} = \widetilde{u_{z2}} =0,\label{eq: pressure 2 is smooth}\end{equation}
meaning that the pressure anisotropy is spatially constant to lowest order, $\widetilde{\Delta_{ 2}}=0$. This
 contrasts with the double-adiabatic result \eqref{eq:pp CGL expansion} and  formally justifies the discussions in the main text 
 regarding the smoothing effects of the heat fluxes.
 
 \paragraph{Order $\epsilon^{2}$.} 

 Again, the  momentum and induction equations \eqref{eq:NDMHD u}--\eqref{eq:NDMHD B} are not useful for our purposes at order $\mathcal{O}(\epsilon^{2})$, so we 
 consider only the pressure equations \eqref{eq:NDMHD pp}--\eqref{eq:NDMHD pl}. These are
\begin{gather}
\partial_{t}p_{\perp 2} + \partial_{z} u_{z2}  + \pi^{-1/2} \beta_0^{1/2}|k_{z}| (p_{\perp 3}-\rho_{3}) = B_{x1}\partial_{z}u_{x1},\label{eq:pp Landau expansion}\\
\partial_{t}p_{\parallel 2} + 3\, \partial_{z} u_{z2} + 2\pi^{-1/2}\beta_0^{1/2}|k_{z}| (p_{\parallel 3}-\rho_{3})= -2 B_{x1}\partial_{z}u_{x1},\label{eq:pl Landau expansion}
\end{gather}
where we also know from order $\mathcal{O}(\epsilon)$ that $\partial_{z} u_{z2} =0$, $\langle p_{\parallel 2}\rangle = p_{\parallel 2}$, and
$\langle p_{\perp 2}\rangle = p_{\perp 2}$ [Eq.~\eqref{eq: pressure 2 is smooth}]. 
Averaging Eqs.~\eqref{eq:pp Landau expansion} and \eqref{eq:pl Landau expansion} and
solving the resulting equation assuming $\Delta_{2}(t=0)=0$ gives, therefore,  
\begin{equation}
\Delta_{2} = \frac{3}{2} \langle B_{x1}(t)^{2} - B_{x1}(0)^{2} \rangle. \end{equation}
This leads to the following wave equation for $B_{x1}$:
\begin{equation}
\partial_{t}^{2} B_{x1} = \partial_{z}^{2} \left[ B_{x1}\left( 1+ \frac{3}{4}\beta_0\langle [ B_{x1}^{2} ]_{0}^{t}\rangle  \right) \right],\label{eq: final Landau1 wave eq}\end{equation}
which is the same as Eq.~\eqref{eq:waves} in the main text, with  Eq.~\eqref{eq:noC Delta} for the anisotropy
and neglecting the $\delta b^{3}$ terms that arise from the magnetic curvature.
The problems with using Eq.~\eqref{eq: final Landau1 wave eq}  to describe the wave near $1+\beta_0 \Delta/2=0$ are immediately 
apparent: no matter how small one takes $\epsilon$, there will always be a time 
for which the higher-order contributions from the magnetic curvature and  spatial 
variation of $\Delta$ play an important dynamical role. 
Indeed, numerical solutions to Eq.~\eqref{eq: final Landau1 wave eq} stay perfectly sinusoidal until 
$1+\beta_0 \Delta/2<0$, at which point small-scale (firehose) fluctuations grow rapidly. There is no tendency for the wave to become square. This issue will be resolved in \ref{app:near interruption} through the use of a different ordering scheme.

\subsubsection{Wave damping.}
A traveling wave, which satisfies $\langle B_{x}^{2} \rangle = \mathrm{const.}$,
propagates linearly, with no nonlinear modification, under Eq.~\eqref{eq: final Landau1 wave eq}. 
Although a continuation of the expansion to higher order is not very useful under this ordering, one
can obtain an estimate of the lowest-order contribution to the damping of the wave energy into thermal energy that occurs for a traveling wave due to the spatial dependence of $p_{\perp 3}$.
In the dimensionless variables \eqref{eq:dimensionless variables}, the kinetic-energy
evolution Eq.~\eqref{eq:kinetic energy} is 
\begin{equation}
\partial_{t}\langle \rho u^{2}\rangle + \partial_{t} \langle B^{2} \rangle = \beta_0 \langle p_{\parallel} \nabla \cdot \bm{u}\rangle - \beta_0 \left< \Delta \frac{1}{B}\frac{dB}{dt}\right>.\label{eq:nondim energy}\end{equation}
The right-hand side of this equation includes compressional and pressure-anisotropy heating, which can cause the 
transfer of mechanical wave energy into thermal energy. 
Using the variable expansions \eqref{eq:Bx expansion 1}--\eqref{eq:pl expansion 1} and Eq.~\eqref{eq: pressure 2 is smooth}, Eq.~\eqref{eq:nondim energy} becomes
\begin{align}
\partial_{t}\langle \rho u^{2} + B^{2}\rangle = &-  \epsilon^{2} \frac{\beta_0}{2}\Delta_{2} \partial_{t} \langle  B_{x1}^{2}\rangle \nonumber \\ &-
\epsilon^{3} \beta_0 \left[ \Delta_{2} \partial_{t} \langle  B_{x1}B_{x2}\rangle + \frac{1}{2}\langle \Delta_{3}\rangle \partial_{t} \langle  B_{x1}^{2}\rangle + \frac{1}{2}\langle \widetilde{\Delta_{3}} \partial_{t}   (B_{x1}^{2})\rangle\right] + \mathcal{O}(\epsilon^{4}),\label{eq:expanded energy conserv}\end{align}
where we have split $\Delta_{3}$ into its mean and spatially varying parts,  $\langle \Delta_{3}\rangle $ and $\widetilde{\Delta_{3}}$. 
The compressional term $\langle p_{\parallel} \nabla \cdot \bm{u}\rangle$ contributes only at order $\mathcal{O}(\epsilon^{5})$ and higher,
because $\partial_{z} p_{\parallel 2} = \partial_{z} p_{\parallel 3} = 0$ [Eq.~\eqref{eq: pl3 equals 0}] and $\partial_{z} u_{z2} = 0$ [Eq.~\eqref{eq: pressure 2 is smooth}].
The  $\mathcal{O}(\epsilon^{2})$ term on the right-hand side of Eq.~\eqref{eq:expanded energy conserv} is zero for a traveling wave, because $\partial_{t} \langle B_{x1}^{2}\rangle=0$, and 
similarly for the second of the order $\mathcal{O}(\epsilon^{3})$ terms. Since we are interested
in the damping of a pure sine wave, we shall also 
ignore the first order $\mathcal{O}(\epsilon^{3})$ term, which is related to 
the development of shape changes in $B_{x}$. This leaves 
us with $(\beta_0 /2) \langle \widetilde{\Delta_{3}} \partial_{t}   (B_{x1}^{2})\rangle$ in the right-hand side of Eq.~\eqref{eq:expanded energy conserv}. This term
describes how the average spatial correlation of $\Delta$ with $B^{-1}dB/dt$ causes a net damping, 
even when the averages of $\Delta$ and $B^{-1}dB/dt$ individually are each zero. 

To calculate $\widetilde{\Delta_{3}}$, consider the spatially varying part of  Eqs.~\eqref{eq:pp Landau expansion} and
\eqref{eq:pl Landau expansion}:
\begin{gather}
 \pi^{-1/2} \beta_0^{1/2}|k_{z}| (p_{\perp 3}-\rho_{3}) = \widetilde{B_{x1}\partial_{z}u_{x1}},\label{app: eq: p rho pert 1 1} \\ 
  2\pi^{-1/2} \beta_0^{1/2}|k_{z}| (p_{\parallel 3}-\rho_{3}) = -2\widetilde{B_{x1}\partial_{z}u_{x1}}.\label{app: eq: p rho pert 1 2}\end{gather}
Using $\widetilde{p_{\parallel3}}=0$ [Eq.~\eqref{eq: pl3 equals 0}], solving for $\rho_{3}$, and inserting this solution into the $p_{\perp}$ equation 
 gives
\begin{equation}
\widetilde{p_{\perp3}} =\widetilde{\Delta_{3}} = 2\sqrt{\frac{\pi}{\beta_0}} |k_{z}|^{-1} \widetilde{B_{x1}\partial_{z}u_{x1}}.\label{eq: D3 explicit}\end{equation}
Therefore, the third-order wave-damping rate is
\begin{equation}
\partial_{t}\langle \rho u^{2} + B^{2}\rangle = 2\sqrt{\pi } \beta_0^{1/2}\langle \partial_{t}(B_{x1}^{2}) \partial_{t}[|k_{z}|^{-1} (B_{x1}^{2})]\rangle.\label{eq:third order wave damping}\end{equation}
For a traveling sinusoidal wave  $B_{x1} = \delta b \cos( z -  t)$, by
carrying out the spatial integrations using $|k_{z}|^{-1}B_{x1}^{2}=\cos(2 z-2t)/4$, we find that the  wave energy is damped at the rate
$\partial_{t}E_{\mathrm{wave}} = \sqrt{{\pi^{3}}} \beta_0^{1/2} \delta b^{4}$,
or, restoring dimensions, \begin{equation}
\partial_{t}E_{\mathrm{wave}} = \frac{\sqrt{\pi}}{8} \omega_{A}\beta_0^{1/2} B_{0}^{2} \delta b^{4} \label{eq:quant traveling damping}\end{equation}
 in Gauss units per unit length. This
provides more formal justification (and the numerical coefficient) for the wave-damping rate [Eq.~\eqref{eq:Ewave noC}] used to derive 
kinetic and magnetic energy damping rates  [Eq.~\eqref{eq:traveling u b decay rates}] in Sec.~\ref{sub:noC traveling}.

\subsection{Landau-fluid closure: Approach to wave interruption}\label{app:near interruption}
In this section, we derive a nonlinear wave equation to describe the
final approach to wave interruption, which is not captured correctly by Eq.~\eqref{eq: final Landau1 wave eq} because it lacks a spatially dependent nonlinearity. Specifically, the distance from marginality, $1+\beta_0 \Delta/2$, is ordered 
as $\mathcal{O}(\epsilon^{2})$ even though $1$ and $\beta_0 \Delta/2$ are each $\mathcal{O}(\epsilon^{0})$.
Since the previous expansion yielded the result that $\widetilde{\Delta}=0$ to lowest order [Eq.~\eqref{eq: pressure 2 is smooth}], 
this may be considered as a re-ordering of the equations, which  becomes valid
when Eq.~\eqref{eq: final Landau1 wave eq} loses its validity because $1+\beta_0 \Delta/2 \ll 1$. 
Under the assumption of small $1+\beta_0 \Delta/2$, we are also forced to assume $u_{x}\sim \epsilon B_{x}$ and 
$\partial_{t} \sim \epsilon$, as should be expected.\footnote{The reader may notice that the 
spatially varying part of $\Delta$ was $\mathcal{O}(\epsilon^{3})$ in the expansion of the previous
section [Eq.~\ref{eq: D3 explicit}], whereas here, by assuming $1+\beta_0 \Delta/2 \sim \epsilon^{2}$ we are effectively 
ordering it to be $\mathcal{O}(\epsilon^{4})$. This apparent discrepancy is resolved by noting that 
as the solutions of Eq.~\eqref{eq: final Landau1 wave eq} evolve towards $1+\beta_0 \Delta/2=0$, the time derivatives
(or equivalently $u_{x}$)
also become one order smaller, meaning that the spatial variation of $\Delta$ is pushed into $\Delta_{4}$ (recall
that $\Delta_{3}$ was determined by the current value of $\partial_{t} B_{x1}^{2}$, not its time history).
Thus, the solutions of Eq.~\eqref{eq: final Landau1 wave eq} will evolve into 
a regime where the expansion discussed in this section in valid. } The
resulting wave equation \eqref{eq:Landau2 final equation} contains spatially dependent nonlinearities
from both the magnetic curvature and the spatial variation of $\Delta$. It thus
contains the terms necessary  to reproduce zig-zag field-line structures seen in solutions of the full LF equations (e.g., Fig.~\ref{fig:noC standing}).

\setcounter{footnote}{0}

Given these considerations, our asymptotic ordering is modified  from Eqs.~\eqref{eq:Bx expansion 1}--\eqref{eq:pl expansion 1} as follows:
\begin{gather}
\partial_{t} f \sim \epsilon f,\label{eq:second expansions 1}\\
\Delta = -\frac{2}{\beta_{0}} + \epsilon^{3}\Delta_{3} + \epsilon^{4}\Delta_{4}+  \mathcal{O}(\epsilon^{5}), \label{eq: Delta in second expansion}\\
u_{x} = \epsilon^{2} u_{x2} + \epsilon^{3}u_{x3} + \mathcal{O}(\epsilon^{4}),\\
u_{z} = \epsilon^{3} u_{z3} + \epsilon^{4} u_{z4}+ \mathcal{O}(\epsilon^{5}),\label{eq:second expansions 2}
\end{gather}
where $f$ represents any variable  [the change in the ordering of $u_{z}$ stems from the time derivative in the 
continuity equation \eqref{eq:NDMHD rho}]. Before embarking on  an order-by-order expansion, we can simplify 
our task significantly by noting that only every second term in each field expansion need be considered, viz.,
$B_{x} = \epsilon B_{x1}+ \epsilon^{3}B_{x3 }+ \mathcal{O}(\epsilon^{5})$,  $p_{\perp} =1+ \epsilon^{2} p_{\perp2}+ \epsilon^{4} p_{\perp 4} + \mathcal{O}(\epsilon^{6})$, $\Delta = -2/\beta_{0} + \epsilon^{4}\Delta_{4 } + \mathcal{O}(\epsilon^{6})$ etc., for all fields. This is justified by the fact that in all equations, the expressions for  order-$n$
quantities depend only on  order-$(n+2)$ quantities, because the  terms that relate to modifications in $B$
all contain $B_{x}^{2}$.\footnote{Note that this was not the case 
in the previous expansion \eqref{eq:Bx expansion 1}--\eqref{eq:pl expansion 1} due to the $\beta_0^{1/2}$ coefficient of the heat fluxes in the pressure equations \eqref{eq:NDMHD pp} and \eqref{eq:NDMHD pl},
whereas now there is an extra $\epsilon$ arising in the time derivatives of $p_{\perp,\parallel}$ in Eqs.~\eqref{eq:NDMHD pp} and \eqref{eq:NDMHD pl} that 
restores this property.}
We must still work through all orders of the equations in $\epsilon$, since 
some fields contain even powers of $\epsilon$ (e.g., $u_{x}$, $\Delta$)  while others contain odd powers (e.g., $B_{x}$, $u_{z}$).

\paragraph{Order $\epsilon^{0}$.} 
This is unchanged from Appendices A.2 and A.3, giving $\widetilde{p_{\parallel 2}}=0$ due to parallel pressure balance. Because $\Delta_{2}=-2/\beta_{0}$ [Eq.~\eqref{eq: Delta in second expansion}], this also implies $\widetilde{p_{\perp 2}}=0$.

\paragraph{Order $\epsilon^{1}$.}
This remains unchanged from \ref{app:initial wave}, giving $\widetilde{p_{\perp 2 }} = \widetilde{\rho_{2}} =0$ due to 
the heat fluxes in the pressure equations \eqref{eq:NDMHD pp}--\eqref{eq:NDMHD pl}. From the $\mathcal{O}(\epsilon^{3})$ part of the continuity equation \eqref{eq:NDMHD rho}, $\rho_{2}=0$  implies $\widetilde{u_{z3}} = u_{z3}=0$.
 
\paragraph{Order $\epsilon^{2}$.}
The parallel momentum equation \eqref{eq:NDMHD u} is 
\begin{equation}
-\frac{\beta_0}{2}\partial_{z}p_{\perp 4} - \frac{1}{2} \partial_{z} B_{x1}^{2} + \frac{\beta_0}{2} \partial_{z} (\hat{b}_{z}\hat{b}_{z} \Delta ) = 0.\end{equation}
Expanding $\hat{b}_{z}$ and using $\Delta = -2/\beta_0 + \mathcal{O}(\epsilon^{4})$, this leads to \begin{equation}
\partial_{z} p_{\parallel 4} = \frac{1}{\beta_0} \partial_{z} B_{x1}^{2}\label{eq: pl4 = dz B2}.\end{equation}
The perpendicular induction equation appears at this order, 
giving
\begin{equation}
\partial_{t}B_{x1} = \partial_{z}u_{x2} \label{eq: induction approach to interruption}\end{equation}
as expected.

\paragraph{Order $\epsilon^{3}$.}
The perpendicular momentum equation, which forms the basis for our wave equation, appears at $\mathcal{O}(\epsilon^{3})$ and reads 
\begin{equation}
\partial_{t} u_{x2} = \partial_{z}\left(  \frac{\beta_0}{2} \Delta_{4}B_{x1}  + B_{x1}^{3} \right),\label{eq:uperp for Landau2}\end{equation}
where we have used $1+\beta_{0} \Delta_{2}/2 = 0$ [Eq.~\eqref{eq: Delta in second expansion}].
The final term in Eq.~\eqref{eq:uperp for Landau2} arises from the $\mathcal{O}(\epsilon^{2})$ contributions to $\hat{b}_{x}\hat{b}_{z}$ from the variation in $B$, 
which is of the same origin as the $1/(1+\delta b^{2})$ in Eq.~\eqref{eq:waves}. To continue, we
 need an expression for $ {\beta_0} \Delta_{4}/2$ in Eq.~\eqref{eq:uperp for Landau2}, which requires the pressure equations \eqref{eq:NDMHD pp}--\eqref{eq:NDMHD pl}.
 
 Because we  assume $\Delta_{2} = -2/\beta_0 $,  the second-order pressures are  constant, $\partial_{t} p_{\perp 2}=\partial_{t}p_{\parallel 2}=0$, and the time derivatives first occur in the pressure equations \eqref{eq:NDMHD pp}--\eqref{eq:NDMHD pl} at order
$\mathcal{O}(\epsilon^{5})$. Noting that $\partial_{z}u_{z3}=0$, the pressure equations \eqref{eq:NDMHD pp}--\eqref{eq:NDMHD pl}
 at $\mathcal{O}(\epsilon^{3})$ are then,
 \begin{gather}
\beta_0^{1/2} \partial_{z}q_{\perp 4} = B_{x1}\partial_{z}u_{x2},\label{eq:pp for Landau2}\\
\beta_0^{1/2} \partial_{z}q_{\parallel 4} = -2 B_{x1}\partial_{z}u_{x2}.\label{eq:pl for Landau2}
\end{gather}
We can obtain useful information from both the spatial average and the spatially varying part of Eqs.~\eqref{eq:pp for Landau2} and \eqref{eq:pl for Landau2}.

A spatial average of
Eqs.~\eqref{eq:pp for Landau2} and \eqref{eq:pl for Landau2}  leads to \begin{equation}
\partial_{t} \langle B_{x1}^{2} \rangle =0 + \mathcal{O}(\epsilon^{5}),\label{eq: B2 av varies slowly}\end{equation}
 which implies that
 $\partial_{t} \langle B_{x1}^{2} \rangle \sim\epsilon^{2} \partial_{t}B_{x1}^{2}$, i.e., that the 
spatial average of $B_{x1}^{2}$ varies in time more slowly than $B_{x1}^{2}$ itself. This can occur, for instance, if  $B_{x1}$ is increasing in some region and decreasing in another, as could occur in the approach to a square wave.

The spatially varying part of Eqs.~\eqref{eq:pp for Landau2} and \eqref{eq:pl for Landau2} can be used to 
solve for $\widetilde{\Delta_{4}}$. We first require the heat fluxes \eqref{eq:NDMHD qp}--\eqref{eq:NDMHD ql}, which are,\footnote{Note that the $p_{\perp}(1-p_{\perp}/p_{\parallel})\nabla_{\parallel}B/B$ part of
$q_{\perp}$ in Eq.~\eqref{eq:NDMHD qp} has made an appearance at this order.}
\begin{gather}
q_{\perp 4} = -\pi^{-1/2} \frac{\partial_{z}}{|k_{z}|} (p_{\perp 4}-\rho_{4}) + \pi^{-1/2} \beta_0^{-1} \frac{\partial_{z}}{|k_{z}|} B_{x1}^{2},\label{app: eq: p rho q pert 2}\\
q_{\parallel 4} = -2\pi^{-1/2} \frac{\partial_{z}}{|k_{z}|} (p_{\parallel 4}-\rho_{4}).\label{app: eq: p rho q pert 3}
\end{gather}
Inserting Eqs.~\eqref{app: eq: p rho q pert 2}--\eqref{app: eq: p rho q pert 3} into Eqs.~\eqref{eq:pp for Landau2}--\eqref{eq:pl for Landau2}, using $\partial_{z} p_{\parallel 4} = {\beta_0}^{-1} \partial_{z} B_{x1}^{2}$ [Eq.~\eqref{eq: pl4 = dz B2}], then solving for $\rho_{4}$ and inserting this into the $p_{\perp 4}$ equation  yields
\begin{equation}
|k_{z}| \Delta_{4} = 2 \sqrt{\pi} \beta_{0}^{-1/2}\partial_{t}\widetilde{B_{x1}^{2}} + \beta_0^{-1} |k_{z}| \widetilde{B_{x1}^{2}}.\label{eq: final D4}\end{equation}
This may be inserted into Eqs.~\eqref{eq: induction approach to interruption} and \eqref{eq:uperp for Landau2} to yield the nonlinear wave equation 
\begin{equation}
\partial_{t}^{2} B_{x} =  \frac{\beta_0}{2} \langle \Delta_{4} \rangle \partial_{z}^{2} B_{x1}
 + \frac{3}{2}\partial_{z}^{2} (B_{x1}^{3}) - \frac{1}{2} \langle B_{x1}^{2}\rangle \partial_{z}^{2}B_{x1}+ \sqrt{\pi}\beta_0^{1/2}\partial_{z}^{2} \left[    B_{x1} \partial_{t}(|k_{z}|^{-1} B_{x1}^{2})\right].\label{eq:Landau2 final equation}\end{equation}

\begin{figure}
\begin{center}
\includegraphics[width=0.5\columnwidth]{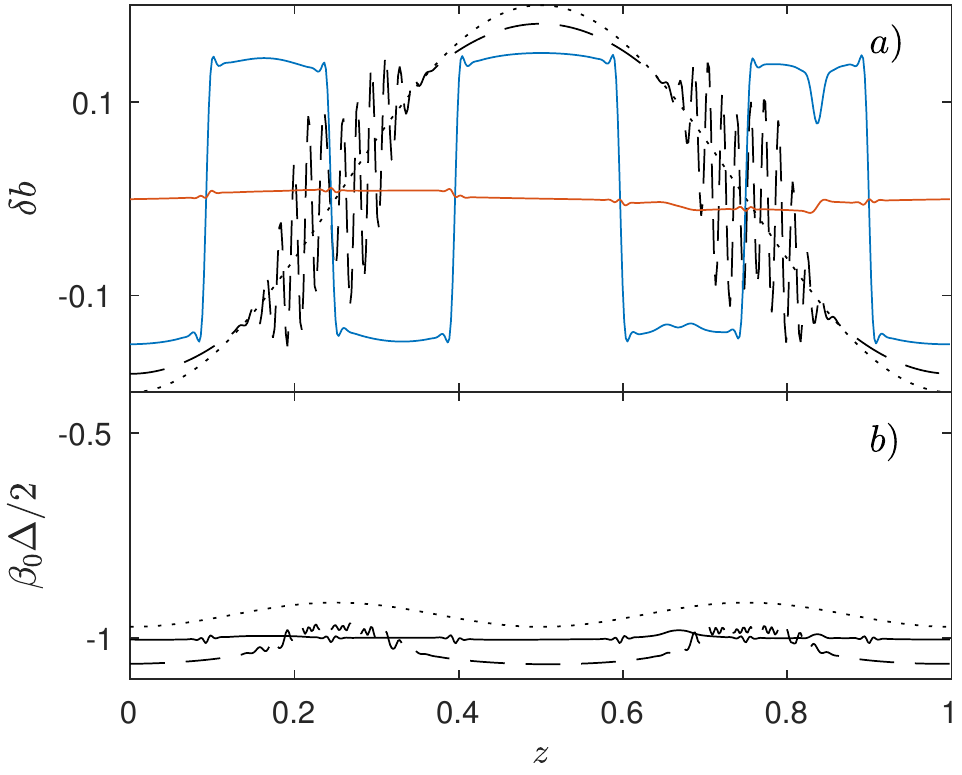}~~\includegraphics[width=0.5\columnwidth]{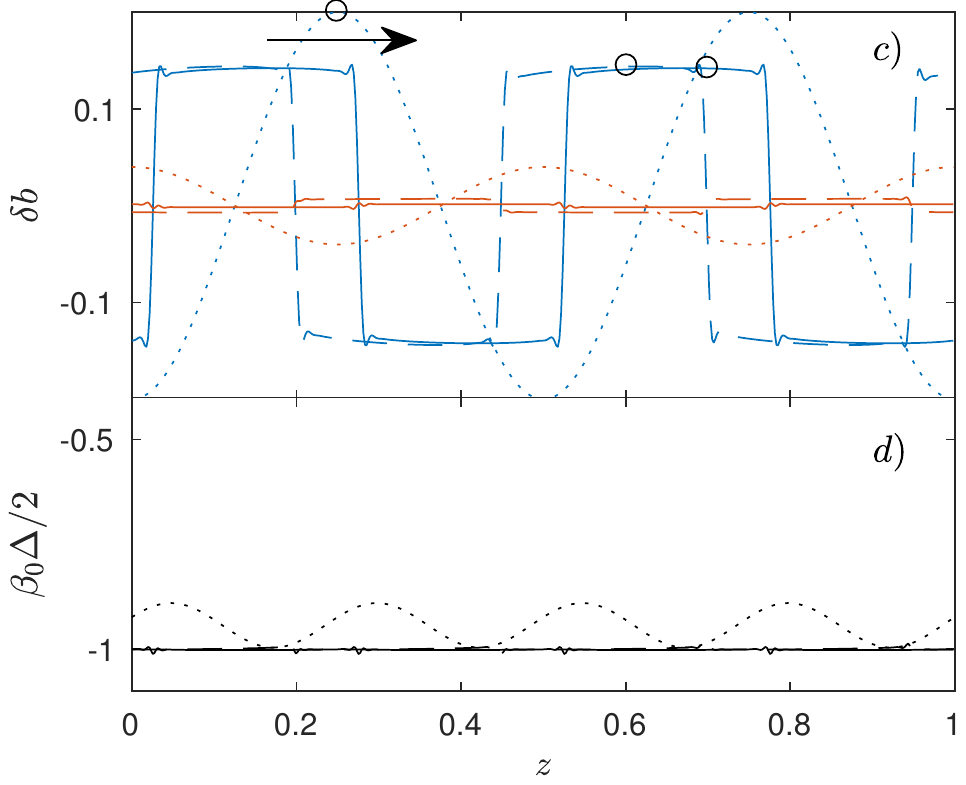}
\caption{Solution of the ``approach-to-interruption'' wave equation \eqref{eq:Landau2 final equation}, using the closure \eqref{eq:PHI ansatz} for $ \langle\Delta_{4}\rangle$, at $\beta_0=100$,
$B_{x1}=-0.2\cos(2\pi z) $, $\langle \Delta_{4}(0)\rangle = -0.2^{2}$. Panels (a)--(b) show a standing wave starting from $u_{x2} = 0.2^{2}\sin(2\pi z)$. Panel (a) shows the field and flow evolution:
 $B_{x1}$  at $t=0$ (black dotted line), $B_{x1} $ at $t=0.3\tau_{A}$ (black dashed line), $B_{x1}$ at $t=3 \tau_{A}$ (blue solid line), and $u_{x2}$ at $t=3 \tau_{A}$ (red solid line). Panel (b) shows the pressure anisotropy, $\beta_{0}\Delta/2$, at $t=0$ (dotted line), $t=0.3\tau_{A}$ (dashed line), and $t=3 \tau_{A}$ (solid line). Panels (c)--(d) illustrate a traveling wave near 
interruption, starting from $u_{x2} = 0.2^{2}\cos(2\pi z)$. Panel (c) shows $B_{x1}$ (blue lines) and $u_{x2}$ (red lines)
at  $t=0$ (dotted lines),  $t=3\tau_{A}$ (dashed lines), $t=6\tau_{A}$ (solid lines), with the circled points showing the 
same point of the wave as it travels to the right. Panel (d) shows the pressure anisotropy at  $t=0$ (dotted line),  $t=3\tau_{A}$ (dashed line), $t=6\tau_{A}$ (solid line). The dynamics are very similar to those seen in the solution of the full Landau fluid equations \eqref{eq:NDMHD rho}--\eqref{eq:NDMHD ql} (cf. Figs.~\ref{fig:noC standing}--\ref{fig:noC traveling}).}
\label{app:fig:LF asym}
\end{center}
\end{figure}

Unfortunately, Eq.~\eqref{eq:Landau2 final equation} still contains the undetermined mean pressure anisotropy $\langle\Delta_{4}\rangle  $. While in principle, one 
can solve for $\langle\Delta_{4}\rangle  $ by considering the mean part of the pressure equations \eqref{eq:NDMHD pp}--\eqref{eq:NDMHD pl}, 
the result contains $B_{x3}$, so $\langle\Delta_{4}\rangle  $ remains unknown. Of course, any attempt to subsequently 
solve for $B_{x3}$ generates dependence on $\Delta_{6}$, leading to a standard closure problem.
Despite this issue, Eq.~\eqref{eq:Landau2 final equation} remains useful 
for a number of reasons. First, the spatially dependent nonlinearities
are interesting: because of the time derivative in  $\widetilde{\Delta_{4}}$ [the first term on the right-hand side of Eq.~\eqref{eq: final D4}], this term has a diffusive  effect  in Eq.~\eqref{eq:Landau2 final equation}, and it can dissipate wave energy into thermal energy. This is not the case for the  the $\partial_{z}^{2}(B_{x1}^{3})$ nonlinearity in Eq.~\eqref{eq:Landau2 final equation},
which arises from spatial variation in the field strength. 
Secondly, the exact form of $\langle\Delta_{4}\rangle  $ plays only a minor role, because it is simply a spatially constant number that must 
decrease as $B^{2}$ decreases (since $\partial_{t}\Delta \sim \partial_{t}B^{2}$). Indeed, it has only one  property that is key to the dynamics described by 
 Eq.~\eqref{eq:Landau2 final equation} -- it must be able to approach 0 and become negative, so as to slow the linear
dynamics and allow the nonlinear terms to dominate. 
Corrections (at order $\epsilon^{4}$)  to the 
exact point at which this zero crossing occurs will presumably not  affect the 
dynamics of the wave strongly. For the purposes of exploring solutions to Eq.~\eqref{eq:Landau2 final equation} numerically (see Fig.~\ref{app:fig:LF asym}),
 we thus
make the simple ansatz 
\begin{equation}
\langle\Delta_{4}\rangle   = \langle\Delta_{4}(0)\rangle + \frac{3}{4}\beta_0 \langle(B_{x1}^{2} - B_{x1}(0)^{2})\rangle.\label{eq:PHI ansatz}\end{equation}
Here $  \langle\Delta_{4}(0)\rangle$ is the initial anisotropy, which would arise through dynamics that satisfy Eq.~\eqref{eq: final Landau1 wave eq}.
The second term comes from the form of $\Delta_{2}$ in Eq.~\eqref{eq: final Landau1 wave eq} [but it is now of order $\mathcal{O}(\epsilon^{4})$ for the same reason that we obtained Eq.~\eqref{eq: B2 av varies slowly}], which is a clear
choice that satisfies the requirements discussed above.

As shown in Fig.~\ref{app:fig:LF asym}, the nonlinear  wave equation \eqref{eq:Landau2 final equation} with $\langle\Delta_{4}\rangle$
given by Eq.~\eqref{eq:PHI ansatz} has solutions that are pleasingly similar to those of the full LF equations (cf. Figs.~\ref{fig:noC standing}--\ref{fig:noC traveling}): we see
a clear approach to zig-zag field lines, a much faster decay of the velocity in comparison to  
magnetic field, the eruption of small-scale firehose modes at early times for a standing wave, and the 
slowing of a traveling wave as the anisotropy approaches the firehose limit.


\section{Asymptotic wave equations -- Braginskii limit}\label{app:asymptotics Brag}

In this appendix, we derive asymptotic wave equations in the Braginskii limit, with $\omega_{A} \ll \nu_{c}$.
As expected, this calculation is significantly simpler than the collisionless cases discussed 
\ref{app:asymptotics}. We find two regimes,  with the dynamics controlled by the parameter $\bar{\nu}_{c}/\beta_0^{1/2}$ (recall that
$\bar{\nu}_{c} \equiv \nu_{c}/\omega_{A}$), which determines whether the effect of heat fluxes is important for the spatial form of the pressure anisotropy.  In the first regime,
when $\bar{\nu}_{c}\gg \beta_0^{1/2}$, one recovers the Braginskii wave equation discussed in the main text [Eq.~\eqref{eq:waves} 
with the closure \eqref{eq:Brag closure}]; 
in the second, when $\bar{\nu}_{c} \sim \beta_0^{1/2}$ (or $\bar{\nu}_{c}<\beta_{0}^{1/2}$), one finds an anisotropy that becomes  smoother in 
space as $\beta_0^{1/2}/\bar{\nu}_{c}$ increases. 
The magnetic field and velocity dynamics  in each regime are generally similar to each other, heuristically 
vindicating the neglect of the heat fluxes in the main text. 
Throughout this appendix, we use the  definitions and dimensionless equations described in \ref{app:equations}.

\subsection{General considerations and ordering}\label{app:Brag general considerations}

For consistency with the collisionless calculations in 
\ref{app:asymptotics}, we again use $\epsilon \sim B_{x}$, with $\delta b/\delta b_{\mathrm{max}} = \delta b\, \beta^{1/2}/\bar{\nu}_{c}^{1/2} \sim\mathcal{O} (1)$, which implies
\begin{equation}
\bar{\nu}_{c}\sim \epsilon^{2} \beta_0.\label{eq: nu = ep2 beta}\end{equation}
Combined with the fundamental  requirement for the Braginskii approximation to be 
valid $\bar{\nu}_{c}\gg 1$, 
Eq.~\eqref{eq: nu = ep2 beta} suggests $\beta_0 \sim \mathcal{O}(\epsilon^{-3})$ or larger. This  contrasts  the collisionless ordering scheme used in \ref{app:asymptotics}, where  we took $\beta_0 \sim \mathcal{O}(\epsilon^{-2})$.
Throughout this section, we shall take $u_{x}\sim B_{x}\sim \mathcal{O}(\epsilon)$, as in Appendices A.2 and A.3.

To  constrain further the ordering of $\bar{\nu}_{c}$ and $\beta_0$ individually, let us consider
the basic scaling of the heat fluxes in the Braginskii regime.
Ignoring -- for reasons that will become clear momentarily -- the effect of the 
collisionality on the heat fluxes, let us consider the pressure equations \eqref{eq:NDMHD pp} and \eqref{eq:NDMHD pl}. 
Since $\bar{\nu}_{c}$ must dominate over $\partial_{t}$, these equations, to lowest order, will 
be dominated by the terms $\beta_0^{1/2} \partial_{z} q_{\perp,\parallel}$, $B_{x}\partial_{z}u_{x}$, and 
$\bar{\nu}_{c} \Delta$. The balance between the
collisions and heat fluxes is thus controlled by $\bar{\nu}_{c}/\beta_0^{1/2}$: if $\bar{\nu}_{c}\sim \beta_0^{1/2}$,
the heat fluxes will enter at the same order as collisional isotropization in the pressure equations, while 
if $\bar{\nu}_{c}\gg \beta_0^{1/2}$, the heat fluxes will simply cause  minor
(higher-order) corrections to the spatial form of $\Delta$. 
Our final equations will thus depend on the ordering 
of $\bar{\nu}_{c}/\beta_0^{1/2}$. This leads us to two natural choices:
the ``high-collisionality regime,''
$\nu_{c} \sim \mathcal{O}(\epsilon^{4})$, $\beta_0 \sim \mathcal{O}(\epsilon^{6})$, and
and ``the moderate-collisionality regime,'' $\nu_{c} \sim \mathcal{O}(\epsilon^{2})$, $\beta_0 \sim \mathcal{O}(\epsilon^{4})$.\footnote{Note that the regime 
in which the heat fluxes dominate, viz., $\bar{\nu}_{c}\ll \beta_0^{1/2}$, will turn out to be a subset of the moderate-collisionality regime in \ref{app: brag moderate coll}. There 
is thus no need to treat it separately.}

In the discussion above, we neglected to mention collisional modifications to the heat fluxes [see Eqs.~\eqref{eq:NDMHD qp}--\eqref{eq:NDMHD ql}]. For the sake of qualitative discussion, this neglect is admissible because collisions reduce the heat fluxes at the same point, $\bar{\nu}_{c}\sim \beta_0^{1/2}$,
as the heat fluxes become subdominant to SA wave dynamics.
This can be seen from the form of the LF heat fluxes, Eqs.~\eqref{eq:NDMHD qp}--\eqref{eq:NDMHD ql}. Expanding these assuming small $p_{\parallel}$ and $\rho$ perturbations, and ignoring numerical coefficients, one finds that the heat fluxes scale as
\begin{equation}
q_{\perp,\parallel} \sim -\frac{1}{|k_{\parallel}| + \bar{\nu}_{c}\beta_{0}^{-1/2}} \nabla_{\parallel}\left(\frac{p_{\perp,\parallel}}{\rho}\right).
\end{equation}
Combined with the discussion of the previous paragraph, this 
shows that in the limit $\bar{\nu}_{c}\gg \beta_0^{1/2}$, where 
the heat fluxes were not important in the pressure equation, the 
heat fluxes are even further reduced, making  their neglect \emph{more}
valid than it would otherwise be. In contrast, when $\bar{\nu}_{c}\sim \beta_0^{1/2}$,
the heat fluxes are only moderately affected by collisionality. 
Thus, the effect of collisionality on the heat fluxes is simply 
to improve the validity of the high-collisionality ordering, while changing 
the moderate-collisionality results by  $\mathcal{O}(1)$ numerical factors.

\subsection{High-collisionality limit}

We now consider the high-collisionality ordering, which is 
\begin{gather}
B_{x}\sim \mathcal{O}(\epsilon),\quad u_{x}\sim \mathcal{O}(\epsilon), \quad \bar{\nu}_{c} \sim \mathcal{O}(\epsilon^{-4}),\quad {\beta_0} \sim \mathcal{O}(\epsilon^{-6}),\nonumber \\ \quad p_{\perp}\sim p_{\parallel}\sim \rho\sim 1+ \mathcal{O}(\epsilon^{6}),\quad u_{z}\sim \mathcal{O}(\epsilon^{6}),\end{gather}
where the ordering of $p_{\perp,\parallel}$ is taken from the requirement that $\beta_0 \Delta\sim 1$
at the  lowest order [or equivalently, from $B_{x}\partial_{z}u_{x}\sim \bar{\nu}_{c}  \Delta$ in the pressure equations \eqref{eq:NDMHD pp}--\eqref{eq:NDMHD pl}].

The $\mathcal{O}(1)$ and  $\mathcal{O}(\epsilon)$ equations under this ordering are effectively the 
same as in the collisionless calculations in \ref{app:CGL} and \ref{app:initial wave} -- 
the parallel momentum equation \eqref{eq:NDMHD u} at $\mathcal{O}(1)$ gives \begin{equation}
\partial_{z}p_{\parallel 6}=0,\end{equation}
while 
the perpendicular momentum equation \eqref{eq:NDMHD u} at $\mathcal{O}(\epsilon)$ gives 
\begin{equation}
\partial_{t} u_{x1} = \partial_{z} \left[B_{x1}\left( 1+ \frac{\beta_0}{2}\Delta_{6} \right)\right].\label{eq:app Brag mom eq}\end{equation}
At order $\mathcal{O}(\epsilon^{2})$, we need to consider only the pressure 
equations \eqref{eq:NDMHD pp}--\eqref{eq:NDMHD pl}, which both give the same result, 
\begin{equation}
\bar{\nu}_{c}\Delta_{6} = B_{x1}\partial_{z} u_{x1}.\label{eq:app Brag Delta 1}\end{equation}
Inserted into Eq.~\eqref{eq:app Brag mom eq}, Eq.~\eqref{eq:app Brag Delta 1}
leads to the expected wave equation
\begin{equation}
\partial_{t}^{2} B_{x1}= \partial_{z}^{2}\left[B_{x1}\left( 1+ \frac{\beta_0}{4\bar{\nu}_{c}} \partial_{t} B_{x1}^{2} \right)\right],\label{eq:Brag high coll}\end{equation}
which, as expected, is identical to the wave equation with studied in the text with a Braginskii closure, viz., Eq.~\eqref{eq:waves} with
the closure \eqref{eq:Brag closure} [aside from the neglect of the $\mathcal{O}(\epsilon^{3})$ correction $1/(1+\delta b^{2})$ 
due to field-strength variation in space].

\subsection{Moderate-collisionality limit}\label{app: brag moderate coll}

As discussed in \ref{app:Brag general considerations}, the most natural moderate-collisionality ($\bar{\nu}_{c}\sim \beta_0^{1/2}$) ordering is 
\begin{gather}
B_{x}\sim \mathcal{O}(\epsilon),\quad u_{x}\sim \mathcal{O}(\epsilon), \quad \bar{\nu}_{c} \sim \mathcal{O}(\epsilon^{-2}),\quad {\beta_0} \sim \mathcal{O}(\epsilon^{-4}), \nonumber\\ \quad p_{\perp}\sim p_{\parallel}\sim \rho\sim 1+ \mathcal{O}(\epsilon^{4}),\quad u_{z}\sim \mathcal{O}(\epsilon^{4}).\end{gather}
Again, the parallel momentum equation \eqref{eq:NDMHD u} leads to $\partial_{z} p_{\parallel 4}=0$ at order $\mathcal{O}(1)$, while the perpendicular momentum equation  \eqref{eq:NDMHD u} at order $\mathcal{O}(\epsilon)$ leads to Eq.~\eqref{eq:app Brag mom eq} with $\Delta_{6}$ replaced by $\Delta_{4}$.
At order $\mathcal{O}(\epsilon^{2})$, the pressure equations contain both the heat fluxes and  
collisional relaxation:
\begin{gather}
\sqrt{\frac{\beta_0}{\pi}}\frac{-\partial_{z}^{2}}{|k_{z}| + a_{\perp}\bar{\nu}_{c}\beta_{0}^{-1/2}}(p_{\perp 4}-\rho_{4}) = B_{x1}\partial_{z}u_{x1} - \bar{\nu}_{c}\Delta_{4},\\
2\sqrt{\frac{\beta_0}{\pi}}\frac{-\partial_{z}^{2}}{|k_{z}| + a_{\parallel}\bar{\nu}_{c}\beta_{0}^{-1/2}}(p_{\parallel 4}-\rho_{4}) = -2 B_{x1}\partial_{z}u_{x1} + 2\bar{\nu}_{c}\Delta_{4},\end{gather}
where the coefficients $a_{\perp}=\pi^{-1/2}$ and $a_{\parallel} = (3\pi/2-4)\pi^{-1/2}$ account for the difference between the collisional parallel and perpendicular heat fluxes \citep{Catto:2004}.
Averaging these equations over space gives 
\begin{equation}
\langle \Delta_{4} \rangle = \bar{\nu}_{c}^{-1}\langle B_{x1}\partial_{z}u_{x1}  \rangle,\label{eq: brag D4 av}\end{equation}
which is also true in the high-collisionality regime [for $\langle\Delta_{6} \rangle$; see Eq.~\eqref{eq:app Brag Delta 1}].
Noting that the parallel momentum equation \eqref{eq:NDMHD u} at $\mathcal{O}(1)$ gives $\widetilde{p_{\parallel 4}}=0$, and that $\widetilde{\Delta_{4}} = \widetilde{p_{\perp 4}}$, we can 
solve for $|k_{z}|\rho_{4}$  to obtain 
\begin{equation}
\widetilde{\Delta_{4}} =  \frac{\Theta \widetilde{B_{x1}\partial_{z}u_{x1}} }{\beta_0^{1/2}\zeta(\bar{\nu}_{c}) |k_{z}| +  \bar{\nu}_{c} \Theta}.\label{eq: Brag D4}\end{equation}
Here  the operator \begin{equation}
\zeta(\bar{\nu}_{c}) = \frac{\pi^{-1/2}}{1+|k_{z}|^{-1}a_{\perp}\bar{\nu}_{c}\beta_{0}^{-1/2}}\end{equation}
 encapsulates the collisional quenching of the heat fluxes, and its effect changes from being a multiplication by $\pi^{-1/2}$ at $\bar{\nu}_{c} \ll \beta_0^{1/2}$, to an operator $\sim \beta^{1/2}\bar{\nu_{c}}^{-1} |k_{z}|$ at $\bar{\nu}_{c} \gg \beta_0^{1/2}$ (but in this limit, this term may be neglected in comparison to $\bar{\nu}_{c}\Theta$). Similarly, the operator \begin{equation}
\Theta = 1+ \frac{|k_{z}| + a_{\parallel}\bar{\nu}_{c}\beta_{0}^{-1/2}}{|k_{z}| + a_{\perp}\bar{\nu}_{c}\beta_{0}^{-1/2}}\end{equation}
is effectively a multiplication by a factor between $2$ (for $\bar{\nu}_{c} \ll \beta_0^{1/2}$) and $\approx 1.71$  (for $\bar{\nu}_{c} \gg \beta_0^{1/2}$), which is necessary due to the numerical difference between perpendicular and parallel collisional heat fluxes.

Put together, Eqs.~\eqref{eq: brag D4 av} and \eqref{eq: Brag D4}, along with the perpendicular momentum equation \eqref{eq:app Brag mom eq}, lead to the wave equation
\begin{equation}
\partial_{t}^{2} B_{x1} = \partial_{z}^{2}\left[ B_{x1}+ B_{x1}\frac{\beta_0}{4 \bar{\nu}_{c}}\left( \partial_{t}\langle B_{x1}^{2} \rangle + \frac{{\bar{\nu}_{c}} \Theta\, }{\zeta(\bar{\nu}_{c})\beta_0^{1/2} |k_{z}| + {\bar{\nu}_{c}} \Theta}\partial_{t} \widetilde{B_{x1}^{2}}  \right) \right].\label{eq:Brag mod coll}\end{equation}
Evidently, this equation includes the high-collisionality limit, Eq.~\eqref{eq:Brag high coll}, when $\bar{\nu}_{c} \gg \beta_0^{1/2}$, viz., Eq.~\eqref{eq:Brag high coll} is the $\bar{\nu}_{c} \gg \beta_0^{1/2}$ limit of Eq.~\eqref{eq:Brag mod coll}. Eq.~\eqref{eq:Brag mod coll} also captures a Braginskii version of the ``spatially-constant-$\Delta$'' limit when $\bar{\nu}_{c} \ll \beta_0^{1/2}$.

\begin{figure}
\begin{center}
\includegraphics[width=0.5\columnwidth]{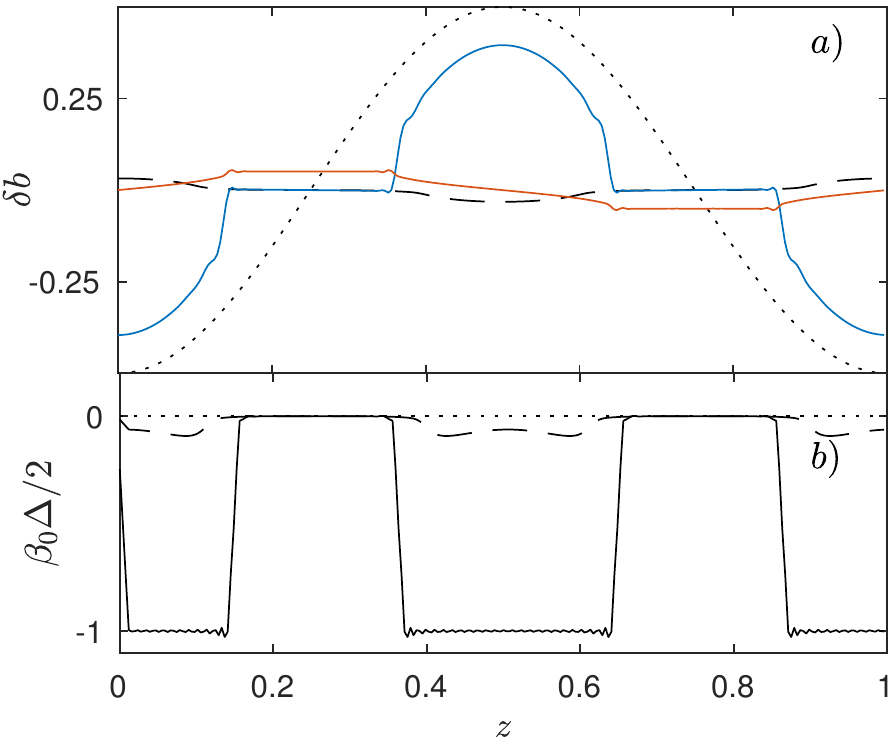}~~\includegraphics[width=0.5\columnwidth]{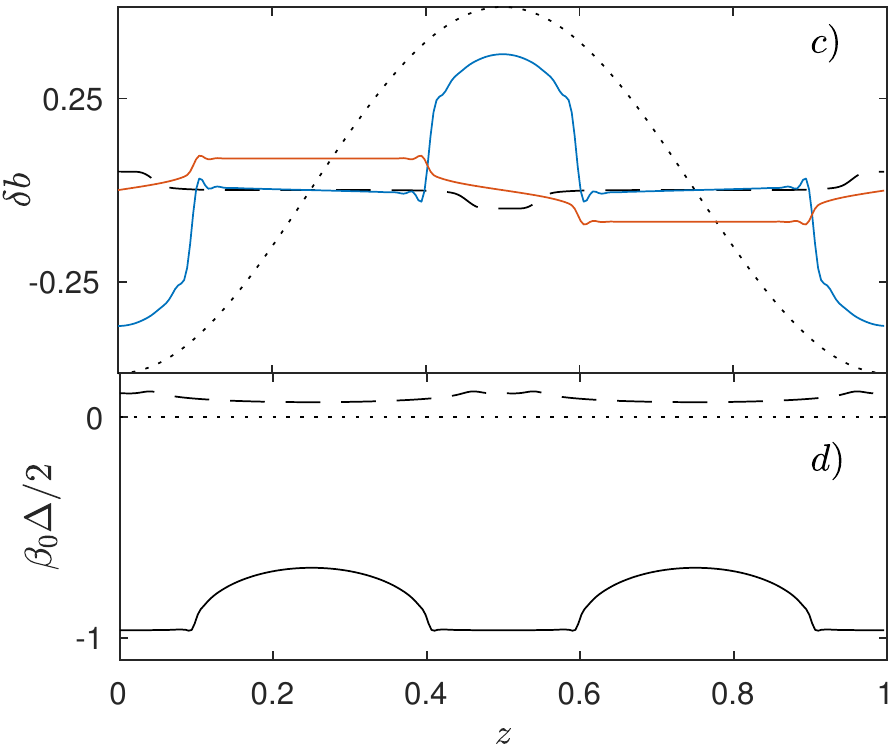}
\caption{Evolution of the Braginskii wave equations \eqref{eq:Brag high coll} and \eqref{eq:Brag mod coll}, 
starting from a sinusoidal magnetic perturbation $B_{x1}=-0.5 \cos(2\pi z)$. In the high-collisionality case shown in panels (a)--(b)
we take $\bar{\nu}_{c}=5^{4}=625$, $\beta = 5^{6}=15625$. In the moderate-collisionality case shown in panel (c)--(d) we 
take $\bar{\nu}_{c}=5^{2}=25$, $\beta = 5^{4}=625$, such that  $\bar{\nu}_{c} = \beta^{1/2}$ (we take $\zeta(\bar{\nu}_{c})=(2\pi)^{-1/2}$ for simplicity). These parameters give the same interruption limit $\sqrt{\bar{\nu}_{c}/\beta} = 0.2$ in both 
cases. In the top panels, we show  $B_{x1}$ at $t=0$ (dotted black line), $B_{x1}$ at $t=0.6 \tau_{A}$ (solid blue line), 
$u_{x1}$ at $t=0.6 \tau_{A}$ (solid red line), and 
$B_{x1}$ at $t=2 \tau_{A}$ (dashed black line), which is after the 
wave has decayed to below the interruption limit. In the bottom panels, 
we show the pressure anisotropy, $\beta_{0}\Delta/2$, at the same times. Although 
the pressure anisotropy profiles are quite different in each case [compare panels (b) and (d)], the dynamics 
of the magnetic perturbation, including the time taken for the wave to decay, are quite similar. This is because
the parts of the wave where $\Delta > -2/\beta$ have $B_{x1}=0$ anyway (see discussion in text).
Note that the sole difference between the calculation shown in Fig.~\ref{fig:Brag standing} and that in panel (a) here is 
the $(1+\delta b^{2})^{-1}$ field nonlinearity term, which is not included here because it is at higher asymptotic order. 
Because the spatially varying nonlinearity due to $\Delta p$ is  larger than that due to $\delta b^{2}$, this term makes
little difference to the dynamics.}
\label{fig:app:Brag solutions}
\end{center}
\end{figure}
Solutions to Eqs.~\eqref{eq:Brag high coll} and \eqref{eq:Brag mod coll} (with $\beta_0^{1/2}=\bar{\nu}_{c}$, and taking $\Theta=2$ and $\zeta(\bar{\nu}_{c})=1$ for simplicity)  are illustrated in Fig.~\ref{fig:app:Brag solutions}.
It is interesting that even with $\beta_0^{1/2}=\bar{\nu}_{c}$  [Fig.~\ref{fig:app:Brag solutions}(b)], when  the heat fluxes 
significantly modify the pressure anisotropy, the dynamics are largely similar 
to the basic high-collisionality Braginskii limit discussed in the main text  [Fig.~\ref{fig:app:Brag solutions}(a); see  also Fig.~\ref{fig:Brag standing}].
The reason is for this is related to the nature of the Braginskii wave decay, as discussed in Sec.~\ref{sec:Braginskii}. Effectively,
the dynamics of the decaying wave separate into regions where $\Delta = -2/\beta_0$ and the field has
curvature, and regions where $\Delta > -2/\beta_0$ and the field has no curvature (i.e., where the perturbed  field $\delta b$ is zero). The primary effect of the  heat fluxes is 
thus to decrease the anisotropy where  $B_{x1}$ is already zero anyway, causing only small modifications to the dynamics of the
standing wave. 
It is worth noting, however,  that because $\Delta p$ is smoothed 
more by the heat fluxes as  $\nu_{c}/\beta_{0}^{1/2}$ decreases [i.e., the ratio of $\widetilde{\Delta_{4}}$ \eqref{eq: Brag D4} to $\langle \Delta_{4}\rangle$ \eqref{eq: brag D4 av} decreases as  $\nu_{c}/\beta_{0}^{1/2}$ decreases], 
the decay rate of a Braginskii traveling wave will be reduced by a factor between $1$ and $\beta_0^{1/2}$ [compared to the estimate in the main text; see Eq.~\eqref{eq:brag wave damping}], when the heat fluxes cause significant smoothing of the pressure anisotropy.

\newcommand{\newblock}{}

\end{document}